\documentstyle [12pt,epsfig]{article} 
\makeatletter
\@addtoreset{equation}{section}
\makeatother
\renewcommand{\theequation}{\thesection.\arabic{equation}}
\newcommand{\ba}{\begin{eqnarray}}
\newcommand{\ea}{\end{eqnarray}}

 \newcommand{\C}{{\rm C}}
\newcommand{\Cp}{{\rm C'}}

\begin{document}
\newcommand{\BS}{\bigskip}
\newcommand{\SECTION}[1]{\BS{\large\section{\bf #1}}}
\newcommand{\SUBSECTION}[1]{\BS{\large\subsection{\bf #1}}}
\newcommand{\SUBSUBSECTION}[1]{\BS{\large\subsubsection{\bf #1}}}

\begin{titlepage}
\begin{center}
\vspace*{2cm}
{\large \bf The physics of space and time II: A reassessment of Einstein's
 1905 special relativity paper}  
\vspace*{1.5cm}
\end{center}
\begin{center}
{\bf J.H.Field }
\end{center}
\begin{center}
{ 
D\'{e}partement de Physique Nucl\'{e}aire et Corpusculaire
 Universit\'{e} de Gen\`{e}ve . 24, quai Ernest-Ansermet
 CH-1211 Gen\`{e}ve 4.
}
\newline
\newline
   E-mail: john.field@cern.ch
\end{center}
\vspace*{2cm}
\begin{abstract}
  A detailed re-examination of the seminal paper on special relativity,
  taking into account recent work on the physical interpretation of the 
  space-time Lorentz transformation as well as the modern understanding
  of classical electromagnetism as a certain limit of the fundamental
  underlying theory ---quantum electrodynamics--- is presented. Many 
   errors both of physical principle and of a mathematical nature are 
   uncovered. The `relativity of simultaneity' and `length contraction'
   effects predicted in the paper are shown to be the spurious consequences of misinterpretations
   of the second postulate and the  Lorentz transformation, respectively.
   The derivation of the latter in the paper is shown to be flawed. In this
   case, and other instances, due to cancellation of mistakes, a correct result
   is obtained in a fallacious manner. Separate lists of the correct and 
    incorrect predictions of the paper are given. Due to the  unique and revolutionary
    nature of its epistemological approach (the use of `thought experiments' and axiomatic
    derivations) and the experimentally verified predictions of time-dilation and
     mass-energy equivalence the fundamental importance of the paper for the
    development of physics is little affected by the many mathematical errors
    and (to modern eyes) physical misconceptions that it contains.

\vspace*{1cm}
\end{abstract}
\end{titlepage}

\SECTION{\bf{Introduction}}
    The present paper is intended as a companion to another paper on space-time
  physics~\cite{JHFST1} by the present author where it is attempted to construct
 the theory of flat~\footnote{That is, for cases in which all gravitational effects
  may be neglected.} space-time on the basis of simpler and more evident axioms than hitherto.
  When this is done, it is seen that certain predictions of standard Special Relativity
  Theory (SRT) --- in particular `relativity of simultaneity' of two supposedly synchronised,
  spatially separated, clocks in the same inertial frame and the associated `length
   contraction' effect, derived by Einstein as a consequence of the Lorentz
    Transformation, in the seminal 1905 paper on SRT~\cite{Ein1}, are spurious
   \footnote{It is important to notice that the relativity of simultaneity effect discussed by
    Einstein in Ref.~\cite{Ein1} and also a popular book.~\cite{EinSGR}, although,
     as will be shown below in Section 4, spurious, is not
    the same as the `text book' relativity of simultaneity effect derived from 
     a misapplication of the space-time Lorentz transformation, described in Section 7 below.
     The source of the relativity of simultaneity effect of Refs.~\cite{Ein1,EinSGR} is more a 
     consequence of a wrong understanding of the operational meaning of Einstein's second
    postulate of SRT, than of that of the  Lorentz transformation.}. They
 result from confusion between arbitary
   clock offsets, completely controlled by the experimenter, and physical time
   intervals observed in different frames, completely controlled by the Lorentz
   transformation. As well as in Ref.~\cite{JHFST1}, how this come about is explained
   in the papers~\cite{JHFUMC,JHFCRCS,JHFLLT}.
    It is also explained,
   in the present paper, in the commentary on Einstein's analysis of the thought
   experiments where such effects were first introduced.
   \par The motivation for writing the present paper is quite clear. As it is claimed
    that some results of Einstein's original 1905 paper, which have since
   been propagated unchanged through several generations of text books, as well as the 
    pedagogical literature~\footnote{Most of this has appeared in two journals:
    The American Journal of Physics and the European Journal of Physics.} during the
  last century, are wrong, it is mandatory to explain just where, and for what reason,
  these conclusions of Einstein are incorrect. 
  \par The corrections to the interpretation of SRT result also in many changes in text-book
   formulae of classical elecromagnetism, all dating from well before the advent of SRT.
   These formulae, many of which occur in Einstein's original SRT paper, are referred to 
    below under the acronym `CEM' for Classical Electro-Magnetism, to be distinguished
   from the recent classical electrodynamic theory~\cite{JHFRCED}, of the present
    author, which is fully consistent with both SRT and, in the appropriate limit, with
   Quantum Electro-Dynamics (QED). This theory is referred to below as `RCED' for
   Relativistic Classical Electro-Dynamics. The salient points concerning RCED are 
    summarised below in the present section, while some detailed differences
    with CEM will become evident to the reader in the discussion of the
    `Electrodynamical Part' of Ref.~\cite{Ein1}. These differences, as well as others, are
     also discussed in Refs.~\cite{JHFRSKO,JHFIND,JHFSTF}.
     \par The century since Einstein's original paper on special relativity
     was written has seen an enormous increase in the understanding of physics.
     Perhaps the qualitatively most important discovery of 20th Century physics,
     and the one with the most important practical ramifications was that matter (and
     anti-matter) can be created and destroyed. The realisation that this was possible followed 
     directly
     from the concept of the equivalence of mass and energy introduced in Ref.~\cite{Ein1}.
     The matter that is created or destroyed consists of elementary particles, the detailed
     properties of which have been revealed by a fruitful symbiosis between experiment and
     theory in the domain of `High Energy Physics' in the second half of the 20th Century.
     By a historical quirk, the theory that describes the creation and destruction of
     elementary particles is called `Quantum Field Theory' (QFT) to distinguish  it from
     `Quantum Mechanics'(QM) in which matter exists permanently.
     \par Indeed, as formulated by Feynman, the original and most successful QFT,
     QED~\cite{Kinoshita} born of the marriage of QM with SRT, and describing the creation
      and destruction of real or virtual photons and electrically charged elementary
      particles, has, as its fundamental concepts, not `fields' but the particles
      (which are what exist) and probability amplitudes that describe the histories,
       in space time, of these existing particles. These are both minimal 
     and sufficient concepts to obtain all predictions of QED.  
      \par The most revolutionary conceptual discovery of 20th Century physics, apart from the
       modification of the concept of time required by SRT, at the basis of
       relativistic kinematics, is that of QM and QFT. Our best
       current fundamental knowledge is that the world is described in a probabilisitic
       fashion by these theories. The `classical' physics of the 19th Century and earlier ones
       is given by certain limits of this quantum description, typically those where
       the number of elementary particles involved is very large or where the kinematical
       properties of objects are such that their corresponding mechanical action is
      much larger than Planck's constant $h$.
       \par The above preamble is relevant to the subsequent discussion of Ref.~\cite{Ein1}
      because of the crucial role of CEM in Einstein's formulation of SRT, in the paper
      with the title `On the Electrodynamics of Moving Bodies'.
      Much of the material in even modern text books on CEM is still presented in terms
       of 19th Century concepts, and, as already mentioned, many standard formulae of the
      subject pre-date the
       discovery of SRT. It will be found that several of the electrodynamic problems
        discussed by Einstein in~\cite{Ein1} can be most easily (and
        always, unlike in some of Einstein's analyses, correctly)
       analysed in the language of `CEM as a limit of QED'. i.e. by introducing
      the photon concept, invented in the same year in an earlier paper~\cite{Ein2} by Einstein
      as the `light quantum', but never used in the SRT paper.
      \par In a series of related papers, the present author has explored two related topics:
      \begin{itemize}
      \item[(i)] What can knowledge of CEM teach about QM?
       \item[(ii)]What can knowledge of QED teach about CEM?
      \end{itemize}
      The question (i) was addressed in a paper in which the comparison by `inverse
     correspondence' of a plane, monochromatic, electromagnetic wave with the equivalent
     mono-energetic, parallel, beam of real photons, provides a simple understanding of many key
     concepts of QM~\cite{JHFEJP}. The question (ii) was considered for the particular
     case of inter-charge forces, mediated by the exchange of virtual photons, in the
     absence of real photon radiation, in Ref.~\cite{JHFRCED}.
     \par The elementary QED process underlying, say the magnetic force between two
       current-carrying conductors, is M\o ller scattering : $e^-e^- \rightarrow e^-e^-$
       It is then assumed that the laws of physics do not change between the situation
      in which just two electrons scatter from each other, or when the conduction 
       electrons in one wire interact with those in another.
       This can be considered as an application of Newton's third rule of reasoning
       in philosophy~\cite{Newton}:
       \par `The qualitities of bodies, which admit neither intension nor remission
       of degrees, and which are found to belong to all bodies within the reach 
        of our experiments are to estimated as the universal properties of all
              bodies whatsoever.'
       \par In the present case, the word `bodies' in this statement 
           should be replaced by `electrons'. Two electrons in free space,
        whose fundamental mutual interaction
        is described by  the M\o ller scattering amplitude,  are the same as, and have the same 
        mutual interaction as, the conduction electrons in a wire. There is no reason to believe
       that this interaction is modified, in any way, when the interacting electrons are separated
       by macroscopic distances.  
      At lowest order,
       the QED momentum-space invariant amplitude for M\o ller scattering
      involves the exchange of a single, space-like, virtual photon. Fourier
       transforming this amplitude, expressed in terms of momenta in the overall
       centre-of-mass (CM) system, into space time, demonstrates the instantaeous, not
       retarded, nature of the inter-charge interaction~\cite{JHFRCED}. This is also
       evident from the relativistic velocity formula $v = p c^2/E$ to be discussed 
       below in Section 13. In the CM frame the energy, but not the momentum, of the
      virtual photon vanishes so that its velocity, $v$, is infinite. A recent experiment measuring
      the distance-dependence of magnetic induction has verified experimentally
      this prediction~\cite{JAP1,JHFAD}.
      \par If a force is transmitted by istotropic emission and subsequent 
       absorption of particles, the inverse square law is a simple consequence
      of conservation of the number of transmitted particles between the
      processes of emission and absorption. For large, space-like, intervals
      $\Delta s$, the Lorentz-invariant Feynman propagator of the corresponding space-like virtual
      particle, of pole mass $m$, is $\simeq \exp[-im\Delta \tau /\hbar]$~\cite{FeynIP}
      where $\Delta \tau \equiv = -i \sqrt{(\Delta x)^2-c^2(\Delta t)^2} = -i\Delta s$.
       Therefore, for space-like intervals with $\Delta x > \Delta t$, the 
      propagator does not vanish as predicted by `causality' (see Section 13 below)
       but is instead exponentially damped $\simeq \exp[-m \Delta s /\hbar]$. 
       For photons (or any other massless particle) for which $m = 0$ this damping
       does not occur, so that conservation of the number of virtual particles
       predicts that the corresponding exchange force law is inverse square. In this
      way, both Coulomb's inverse square law and the instantanteous nature of the intercharge force are
      first-principle predictions of QED.

\begin{figure}[htbc]
\begin{center}\hspace*{-0.5cm}\mbox{
\epsfysize15.0cm\epsffile{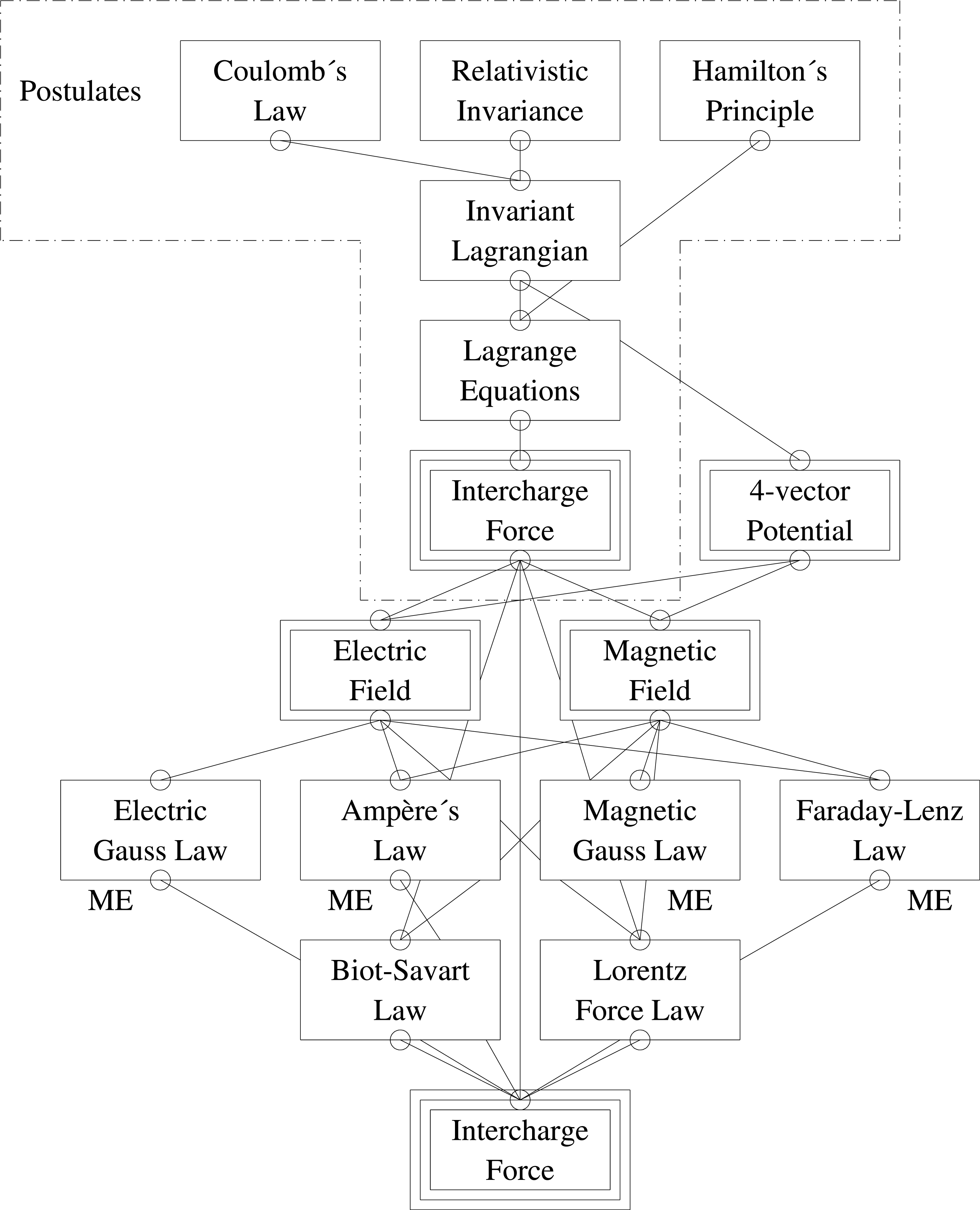}}
\caption{{\em Concept flow diagram of RCED. The steps in the derivation of the intercharge
 force equation are inside the dot-dashed line. Downwards-directed lines indicate concept transfer or
  derivation, e.g. in the case of the 4-vector potential or electric or magnetic fields, by mathematical
  substitution in a previously derived equation. The `Intercharge Force' is given by Eq.(3.4) below.
  Maxwell's Equations, which are all derived, to lowest order in $\beta$, in this approach
   are denoted by {\it ME}. The lower part of the figure, outside the dot-dashed line,
   where the `Intercharge Force' is obtained from Maxwell's Equations and/or the Biot-Savart and
   Lorentz Force Laws, corresponds  closely with the concept flow of 19th Century CEM shown in Fig.2.
 }}
\label{fig-fig1}
\end{center}
\end{figure}

\begin{figure}[htbc]
\begin{center}\hspace*{-0.5cm}\mbox{
\epsfysize10.0cm\epsffile{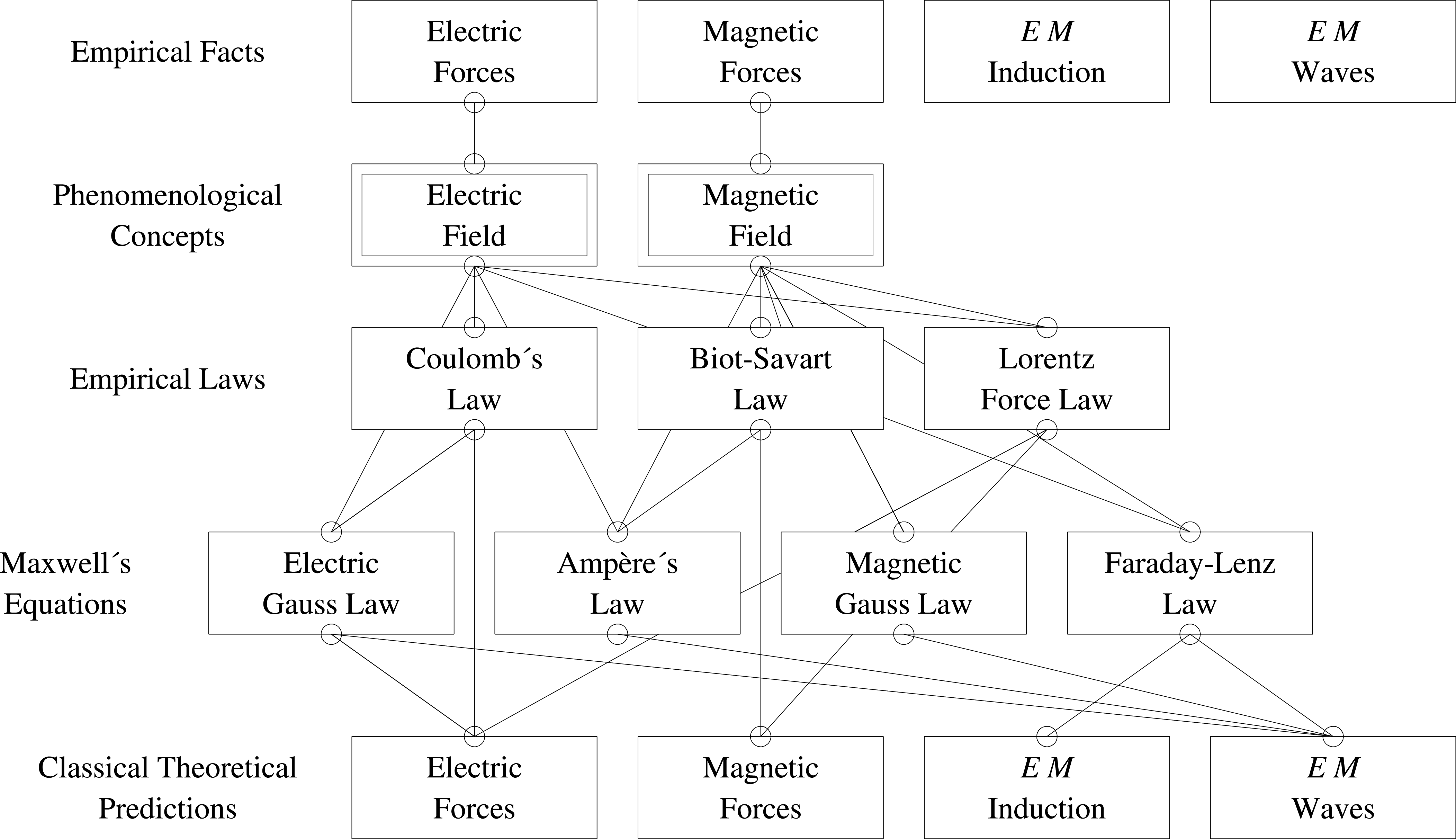}}
\caption{{\em Concept flow diagram of the 19th Century development of CEM. 
 All `Empirical Laws' and `Classical Theoretical Predictions', with the exception of `{\it EM} waves',
 are included in the single box  `Intercharge Force' of Fig.1.}}
\label{fig-fig2}
\end{center}
\end{figure}

      \par  An explicit relativistic formula for the inter-charge force in RCED ({\rm E}qn(3.4) below) has
        been obtained in
       Ref.~\cite{JHFRCED} from the three postulates: an instantaneous Coulomb Law 
       force in the non-relativistic limit, special relativistic invariance and Hamilton's
      Principle. A Lorentz-invariant Lagrangian for two interacting charges is 
      constructed from the space-time and velocity 4-vectors of the two charges
      by imposing consistency with the well-known Lagrangian for the two-body 
       central-force problem in the non-relativistic limit~\cite{Goldstein}. 
       A logical `concept-flow' diagram of the derivation is shown in Fig.1.
       All essential physics is contained in the six upper boxes, surrounded by the
       dot-dashed line. It is obtained without the introduction of any
       `force' or `field' concept. However, as shown by the boxes outside the
       dot-dashed line, all the concepts and laws of CEM developed during the
       19th Century are derived by simple mathematical substitutions, either
       in the invariant Lagrangian or in the intercharge force equation. It is
      interesting to compare Fig.1 with the concept-flow diagram of Fig.2 which
      shows the logical development of the ideas of CEM during the 19th Century. 
       The theoretical starting points here are the phenomenological concepts of electric
      and magnetic fields, introduced by Faraday and Maxwell and operationally defined
      in terms of forces exerted on a test charge. These are the concepts in terms
      of which Einstein originally formulated SRT. Note the similarity between Fig.2
      and the bottom half, below the dot-dashed line, of Fig.1. However all the
    `Empirical Laws' and `Classical Theoretical Predictions' of Fig.2, with the 
     exception of `{\it E M} waves', are included in the single box `Intercharge Force'
     of Fig.1.
     \par The RCED predictions of Ref.~\cite{JHFRCED} for the electric field
    of a uniformly moving charge and its associated magnetic field, Eqs.~(9.19) 
    and (9.20) below,
    do not agree with the
    pre-relativistic Heaviside formulae~\cite{Heaviside}.
     This has important consequences for the discussion of
    the formulae in the `Electrodynamical Part' of Ref~\cite{Ein1}. As will be discussed
    in Section 3 below, the Heaviside formula has been demonstrated to be 
    erroneous by calculating the electromagnetic induction effect
    produced by a simple two-charge `magnet' in different reference frames~\cite{JHFIND}.
     Also, in Ref.~\cite{JHFSTF}, it is shown that there is an important mathematical
     error in the derivation if the retarded Li\'{e}nard-Wiechert~\cite{LW} potentials 
     from which the `present time' Heaviside formula for the retarded fields may
    be obtained. That is, it is not only that the intercharge force is instantaneous, not
     retarded, but also that the Heaviside formula does not even correctly describe a hypothetical
    retarded intercharge force. Correct formulae for the retarded fields of a uniformly-moving
   charge in the `present time' form, directly comparable with the Heaviside formula may be
    found in Ref.~\cite{JHFRF}.       
    \par In Ref.~\cite{JHFRCED} the simplest possible system --just two interacting charges
    was considered. However in the `Electrodynamic Part' of Ref.~\cite{Ein1}, the electric
     and magnetic fields introduced exist {\it a priori} and nowhere is there any discussion
    of how they are produced. As aleady pointed out in Ref~\cite{JHFSTF} this has
    important consequences both on the very nature, in a mathematical sense, of the electromagnetic
    fields, as well as their transformation laws and the status of their relativistic
   covariance in the presence of interactions. This important point is discussed 
    in some detail in Section 9 below. 

    \par Also, in common with most authors of text books on CEM,
     Einstein makes no distinction between the `force fields' which correspond to
    the `Intercharge Forces' of Fig.1 and `Electric Forces', `Magnetic Forces' and 
      `{\it E M} Induction' of Fig.2 and, `radiation fields', not considered Fig.1,
      and describing `{\it E M} Waves' in Fig.2, using identical mathematical 
      symbols to describe them both. Actually  the operational meanings of the two types 
     of fields are quite distinct~\cite{CSR,JHFRCED,JHFSTF}. This difference will be
      taken into account in the discussion of the various sections of the
       `Electrodynamic Part' of Ref.~\cite{Ein1}.
    \par Turning now to the `Kinematical Part' of Ref.~\cite{Ein1} it is clear that Einstein's
      presentation of SRT results in a strong entanglement (in the common, not the quantum
      mechanical technical sense) of concepts from space-time physics and CEM. In Fig.3 is
      shown a concept flow diagram of the Larmor-Lorentz~\cite{Lar1900,Lor} 
\footnote{It is difficult to understand why Larmor's
    work, published in 1900, was not cited in Lorentz' 1904 paper where identical
    transformation equations were given.The appellation `Lorentz Transformation' is due to Poincar\'{e}~\cite{HP1905}.
    See Ref.~\cite{CK} for a discussion of priority issues related to the discovery of the Lorentz Transformation.}
    and Einstein derivations of the LT and
      of Voigt's derivation of his earlier, and closely
        related\footnote{The Voigt transformation is obtained from
      the LT by by multipying the right sides of all  the equations of the latter
      by the factor $\sqrt{1-(v/c)^2}$.} transformation~\cite{Voigt}. Of particular importance
       for Einstein's formulation of SRT is the light-signal synchronisation procedure
        for spatially separated clocks, that assumes spatial isotropy of the speed of light.
        Relaxing the latter postulate leads to the `conventionality' of clock
        synchronisation by this method, a subject on which a vast literature exists~\cite{AVS}.

\begin{figure}[htbc]
\begin{center}\hspace*{-0.5cm}\mbox{
\epsfysize8.0cm\epsffile{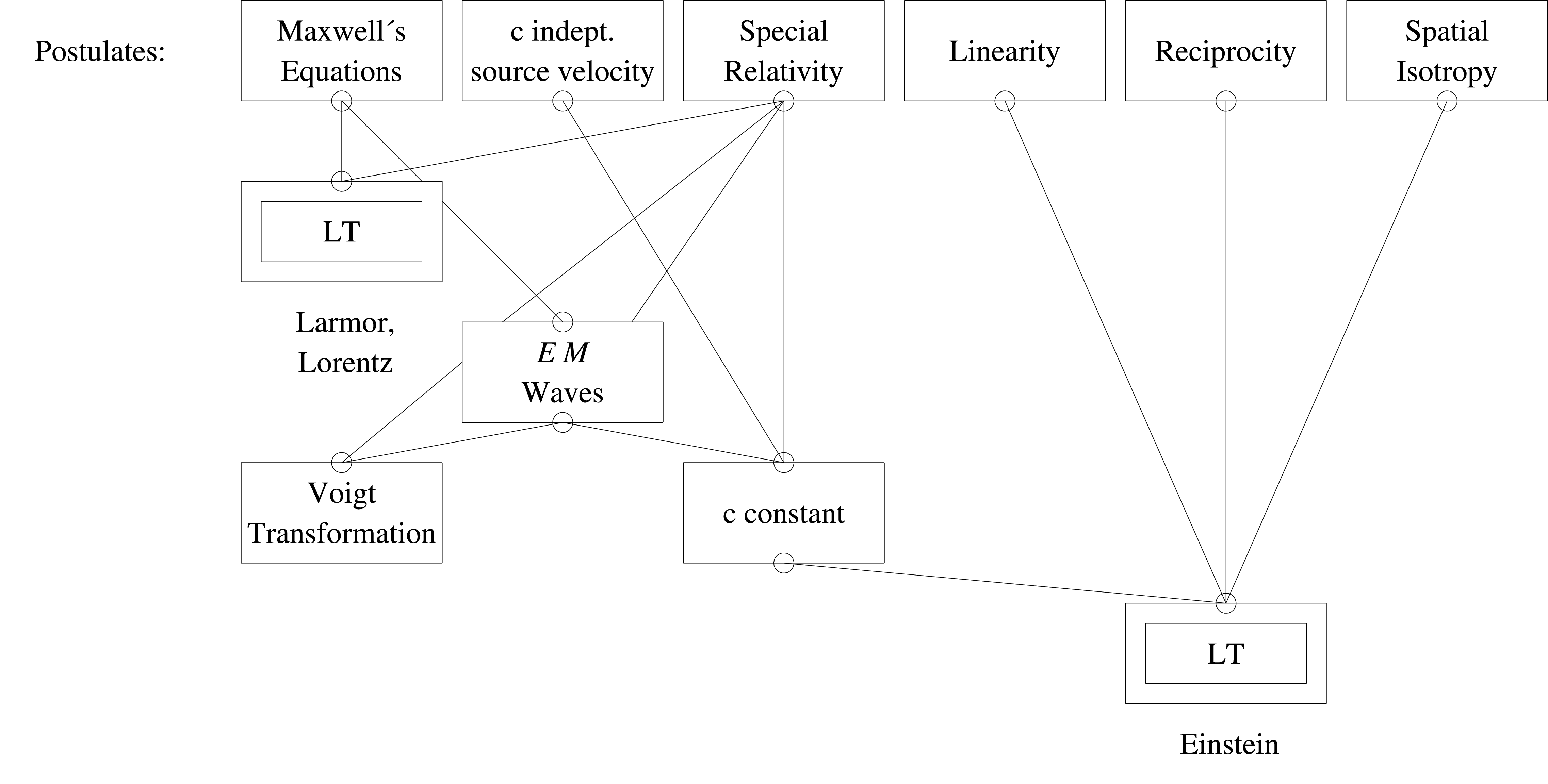}}
\caption{ {\em Concept flow diagram of the derivation of the Voigt Transformation, and of the
    Larmor-Lorentz and Einstein derivations of the Lorentz Transformation (LT). Einstein's second
   postulate stated only the source velocity independence of the speed of light. The constancy
  of the speed of light in all inertial frames was considered by Einstein to be a `Law of Nature'
   and to therefore
   be a direct consequence of the Special Relativity Principle. Additional  necessary assumptions 
  for Einstein's derivation are linearity of the equations, the Reciprocity Principle
    (if the velocity of the inertial frame S' 
   relative to S is $\vec{v}$, then the velocity of S relative to S' is -$\vec{v}$) and spatial
    isotropy. }}
\label{fig-fig3}
\end{center}
\end{figure}

\begin{figure}[htbc]
\begin{center}\hspace*{-0.5cm}\mbox{
\epsfysize12.0cm\epsffile{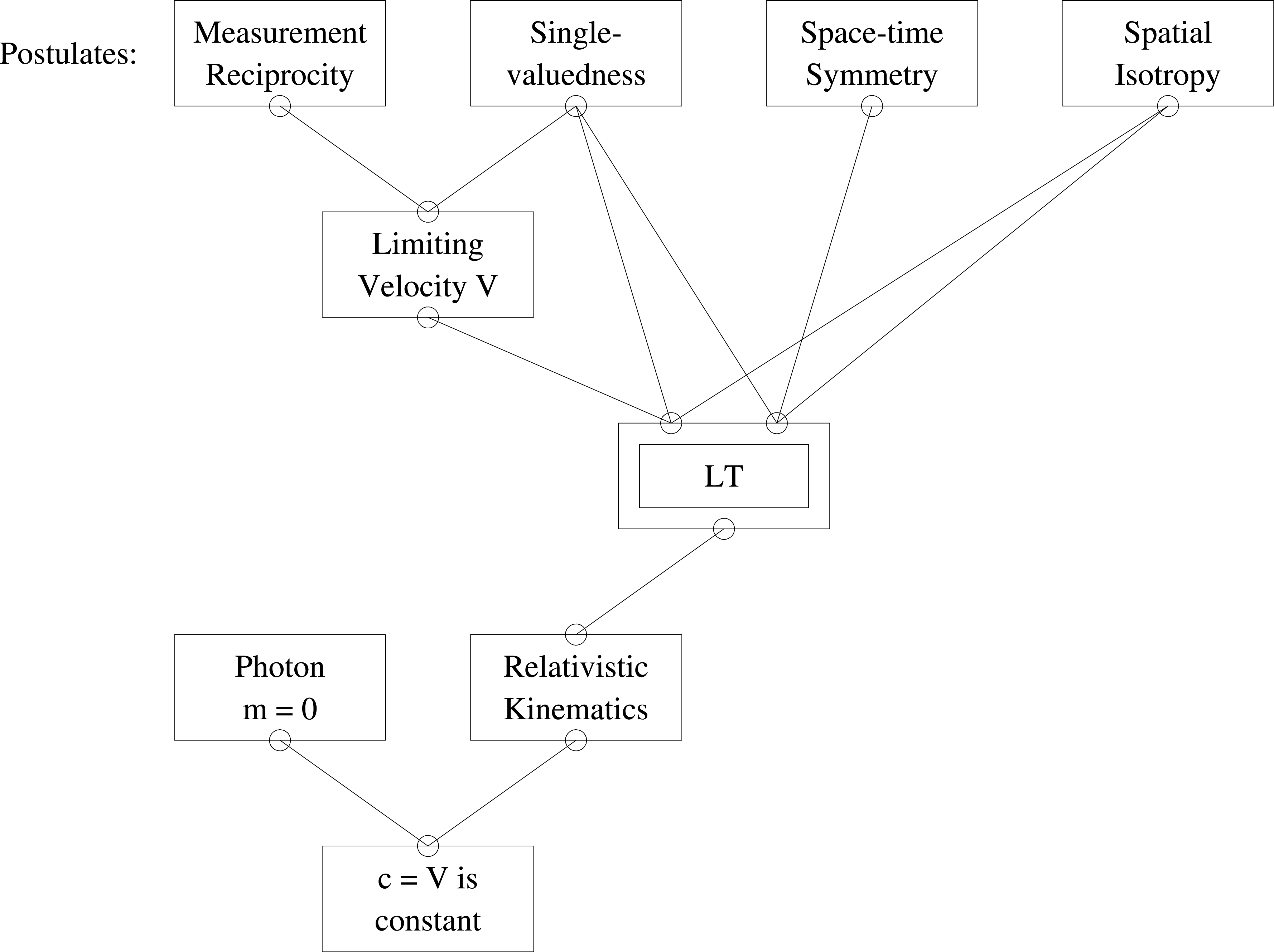}}
\caption{ {\em  Concept flow diagram of the derivations of the LT in Refs.~\cite{JHFLT1,JHFLT2}.
      No electrodynamical concepts or other dynamical laws are invoked. Einstein's second
     postulate, as well as the constancy of the speed of light, are consequences of relativistic kinematics
    and the masslessness of the photon. The crucial derivation of the necessary existence of the limiting 
      velocity, $V$, was already done by Ignatowsky~\cite{Ignatowsky} in 1910.}}
\label{fig-fig4}
\end{center}
\end{figure}

     \par However, applying Occam's razor to the derivation of the LT, in a similar way to 
        that in which it was applied to CEM in Ref.~\cite{JHFRCED} to obtain RCED, is there, in fact,
          any necessary
        connection between SRT and CEM?  It was realised very early that the answer to this 
        question is `No'. Of particular importance in this context is the 1910 paper of
         Ignatowsky~\cite{Ignatowsky} in which the necessary existence of a limiting
         relative velocity, $V$, between any two inertial frames was pointed out. Indeed,
         a very extensive literature on `lightless' derivations of the LT exists. See Ref.~\cite{BG}
       for citations of work prior to 1968, and the paper published some years ago by the present
       author~\cite{JHFLT1} giving later citations, and in which the one-dimensional LT was 
       derived from just two postulates,
       Single-Valuedness(SV) and the Measurement Reciprocity Postulate (MRP):
        \begin{itemize}
        \item [SV:] Each Lorentz transformation equation must be a single-valued
            function of its arguments.
         \item [MRP:] Reciprocal space-time measurements of similar rulers and clocks
          at rest in two different inertial frames: S,S' by observers at rest in S',S, 
          respectively, give identical results.
      \end{itemize}
       Another, much shorter, derivation can be found in Ref.~\cite{JHFLT2} based on
       the postulates SV and Space-Time Exchange Invariance (STEI):
       \begin{itemize}
          \item [STEI:] The equations describing physical laws are invariant with
        respect to the exchange of space and time coordinates, or more generally, 
        with respect to the exchange of spatial and temporal components of
        4-vectors. 
     \end{itemize}
          In order to derive the full three-dimensional LT the additional postulate of
        spatial isotropy is required in both cases. 
       In the derivation from SV and the MRP  the limiting velocity $V$ appears naturally in
       the course of the derivation, as was the case in Ref.~\cite{Ignatowsky}. In the one based
       on SV and STEI the universal parameter, $V$, with the dimensions of velocity,
        must be introduced, for dimensional reasons, in order to define the space-time exchange
       operation, that requires unidimensional space and time coordinates. The concept flow
        diagram of the LT derivations of Refs.~\cite{JHFLT1,JHFLT2} in Fig.4 may be compared
       with that in Fig.3. No appeal is made to CEM, or any other dynamical theory. As discussed
       below, Einstein's second postulate of the constancy of the speed of light
       is {\it derived} (see  Fig.4) from relativistic kinematics and the 
       massless nature of the photon~\cite{JHFLT1}. The LT derivations of Refs.~\cite{JHFLT1,JHFLT2} are 
       recalled in the Appendix of Ref.~\cite{JHFST1}.  
      \par In Ref.~\cite{JHFST1}, several clock synchronisation procedures, not making any use of
       light signals are introduced. The crucial point (as clearly stated by Einstein himself,
       although never correctly applied by him,
       or by anyone else) is that, as will be shown in the following section,
       the space-time LT describes, fundamentally, how times 
       recorded by clocks in uniform motion appear to a `stationary' observer.
       If there are several such synchronised clocks in the same moving inertial frame,
       at different spatial positions, suitable constants must be added, for each such clock, to the right   
       sides of the equations of the conventional space time LT. The latter describes correctly
       only a synchronised
       clock at the origin of the moving frame. Until now, this has not been done,
       resulting, as described  previously in Refs ~\cite{JHFST1,JHFUMC,JHFCRCS,JHFLLT}, in the 
       spurious `relativity of simultaneity' (RS) and `length contraction' (LC) effects of
       conventional SRT. 
       \par The following section contains, for the reader's convenience, a brief summary of the
         correct physical interpretation of the space-time LT given in Refs.~\cite{JHFST1,JHFUMC,JHFCRCS,JHFLLT}.
        In the subsequent sections the spurious arguments by which Einstein arrived at the
         predictions of RS and LC are examined in detail. Sections 3-13 contain
       commentaries on \S 1-\S 10 of Ref.~\cite{Ein1}. Section 14 contains a summary that includes a list
       of Einstein's mistakes,
        and the concluding Section 15 discusses the legacy of the 1905 special relativity
        paper in the light of both the findings of the present reassessement and its 
        historical context.
         \par Since Einstein did not
        number the equations of Ref.~\cite{Ein1} (there are $\simeq 140$ in the paper) a complete 
       numbered list is presented, section-by-section, in the Appendix.
      They are numbered with an `E' in correspondence with the numbered sections
       of  Ref.~\cite{Ein1}. e.g., ({\rm E}10.2) is the second equation of \S 10.
        Although all the equations in the Appendix are fair
       copies of those in the English translation of Ref.~\cite{Ein1}, in places the notation
         has been changed to be more in conformity with modern usage. For example
         $\gamma$ for $1/\sqrt{1-(v/c)^2}$ and $\beta$ for $v/c$, as well as 3-vector
            notation for components of electric and magnetic fields. Also, depending
         on what is most convenient, in some places Einstein's original
          notation for space-time coordinates, and in others that of Section 2 below, will
          be employed. The only ambiguous symbol is $\tau$ which stands, in Einstein's notation, for the
          apparent time of clocks in the `moving' frame S' as viewed from the `stationary' frame S,
          and in the notation of Section 2 for the proper time of clocks at rest in S as viewed in this frame.
        Forewarned of this, the reader should not be too confused by the use of
         the two different nomenclatures.  For clarity, the
        most important E-numbered equations are repeated in the text at the appropriate
        place. All other equations are numbered according to the section of the present
        paper in which they occur.

\SECTION{\bf{Description of a clock in uniform motion by the Lorentz or Galilean transformation}}
 The space-time LT connects observations of the same space time event in
 different inertial frames. If the frame S' moves with speed $v$ along the $x$-axis
  of the frame S, and if
 the Cartesian axes of the frames are parallel, the LT is written as:
 \begin{eqnarray}
  x' & = & \gamma[x-vt], \\
  t' & = & \gamma[t-\frac{\beta x}{c}], \\
   y' & = & y, \\
   z' & = & z
 \end{eqnarray}
 where $\beta \equiv v/c$, $\gamma \equiv 1/\sqrt{1-\beta^2}$ and $c$ is the speed of light in vacuum.
  The space-time coordinates of the transformed events are ($x$,$y$,$z$,$t$) in S and ($x'$,$y'$,$z'$,$t'$) in S'.
  In this section, for simplicity, only events lying on a common $x$-$x'$ axis are considered.
 \par To give an operational meaning to the time coordinates,it is necessary to introduce clocks; say C at rest in S and
  C' at rest in S'. The time symbols in the LT are then identified with the readings of these clocks. The LT can
   then describe two different experiments: 
   \begin{itemize}
  \item[(i)] C' (in motion) is viewed from S 
  \item[(ii)] C (in motion) is viewed from S'
  \end{itemize}
  The appropriate $x$-$t$ LT equations to describe these two possiblities are,
   for experiment (i):
 \begin{eqnarray}
 x'(\Cp) & = & \gamma[x(\Cp)-v\tau] = 0, \\
  t' & = & \gamma[\tau-\frac{\beta x(\Cp)}{c}] 
 \end{eqnarray}
  and for experiment (ii):
 \begin{eqnarray}
 x(\C) & = & \gamma[x'(\C)+v\tau'] = 0, \\
  t & = & \gamma[\tau'+\frac{\beta x'(\C)}{c}]. 
 \end{eqnarray}
 Note that (2.7) and (2.8) are not simply the inverse transformation of
 (2.5) and (2.6) but four different time symbols appear in the equations with 
 different operational meanings:
   \begin{itemize}
  \item[$\tau$:]  proper time recorded by C in S
\item[$\tau'$:]  proper time recorded by C' in S'
\item[$t$:]  apparent time of C as viewed from S'
\item[$t'$:]  apparent time of C' as viewed from S
  \end{itemize}
  Also four different space coordinates: $x(\C)$,  $x(\Cp)$,  $x'(\C)$ and $x'(\Cp)$ occur. 
 Inspection of (2.5) and (2.6) shows that $t' = \tau = 0$ when $x = 0$. This means, operationally,
   that the clocks C and C' are synchronised (set to zero) at the instant that the clock C' at $x' =0$  is
  at the $x$-origin in S. If it is now desired to synchronise, in a similar manner, the 
  clocks C and C' when C' is at $x' = L'$, and its $x$ coordinate in S is $L$, the equations
  (2.5) and (2.6) require modification. As correctly pointed out by Einstein~\cite{Ein1A}:
  \par {\tt If no assumption whatever be made as to the initial position of the moving system
   and as the the zero point of $\tau$,} ($t'$ in the notation used above){\tt  an additive constant
   is to be placed on the right side of each of these}~(the LT (2.1)-(2.4)) \newline
{\tt equations.}
   \par A suitable modification of (2.5) and (2.6) is:
 \begin{eqnarray}
 x'(\Cp) & = & \gamma[x(\Cp)-v(\tau- \tau_0)], \\
  t'-t'_0 & = & \gamma[\tau- \tau_0 -\frac{\beta x(\Cp)}{c}] 
 \end{eqnarray}
   where the constants $\tau_0$ and $t'_0$ are to be chosen in such a way that
   $t' = \tau = 0$ when $x(\Cp) = L$, $x'(\Cp) = L'$. This requires that:
 \begin{eqnarray}
 L' & = & \gamma[L+v \tau_0], \\
 t'_0(L) & = & \gamma[\tau_0+\frac{\beta  L}{c}]. 
 \end{eqnarray}
 Solving these equations gives:
 \begin{eqnarray}
 \tau_0 & = & \frac{1}{v}(\frac{L'}{\gamma}-L), \\
   t'_0(L) & = & \frac{1}{v}(L'-\frac{L}{\gamma}).
 \end{eqnarray}
 Substituting these parameters in (2.9) and (2.10) gives:
 \begin{eqnarray}
 x'(\Cp)-L' & = & \gamma[x(\Cp)-L-v\tau], \\
  t' & = & \gamma[\tau-\frac{\beta}{c}(x(\Cp)-L)]. 
 \end{eqnarray}
 Since $L \equiv x(\Cp,\tau = 0)$ and  $L' \equiv x(\Cp,t')$, for all values of $t'$
 are constants, independent of $v$, depending only on the choice of spatial coordinate systems
in S and S' respectively, Eq.~(2.15) holds for all values of $v$. In particular,when $v = 0$,
  so that S $\rightarrow$ S', $\gamma \rightarrow 1$  and $x \rightarrow x'$,it gives:
    \begin{equation}
  x'(\Cp)-L' =  x'(\Cp)-L
  \end{equation}
   or 
 \begin{equation}
  L' = L.
  \end{equation}
  Since $x'(\Cp) = L'$, (2.15) gives, for the equation of motion of $\Cp$ in S: 
  \begin{equation}
  x(\Cp) =  L +v \tau. 
  \end{equation}
  Substituting (2.19) into (2.16) and recalling the definition of $\gamma$
 gives the time dilation (TD) formula:
    \begin{equation}
    \tau = \gamma t'. 
   \end{equation}
 Equations (2.19) and(2.20) give the complete space-time description of the clock C' at $x' = L$,
  as viewed from S.
  \par Note that (2.19) is also valid for the Galilean transformation (i.e. the limit of the 
   LT when $c \rightarrow \infty$), whereas the latter gives, instead of (2.20),
   $\tau = t' \equiv T$ where $T$ is a universal
   (Newtonian) time.
    \par Repeating the above chain of arguments for the experiment (ii), where the clock C is viewed 
   from the frame S', gives the equations describing this (reciprocal) experiment:
  \begin{equation}
  x'(\C) = L-v \tau', 
  \end{equation}
    \begin{equation}
    \tau' = \gamma t. 
   \end{equation}
 \par Introducing now two clocks in S', $\Cp_A$ at $x' = 0$ and $\Cp_B$ at $x' = L$ (2.19) and (2.20) give:
   \begin{eqnarray}
    x(\Cp_A) & = & v \tau,  \\
    x(\Cp_B) & = & L+v \tau,  \\
   \tau & = & \gamma t'(\Cp_A), \\
   \tau & = & \gamma t'(\Cp_B).
  \end{eqnarray}
 Eqs.~(2.23) and (2.24) give:
  \begin{equation}
  x(\Cp_B)- x(\Cp_A) = L = x'(\Cp_B)- x'(\Cp_A) 
  \end{equation}
  so there is no `length contraction' (LC) effect, while (2.25) and (2.26) give:
  \begin{equation}
    t'(\Cp_A) =  t'(\Cp_B) \equiv t'
  \end{equation}
  where $t'$ is the apparent time, as viewed in S, of {\it all} synchronised
  clocks in S'. There is therefore no `relativity of simultaneity' effect
  in this case. How these effects arise from misuse of the LT (2.1)-(2.4) has been previously
   explained ~\cite{JHFST1,JHFUMC,JHFCRCS,JHFLLT} and will become
  clear in the following sections of
   the present paper where Einstein's predictions of these effects are critically examined.
   \par It  will become apparent in the following discussion of Ref.~\cite{Ein1} that Einstein
  did not take into account the existence of the four operationally distinct time symbols
  appearing in the LTs (2.5)-(2.8). All discussions are based on (2.1) and (2.2) where
   $t$ is identified with the proper time in S and $t'$ with the proper time in S'. 
   This is equivalent to making the substitutions $t \rightarrow \tau$ in Eq.~(2.8) and  
   $t' \rightarrow \tau'$ in Eq.~(2.6). That this is a logical absurdity becomes evident on
  making the same substitutions in (2.20) and (2.22) leading to the equations:
   \begin{eqnarray}
   \tau & = & \gamma \tau', \\
   \tau' & = & \gamma \tau.
  \end{eqnarray}
   These equations require that $\gamma^2 = 1$ or $v = 0$, in contradiction
  to the initial hypothesis that the the velocity, $v$, of S' relative to S, is 
  non-vanishing.
   \par In some applications it is convenient to introduce an
     alternative notation for the apparent time intervals $t'$ and $t$:
     \begin{eqnarray}
     t' & \equiv & \tau'(\Cp), \\
     t & \equiv & \tau(\C).
    \end{eqnarray}
 Here, $\tau'(\Cp)$, for example, is a proper time interval recorded by the clock
   C', at rest in S', in an experiment in which C' is in motion. This is to be contrasted 
   with $\tau'$ in Eq.~(2.22) that is, instead, the proper time interval in
    S' recorded by C' in an experiment in which this clock is at rest. Use of the symbols $\tau$, $\tau'$. $t$ and $t'$
     has the advantage that they all represent direct observations of time intervals
     of clocks, either at rest ($\tau$, $\tau'$), or in motion ($t$,$t'$). Since
     these times are the same for all synchronised clocks in the frames, no clock labels
     are necessary. If (2.31) and (2.32) are used, all symbols specify proper times
     but additional clock labels are required to distinguish, for example, $\tau$ (time recorded by C in an experiment
     in which it is at rest)
     and $\tau(\C)$ (time recorded by C in an experiment in which it is in motion). 
     \par An important application of (2.31) occurs in particle physics where the clock
      C' is identified with a moving unstable particle P, and $\tau'(\Cp \equiv {\rm P})$
      with its proper decay time. In this case (2.20) and (2.29) give:
      \begin{equation}
      \tau'(\Cp \equiv {\rm P}) = \frac{\tau}{\gamma}
       \end{equation}
       where $\tau$ is the observed decay lifetime of the particle in the 
        laboratory system and the time dilation factor, $\gamma$, is derived from the 
        measured speed, $v$, of the particle in the laboratory system. In this case
       the proper time interval $\tau'(\Cp \equiv {\rm P})$ is not directly 
       observed, but deduced from the TD relation (2.33) and the observed laboratory
       lifetime $\tau({\rm P})$.
       \par Another application of Eq.~(2.31) occurs in the discussion of relativistic
          velocity addition in Section 8 below.   
\SECTION{\bf{Einstein's Introduction: The physics of electromagnetic induction in different
  reference frames}}
   In the introductory section of Ref.~\cite{Ein1} Einstein discusses the phenomenon of
  electromagnetic induction in two reference frames: the first in which a magnet is in motion
  and a test charge is at rest, and the second in which the magnet is at rest and the
   test charge in motion. For the present discussion, the test charge can be conveniently considered
   to be at rest in the frame S', while S is the magnet rest frame. Einstein states that in the
    first case an electric field acts on the test charge, but did not attempt to calculate
   this field. As shown below, if he had, he might have obtained a very surprising result!
    In the case that the magnet is at rest and the test charge in motion, the latter moves
   in a inhomogeeous but static magnetic field and is subjected a transverse Lorentz force which, in this
   case, explains the induction effect. Einstein states that, in this second case, `no electric field 
   occurs in the neighbourhood of the magnet'. Later in Ref.~\cite{Ein1}, after discussing the
    transformation laws of electric
   and magnetic fields, it is pointed out that the magnetic (Lorentz) force in one frame (that
   in which the test charge is in motion) becomes an electric force in the frame in which the test 
   charge is at rest, as a direct consequence of the transformation laws of electric and magnetic
   fields. Thus the magnetic and electric forces are just the descriptions in different frames
   of the force on the test charge responsible for electromagnetic induction. The two separate
   phenomena discussed by Einstein are thus unified in special relativity by the transformation
   laws of electric and magnetic fields. However, explicit calculation of the force on the
    test charge in the two cases, using standard formulae of CEM, 
     will now be shown to lead to a less satisfactory conclusion.
    \par Consider a test charge, $q$, near an elementary `magnet' consisting of two
    charges of magnitude $Q$ situated, at some instant, at the points U and D (See Fig.5).
    The points U and D lie along the $y$ axis equidistant from the origin, whereas the charge $q$
     lies in the $x$-$y$ plane displaced, by a small distance, from the $x$-axis. 
    In Fig.5a the charges at U and D move parallel to the positive
   and negative $z$ axes, respectively, with speed $u$. The velocity vectors of the charges are
    $\vec{u}_U$ and $\vec{u}_D$, where  $\vec{u}_U = -\vec{u}_D$ and $|\vec{u}_U| =|\vec{u}_D| = u$.
  The test charge moves with speed $v$ in the positive
     $x$-direction. An imaginary rectangular contour, abcd,  parallel to the $y$-$z$ plane moves
    together with the test charge. Fig.5a corresponds to the second case considered by Einstein
    ---stationary magnet and test charge in motion. Because of the displacement of the test 
    charge from the symmetry axis, O$x$, there is a net magnetic field component parallel to the $y$-axis.
    The corresponding Lorentz force is in the direction of the negative $z$-axis and explains
    the induction effect in the rest frame of the magnet.

\begin{figure}[htbc]
\begin{center}\hspace*{-0.5cm}\mbox{
\epsfysize15.0cm\epsffile{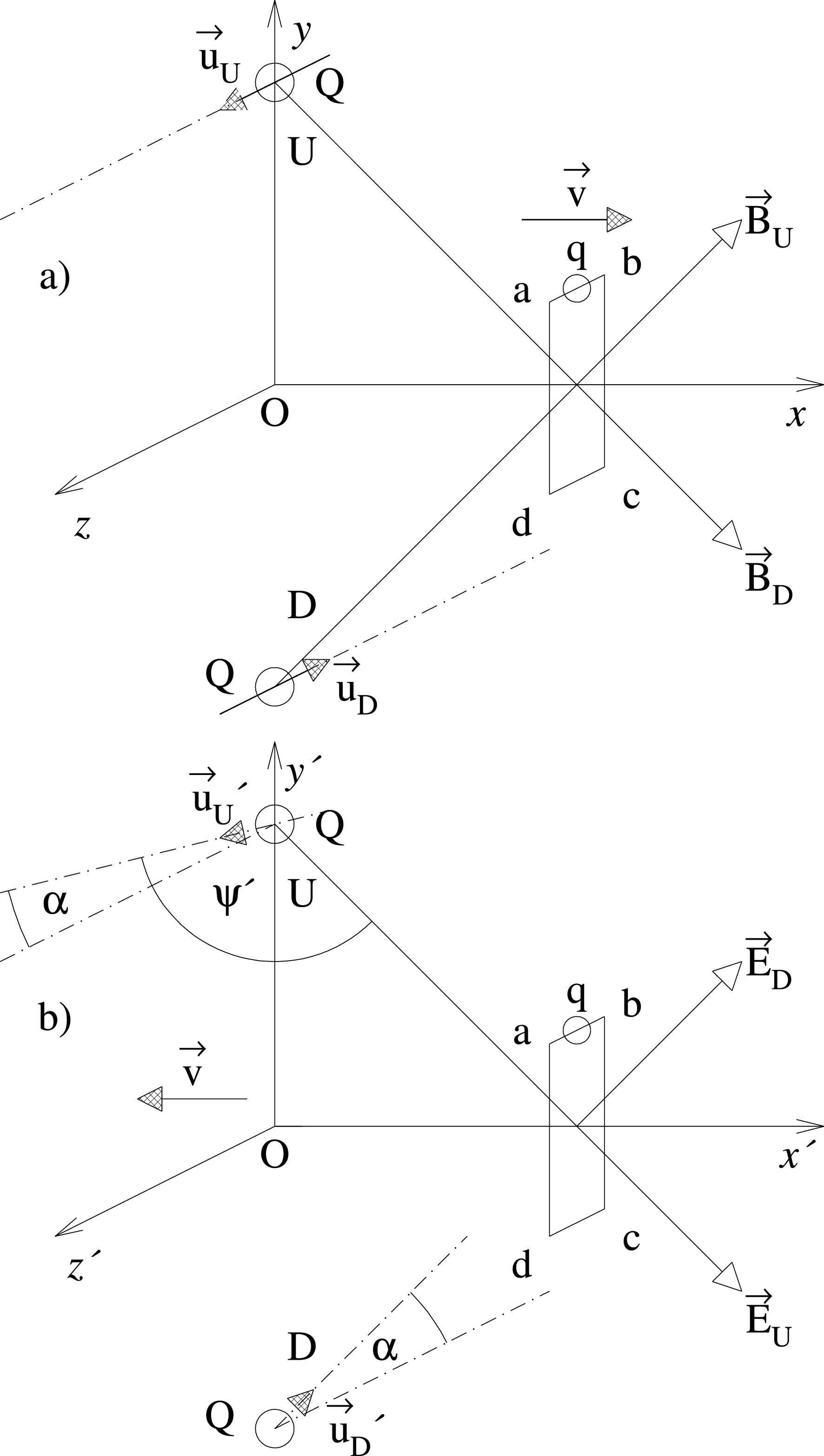}}
\caption{ {\em  Electromagnetic induction due to a simple two charge `magnet' in: a) the frame in which the
     magnet is at rest and the test charge $q$ is in motion and: b) the frame in which the magnet
    is in motion and the test charge is at rest. In b) the Heaviside formula (3.1) predicts that the 
     electric field of each source charge is radial and so confined to the $x'y'$  plane at the position
      of the test charge. Since there is then no electrical force in the $z'$ direction the induction 
      effect vanishes. Calculation of the electric force parallel to the $z'$ axis using the RCED
      formula (3.4) gives a result consistent with the Lorentz force in a) or by use of the
      Faraday-Lenz Law in either a) or b)~\cite{JHFIND}.}}
\label{fig-fig5}
\end{center}
\end{figure}

     \par In Fig.5b is shown the corresponding configuration in the frame S' where the test charge
   is at rest and the magnet moves with speed $v$ along the negative $x'$-axis. The velocity
   vectors  $\vec{u}'_U$  and  $\vec{u}'_D$ of the charges at U and D are now at an angle
   $\alpha$ to the $z'$-axis where $\sin \alpha = \beta/\beta_{u'} = \gamma \beta /\sqrt{\beta_u^2+ \gamma^2 \beta^2}$ and
   $\beta_u = u/c$. Since the test charge is at rest there is no Lorentz force, so, as pointed
   out by Einstein, electromagnetic induction can be produced in this frame only
   by an electric field. The electric field vectors $\vec{E}_U$ and  $\vec{E}_D$, at the point
   where the plane abcd cuts the $x'$-axis, are shown as predicted by the Heaviside formula~\cite{Heaviside}
    for the electric field of a uniformly moving charge: 
  \begin{equation}
  \vec{E}({\rm CEM}) =\frac{Q \vec{r}}{r^3 \gamma_{u'}^2(1- \beta_{u'}^2 \sin^2 \psi')^{\frac{3}{2}}}.
  \end{equation}
   The radius vector $\vec{r}$ is that from the source charge to the field point (so the
   electric field is predicted to be radial) and $\psi'$ is the angle between the radius vector and the
    source charge velocity vector $\vec{u'}$. Because of the radial character of the
    electric field, the vectors $\vec{E}_U$ and $\vec{E}_D$ associated with the source charges U and D,
 lie in the $x'$-$y'$ plane at every point in this plane, including that at the position of the
   test charge. There is therefore no electric field in the $z'$-direction, and so, according to Eq.~(3.1),
    no electromagnetic induction effect in the frame S'. The Heaviside Formula (3.1) is therefore 
   inconsistent with SRT for the configuration of moving charges shown in Fig.5.     
     \par In the recent paper by the present author~\cite{JHFRCED}, mentioned in the Introduction,
     a different formula to (3.1) for the electric field of a moving charge
    was obtained:
  \begin{equation}
  \vec{E}({\rm RCED}) =\frac{Q}{r^2}\left[ \frac{\hat{u}' \cos \psi'}{\gamma_{u'}} +
   \hat{t}'\gamma_{u'}\sin \psi' \right]
  \end{equation}
  where $\hat{u}'$ is a unit vector parallel to the source charge velocity and $\hat{t}'$ is a unit vector
   perpendicular
   to  $\hat{u}'$ lying in the plane defined by  $\hat{u}'$ and the radius vector $\vec{r}$
   specifying the field point, on the same side as the field point. The electric field defined by (3.2)
   is not radial and calculation~\cite{JHFIND} in the frame S' shows agreement at O($\beta^2$)
    \footnote{Due to the lack of covariance of the longitudinal electric field (see Section 6 below)
    the induction force differs between the frames S and S' by terms of O($\beta^4$) and higher.} 
   with the induction
   effect calculated in the frame S, either by explicitly evaluating the Lorentz force in this frame,
    or by use of the integral form of the Faraday-Lenz Law in either S or S'.
   \par It is interesting to note, in connection with the previous remark, that the problem of the
   apparent inconsistency of the description
   of induction in different frames, posed by Einstein, does not arise if the integral form of the
    Faraday-Lenz law:
  \begin{equation}
  -\frac{d \phi}{d t} = \int \vec{E} \cdot d \vec{S}
  \end{equation}
  is used to solve the problem. In Fig.5a, the flux $\phi$ of magnetic field threading the contour
   abcd changes because the contour moves in a static, but non-uniform, magnetic field, whereas in
   Fig.5b, the contour is stationary but $\phi$ changes because of the motion of the magnet.
   Typically, in text books, it is pointed out that the Faraday-Lenz law can be used to derive
   electromagnetic induction in either the frame S or S', whereas a fundamental dynamical 
   understanding (the Lorentz force) is available only in the frame S. This situation
   provoked the following statement by Feynman~\cite{FeynFLP}:
   \par {\tt We know of no other place in physics where such a simple and accurate \newline
   general principle requires for its real understanding an analysis in terms of}
   ~{\it two different phenomena}~{\tt Usually such a beautiful generalisation is found to stem from 
   a single}~{\it deep underlying principle.}~{\tt Nevertheless, in this case there does not
    appear any such profound implication. We have to understand the\newline rule as the combined effects
    of two quite separate phenomena.} (Italics in the original).
    \par The `separate phenomena' referred to are the Lorentz force law and the Faraday-Lenz law (3.3).
     In fact a completely unified description of electromagnetic induction as well all the other
      mechanical effects\footnote{These include the relativistic generalisations of
   Coulomb's Law and the Biot and Savart Law as well as the Lorentz force equation.}
 of classical electromagnetism 
    is provided by the following formula~\cite{JHFRCED}
     containing no fields, that gives the force on a test charge
      due to a source charge in arbitary motion: 
      \begin{eqnarray} 
      \frac{d\vec{p}}{dt} &  = & \frac{q}{c}\left[\frac{ j^0\vec{r} +  \vec{\beta} \times
     (\vec{j} \times \vec{r})}{r^3} -\frac{1}{c r}\frac{d \vec{j}}{d t}-\vec{j}
     \frac{(\vec{r} \cdot \vec{\beta}_u)}{r^3}. 
      \right]  
     \end{eqnarray} 
   Here $q$ and $\vec{\beta}c$ are the charge and velocity of the test charge of Newtonian mass $m$ and
    relativistic momentum $\vec{p}= \gamma \vec{\beta} m  c$,  and
    $j \equiv Q(\gamma_u c,\gamma_u \vec{u})$ is the 4-vector current of the source charge.
    The unified description of of electromagnetic induction provided by Eq.~(3.4) would no doubt have 
   pleased both Einstein and Feynman. As discussed in Ref.~\cite{JHFRCED} `the single deep
    underlying principle' underlying Eq.~(3.4) is the Feynman path integral formulation
    of quantum mechanics, applied to space-like virtual photon exchange processes, in the classical
    ($h \rightarrow 0$) limit. This is the fundamental `constructive physics' underlying both the
     Coulomb force law and Hamilton's Principle, from which Eq.~(3.4) is derived.
    \par It has been pointed out in Ref.~\cite{JHFSTF} that there is an important mathematical error
    \footnote{After completing the work in Ref.~\cite{JHFSTF} I was informed~\cite{McAndrew} that the same error
      had been pointed out previously by Whitney~\cite{Whitney}.}
   in the derivation of the Li\'{e}nard-Wiechert retarded potentials~\cite{LW} from which
   the `present time' Heaviside formula (3.1) is obtained. When this error
    is corrected then, up to an overall multiplicative factor, the formula (3.2)
    (but with a retarded time argument) is recovered.
   \par After the discussion of electromagnetic induction, Einstein introduces the postulates on which his 
   analysis of space-time physics is based: The Relativity Principle and the postulate that the
    speed of light does not depend upon that of the emitting body. Here, and later when explicitly
    defining this postulate it is not stated that the speed of light as measured in {\it any} inertial
   frame is constant. Einstein refers only to the `stationary' frame S. Also, in the introduction,
    the  Relativity Principle is applied only to the description of electromagnetic and optical
   phenomena in different reference frames. In accordance with the title, it is stated 
   that the aim of the paper is to provide `a simple and consistent theory of the electrodynamics
   of moving bodies based on Maxwell's theory of stationary bodies'. In this theory the
   `luminiferous aether', such an important concept of 19th Century electromagnetic theory, will
    become superfluous.
    \par In the final paragraph of the introduction, Einstein stresses the importance of actual
    measurements of length intervals with rulers, and time intervals with clocks, in constructing
     the theory. This is certainly a very important remark. However, as shown in Ref.~\cite{JHFST1}, the
    consideration of electromagnetic and optical pheonomena, as in Einstein's paper, is inessential
    to establish the foundations of space time physics in flat space, which are the
     space-time LT and its correct physical interpretation.
       The last sentence of the introduction is:
     \par {\tt Insufficient consideration of these circumstances}~(i.e. actual, experimental, space and
      time measurements)~{\tt lies at the root of the difficulties which the electrodynamics of
      moving bodies at present encounters.}
     \par As will be demonstrated below, this sentence remains valid one century after the work
     presented in Ref.~\cite{Ein1} if the words `the physics of space time' are substituted for
    `the electrodynamics of moving bodies' in the sentence just quoted. 
   
\SECTION{\bf{\S 1. Definition of Simultaneity}}
  At the beginning of this section, Einstein stresses again the importance of actual clock readings
  and the concept of simultaneous events in establishing the meaning of physical time:

  \par  {\tt We have to take into account that all our judgements in which time
     plays a part are always judgements of} {\it simultaneous events}. {\tt If, for
     instance, I say ``That train arrives here at 7 o'clock'', I mean
     something like this: \newline ``The pointing of the small hand of my watch to 7 and the
     arrival of the \newline train are simultaneous events''.}
     \par What Einstein is stating here is that any actual time measurement (and therefore any measurement
     of a time interval) is built up from pointer-mark coincidences, which constitute the irreducible
     experimental raw data, and the concept of simultaneity. In the above example two, simultaneous,
     pointer-mark coincidences (between the hand of the clock and a mark on its dial, and between the
     front of the train and the position on a platform of Bern railway station where  Einstein happened
    to be standing) are involved. In a similar way the raw experimental data of data of any spatial
    measurement involves, or is equivalent to, the spatial coincidence of a pointer specified by
   the geometry of some
   physical object and a mark on a ruler. A mathematically rigorous calculus of such pointer-mark
   coincidences in constructing  measurement of space and time intervals is developed by the
   present author in Ref.~\cite{JHFST1}.
   \par Einstein then addresses the problem of establishing the simultaneity of spatially separated
   events in a given reference frame, and proposes the light-signal clock synchronisation procedure of
    Eqs.~({\rm E}1.1) and ({\rm E}1.2). This procedure immediately couples the concept of simultaneity
   with the physics of light propagation. In particular, Einstein's definition 
   of simultaneity (or equivalently of clock synchronisation) assumes spatial isotopy of
    the speed of light. However, as shown in Ref.~\cite{JHFST1} it is perfectly possible
    to synchronise spatially separated clocks without using light signals, and so independently
    of the assumption of light-speed isotropy. Two such procedures: `Pointer Transport' and
    `Length Transport' are described in Ref.~\cite{JHFST1}. Pointer Transport can be used to synchronise
   an arbitary number of spatially-separated clocks in the same reference frame, whereas Length Transport
   synchronises four clocks, two in each of two different inertial frames. Given the isotropy
   of the speed of light, Einstein's light-signal synchronisation procedure is, of course, a perfectly
    valid one. However applications of this idea to the discussion of the relativity of length and time
   intervals and the derivation of the LT in subsequent sections of Ref.~\cite{Ein1} lead to 
   serious errors in the interpretation of the physical meaning of Einstein's second postulate.

\SECTION{\bf{\S 2. On the Relativity of Lengths and Times}}

   At the begining of this section Einstein restates his initial postulates\footnote{As will be seen, others
   are introduced, either explicitly or tacitly during the derivation of the LT. See Fig.3.} 
   of special relativity:
 \par {\tt 1. The laws by which the states of physical systems undergo change are \newline not  affected whether
       these changes of state be referred to one or the other of two systems of coordinates in uniform
       translatory motion.}

   \par {\tt 2.Any ray of light moves in the ``stationary'' system of coordinates with the 
    determined velocity, c, whether the ray be emitted by a stationary or a moving body. Hence:}

   \[~~~~~~~~~~~~~~~~~~~{\tt velocity} = \frac{{\tt light~path}}{{\tt time~interval}} = c~~~~~~~~~~~~~~~~~~~~~~~~~~~~~({\rm E}2.1) \]
     {\tt where `time~interval' is to be taken in the sense of the definition in
    \newline \S 1\footnote{That is, as defined by appropriate clocks synchronised by the
    light signal procedure defined in \newline  \S 1 of Ref.~\cite{Ein1}.}.} 

   \par Postulate 1 applies equally in classical (Newtonian) mechanics and was well known to both
    Galileo and to Newton. The postulate 2 was introduced, for the first time, by Einstein and is
   the basis of his formulation SRT. The above statement of this postulate
   refers only to Einstein's `stationary' coordinate system (S in the notation of the present paper).
    It is only when the relation ({\rm E}2.1) is further assumed to hold in an {\it arbitary} inertial
    frame that its highly non-intuitive nature becomes evident. The latter was discussed in some
    detail in Einstein's later popular book on relativity~\cite{EinSGR}. However, it is quite
   unnecessary to introduce Eq.~({\rm E}2.1) as a stand-alone postulate. Given that light consists of
   massless particles, photons (for the discovery of which, in another paper~\cite{Ein2}, published earlier
  in 1905, Einstein was awarded the Nobel Prize) Postulate 2 follows directly from relativistic kinematics
 ~\cite{JHFLT1}. The definitions of the relativistic energy and momentum of a physical object
    of Newtonian mass $m$:
  \begin{eqnarray}
   E & \equiv & \gamma m c^2, \\
 \vec{p} & \equiv &  \gamma m \vec{v}
  \end{eqnarray}
  and of the relativistic parameter $\gamma$ give:
  \begin{equation}
    v =\frac{p c^2}{E} = \frac{p c^2}{(m^2c^4+p^2c^2)^{\frac{1}{2}}}.
  \end{equation}
 Hence, any massless particle has the velocity $v = c$ in any inertial reference frame. This
  is exactly the relation ({\rm E}2.1), not only in the `stationary', but in {\it any} inertial
  reference frame. The kinematical relations (5.1)-(5.3) are derived and discussed for the case
   of a object subject to electromagnetic forces in Section 13 below.
  \par After describing carefully and correctly how the length of a measuring rod is defined
    both in the frame S', in which it is rest, and that, S, in which it is in uniform motion,
    Einstein introduces a thought experiment based on the light signal clock
    synchronisation procedure (LSCSP) introduced in the previous section. This experiment is
    purported to give a direct demonstration of `relativity of simultaneity' (RS):
   \par {\tt We imagine further that at the two ends A and B of the rod, clocks are \newline placed
    which synchronise with clocks of the stationary system, that is to say that their
    indications correpond at any instant to the ``time in the \newline stationary  system''
    at the places that the happen to be. These clocks are therefore ``synchronous in the
     stationary system.}
    \par{\tt  We imagine further that with each clock there is a moving observer and \newline that these
    observers apply to both clocks the criterion established in \S 1 \newline for the synchronisation
    of two clocks.}
     \par The clocks are situated at the ends of the rod and so are at rest in S'. It follows
    from the TD formula (2.18) that these clocks must be running faster than a clock at rest in
    S by the factor $\gamma$ in order to remain synchronous with it. It is clear however from
    the discussion of Section 2 above, in particular Eq.~(2.26), that these clocks must
    also be synchronous for an observer in S'. Einstein states just the contrary. Why?
     In fact Einstein does not discuss the LSCSP in the frame S', as performed by the
     `moving observers', but instead considers
    a series of events defined in, and observed from, the frame S:
     \par{\tt Let a ray of light depart from A at the time $t_A$ where `time' here denotes
     `time of the stationary system' and also `position of the hands of the moving clock situated
    at the place under discussion'. Let the ray be reflected at B at time $t_B$ and reach
    A again at the time $t'_A$. Taking into consideration the principle of the
     constancy of the speed of light we find that}

  \[~~~~~~~~~~~~~~~~~~~~~~~~~~~~~~~~~~~~~~~~~~~~~~t_B -t_A = \frac{r_{AB}}{c-v},~~~~~~~~~~~~~~~~~~~~~~~~~~~~~~~~~~~~~({\rm E}2.2) \]
  \[~~~~~~~~~~~~~~~~~~~~~~~~~~~~~~~~~~~~~~~~~~~~~~t'_A -t_B = \frac{r_{AB}}{c+v}~~~~~~~~~~~~~~~~~~~~~~~~~~~~~~~~~~~~({\rm E}2.3) \]
    {\tt where $r_{AB}$ denotes the length of the rod --measured in the stationary system.}
    \par All of the obove times are specified in the frame S, whereas the LSCSP for clocks at rest
   in S', as specified according the the prescription of \S 1 is based on the equations
 \[~~~~~~~~~~~~~~~~~~~~~~~~~~~~~~~~~~~~~~~~~~~~~~t_B -t_A = t'_A -t_B,~~~~~~~~~~~~~~~~~~~~~~~~~~~~~~~~~~~~~~~~~~~({\rm E}1.1) \]
 \[~~~~~~~~~~~~~~~~~~~~~~~~~~~~~~~~~~~~~~~~~~~~~~\frac{2 AB}{t'_A -t_A} = c~~~~~~~~~~~~~~~~~~~~~~~~~~~~~~~~~~~~~~~~~~~~~~({\rm E}1.2) \]
    where all times are those of clocks at rest in the frame S', the proper frame of the rod, and
     AB is the length of the rod in this frame.
    Einstein's conclusion:
    \par{\tt Observers moving with the moving rod would thus find that the two clocks were
    not synchronous while observers in the stationary system would declare the clocks
    to be synchronous.}
    \par is false. This is because the events observed in S appearing in Eqs.~({\rm E}2.2) and ({\rm E}2.3)
     are not the Lorentz-transformed LSCSP events defined in the frame S'.  These latter events as
     observed either in S ' or S, as well as the events in S considered by Einstein are shown in Table 1
     and Fig.6. 

\begin{table}
\begin{center}
\begin{tabular}{|c||c|c|c|c|}  \hline
 Event & $\tau' = \tau'_A/\gamma =  \tau'_B/\gamma$ & $x'$ & $\tau$  & $x$  \\ \hline
 Light signal at A & $0$ & $0$ & $0$ & $0$ \\ \hline
 Reflection from M' & $L/c$ & $L$ & $\gamma L/c$ & $L(1+\gamma \beta)$ \\  \hline
 Reflection from M & $L/[\gamma(c-v)]$ & $L[2-\gamma(1+\beta)]$ & $L/(c-v)$ & $cL/(c-v)$ \\  \hline
 Return to A in S' & $2L/c$ & $0$ & $2 \gamma L/c$ & $2\gamma \beta L$ \\  \hline
 Return to A in S & $2 \gamma L/c$ & $-2(\gamma-1)L$ & $2 \gamma^2 L/c$ & $2 \gamma^2 \beta L$  \\ \hline
\end{tabular} 
\caption[]{{\em  Times and positions in S ($\tau$ and $x$) and in S' ($\tau'$ and $x'$)
  for the events shown in Fig.6}}     
\end{center}
\end{table}

\begin{figure}[htbc]
\begin{center}\hspace*{-0.5cm}\mbox{
\epsfysize15.0cm\epsffile{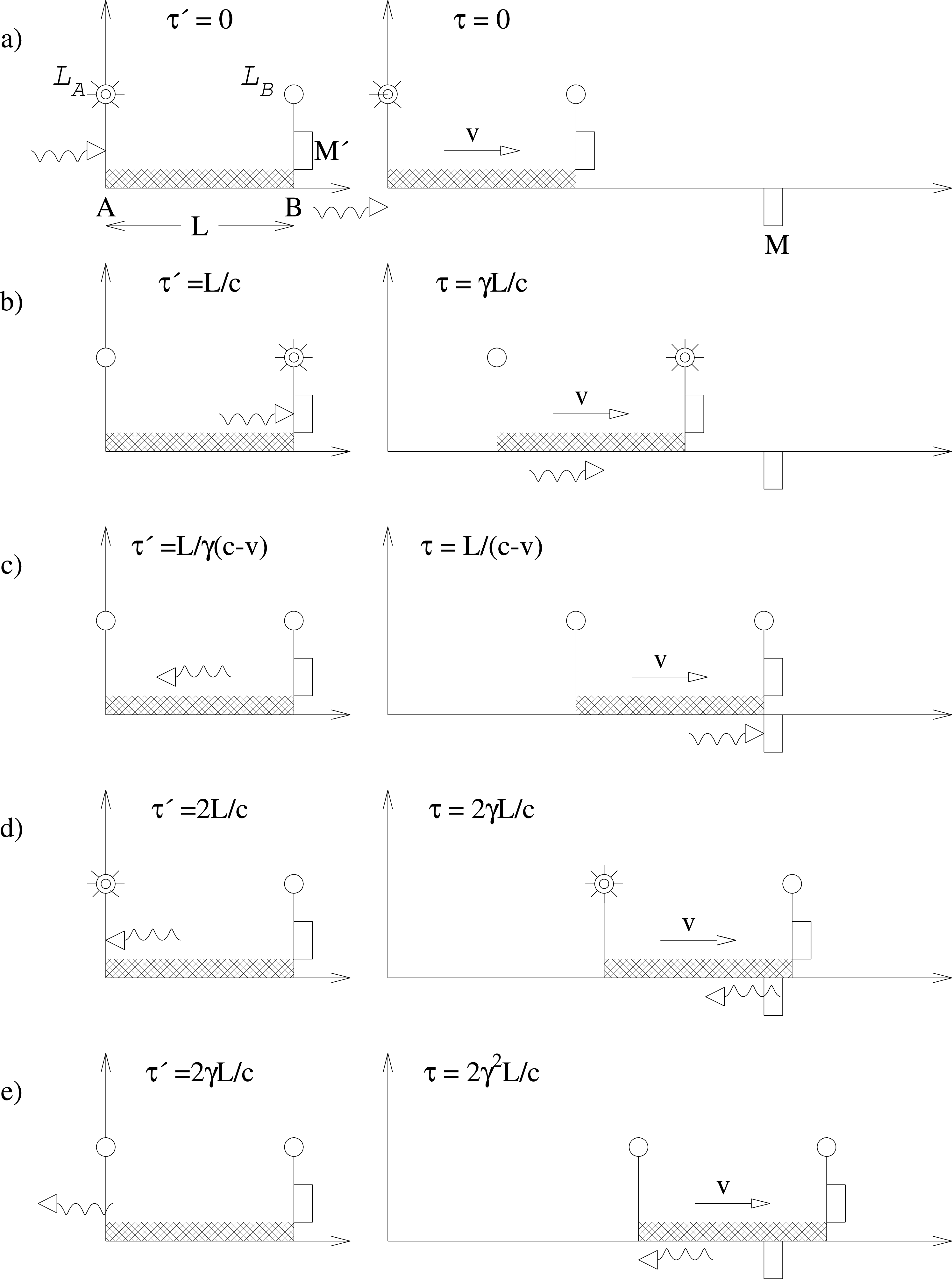}}
\caption{ {\em Einstein's light synchronisation procedure is carried out in the frame 
    S' (left hand figures) and observed in the frame S (right hand figures).
 The lamps {\it L}$_A$ and {\it L}$_B$ in S' are triggered by the passage of the light signal in S'
   and observed from S. a) left, $\tau' = 0$, light signal passes A,  {\it L}$_A$ fires.
   b) left  $\tau' = L/c$, light signal is reflected at the mirror at B,  {\it L}$_B$ fires.
   d)  left  $\tau' = 2L/c$, light signal arrives back at A,  {\it L}$_A$ fires a second time.
      The events observed in S at the corresponding times $\tau = 0$, $\tau = \gamma L/c$,
       $\tau' = 2 \gamma L/c$ are shown in the right-hand figures in a), b), d), respectively.
      Also shown in  a) c) and e) right are the light signal events in the frame S considered by Einstein, together
     with (left) the positions of the light signal in S', considered in a), b) and d), at the corresponding times. The position
   of the light signal is at the head of the wavy arrow in each figure. In this figure $\beta = c/2$,
    $\gamma = 2/\sqrt{3}$.}}
\label{fig-fig6}
\end{center}
\end{figure}

     \par It is assumed that, in addition to the mirror, M', at B, there are also light detectors
    (not shown) at A and B connected to the lamps {\it L}$_A$ and {\it L}$_B$ which flash
   when photons are detected near A and B. For example, in Fig.6a, {\it L}$_A$ flashes when the light
   signal is close to A. The flashes of the lamps are visible to observers in both S and S'
   and so establish the correspondence between the same event seen in the two frames.
   Events as observed in the frame S' are shown on the left side of Fig.6, the same events, as
   observed in S, on the right side. Figs.6a, 6b and 6d show the events of the light signal 
    synchronisation procedure as performed in the frame S', while Figs.6a, 6c and 6e show the 
   events considered by Einstein in the frame S. In the frame S, the light signal is reflected from
   the mirror M that is in spatial coincidence with M' at the instant of reflection in S (Fig.6c, right).
   \par Inspection of Table 1 and Fig.6 shows that there is no correlation between the events introduced
     by Einstein
   in the frame S, with times calculated according to the relations ({\rm E}2.2) and ({\rm E}2.3) and the clock
    synchronisation events in the frame S'. The times indicated in Fig.6 are the proper times, as recorded
    by identical clocks, in these frames. The clocks at A and B introduced by Einstein, that are synchronous
    with a clock in S, record times: $\tau'_A = \tau'_B = \gamma \tau'$ where $\tau'$ is the proper time in 
    S' as recorded by a clock identical to the one in S. Using Einstein's notation of ({\rm E}1.1) for the times 
   of events in S' (that is, setting, here, $\tau' = t$) the events shown in Table 1 and Fig.6 give:
    \begin{eqnarray}
      t_A &=& 0 = \tau'_A, \\
      t_B &=& L/c = \tau'_B/\gamma, \\
      t'_A &=& 2L/c = (\tau'_A)'/\gamma.
    \end{eqnarray}
  These equations give:
   \begin{equation}
        t_B-t_A = t'_A-t_B = L/c.
   \end{equation}
    and 
     \begin{equation}
        \tau'_B-\tau'_A = (\tau'_A)'-\tau'_B = \gamma L/c
   \end{equation}   
   The last equation shows that the clocks at the ends of the rod
   that are synchronous with the clock in S will also be seen to be
   synchronous, according to Einstein's light signal procedure, by observers in the frame S'. This conclusion is
  contrary to Einstein's one. 
  \par Einstein's claim of non-simultaneity  is based on ({\rm E}2.2) and ({\rm E}2.3), referring to the events in S shown 
   in Figs.6c and 6e,  and for which:
    \[ t_B-t_A \ne t'_A-t_B. \]
    Einstein incorrectly identifies these events with those of the synchronisation procedure in S'
    and so reaches the false conclusion that the clocks at A and B are not seen to be synchronous in S'.
   \par It is clear however, from the very concept of Einstein's synchronisation procedure that both
   the photon source and the mirror must be at rest in some frame. Only then is the relation ({\rm E}2.1) valid.
   The difference between $t_B-t_A$ and $t'_A-t_B$ for the events shown in Figs.6c and 6e has a trivial,
   and purely classical, explanation. The relative velocity of the light signal and the mirror M', as
   observed in S, is $c-v$ before reflection and $c+v$ after reflection, whereas formula ({\rm E}2.1), which is
   the basis of Einstein's synchronisation procedure requires the `light path' to be fixed
    in some frame, not in motion, as is the case for equations ({\rm E}2.2) and ({\rm E}2.3). There is confusion
   here between the `relative velocity' between a moving object and a light signal
   as observed in some frame, which can be greater of less than $c$, and
    the `speed of light', defined by ({\rm E}2.1), which is the constant $c$ in any inertial frame.
 
  \par Einstein mentions `relativity of length' in the title of this section and makes the statement that a moving rod
   will be measured to have a different length from a stationary one. However no arguments are given to
   support this assertion, or mention made of the following \S 4, where Einstein claims to derive the
   `length contraction' effect.

\SECTION{\bf{\S 3. Theory of the Transformation of Co-ordinates and Times from a Stationary
 System to another System in Uniform Motion of Translation Relative to the Former}}
 In this section Einstein's notation for space-time coordinates is retained. However, as is now conventional,
 Einstein's frames K, k  are denoted, as above, by S, S' respectively. An event as observed in S, S' then has 
   space-time coordinates ($x$,$y$,$z$,$t$), ($\xi$,$\eta$,$\zeta$,$\tau$) respectively.
 \par Einstein's derivation of the LT is based on the LSCSP introduced in \S 1. However, the same false connection
  between events in the frames S and S' is assumed as in the discussion of relativity of simultaneity in \S 2.
  In spite of this, Einstein does succeed to derive the correct LT, at least the one appropriate to
  a synchronised clock situated at the origin of S'. Einstein assumes that the LT is linear in the space-time
  coordinates and seeks first to derive the time transformation equation. It would seem natural to assume
  the functional dependence $\tau = \tau(x,y,z,t)$ for this equation. Einstein however does not do this.
   It is assumed instead that $\tau = \tau(x',y,z,t)$ where:
  \[~~~~~~~~~~~~~~~~~~~~~~~~~~~~~~~~x' \equiv x-vt = x(t=0) = {\rm~constant}~~~~~~~~~~~~~~~~~~~~~~~~~~~~({\rm E}3.1) \] 
  even though the final result ({\rm E}3.7) has the functional dependence $\tau = \tau(x,t)$. This, unlike the initial
   ansatz, does depend on $x$, (as it should), whereas the $x'$ dependence (which must also be there) has been
    lost. How this comes about is explained below. Actually, for synchronised clocks, the time transformation
    ({\rm E}qn(2.18), in Einstein's notation ):
    \begin{equation}
      \tau = t/\gamma
    \end{equation}
     depends on neither $x'$ nor $x$. It would seem however to be impossible to derive this equation
     by assuming simply $\tau = \tau(t)$ as an ansatz. Actually, (6.1) has been derived by combining
      (2.15) and (2.16). The latter equation has the functional dependence (in Einstein's notation)
      of $\tau = \tau(x',x,t)$ as it must have in order to correctly describe synchronised clocks with
      different values of $x'$. Note that in (2.15) and (2.16) Einstein's constant $x'$ is $L$.
       Einstein's neglect of the essential $x$ dependence of the LT in his initial
    ansatz is the first error in his derivation. Other, compensating, errors enable the correct 
    transformation to be finally obtained.
    \par Einstein next considers a light signal emitted from the origin of S'; parallel to the $\xi$ axis,
   that is reflected straight back by a mirror, and observed again at the origin of S'. This is just the sequence
   of events shown in Figs.6a, 6b, and 6d for which the space-time coordinates in the two frames are
    presented in the first, second and fourth rows of Table 1. However Einstein assumes, following the
    argument presented in \S 2, that the reflection event in S' is that shown in Fig.6c and the third row
     of Table 1, while the return-to-origin event in S is that in Fig5e and the fifth row of Table 1.
     \par The LSCSP relation ({\rm E}3.2) that is the basis of Einstein's derivation of the LT will now be discussed,
      first inserting into it the correct times and positions in the frames S and S' from Table 1, and then, 
    following Einstein's argument, with the functional dependence $\tau = \tau(x',y,z,t)$, and assuming that the
    corresponding events in S are those in the third and fifth rows of Table 1.  
     \par Since the light signals have $y = \eta = z  = \zeta = 0$, the corresponding time transformation equation
     has, in general, the functional dependence  $\tau = \tau(x',x,t)$.  The LSCSP relation in the frame S':
     \begin{equation}
       \frac{1}{2}[ \tau_A + \tau'_A] = \tau_B
     \end{equation}
     gives, noting that (see Fig.6) $x' = 0$ for the clock at A and $x' = L$ for the clock at B:
     \begin{equation}
       \frac{1}{2}[ \tau(0,0,0) + \tau(0,2\beta \gamma L,2 \gamma L/c)] = \tau(L, (1+\gamma \beta)L, \gamma L /c).
     \end{equation}
     Assuming linearity of the equations and partially differentiating (6.3) with respect to its arguments
    \footnote{Einstein here makes the unnecessary assumption, in view of the postulated
       linearity of the equations that $x'$ is infinitesimally small.} gives:
    \begin{equation}
     \gamma \beta \frac{\partial \tau}{\partial x} +\frac{\gamma}{c} \frac{ \partial \tau}{\partial t} =
       \frac{\partial \tau}{\partial x'}  
     +(1+\gamma \beta)\frac{\partial \tau}{\partial x} + \frac{\gamma}{c} \frac{ \partial \tau}{\partial t} 
    \end{equation}
      or
    \begin{equation}
       \frac{\partial \tau}{\partial x}  = - \frac{\partial \tau}{\partial x'} 
   \end{equation}
    consistent with Eq.~(2.16) after the substitutions, to recover the notation of Section 2, 
    $\tau \rightarrow t'$,   $t \rightarrow \tau $ and $x'\rightarrow L$. 
     \par With the functional dependence  $\tau = \tau(x',t)$ assumed by Einstein, and taking the corresponding
      events in S from the third and fifth rows of Table 1, the LSCSP relation (6.2) is written as:
\[~~~~~~~~~~~~~~~~~~~~~~~~~~~~~\frac{1}{2}[ \tau(0,0) + \tau(0,\frac{x'}{c-v}+ \frac{x'}{c+v})] = \tau(x',\frac{x'}{c-v}).~~~~~~~~~~~~~~~~~({\rm E}3.3) \]
       Partial differentiation, assuming linearity gives: 
 \[~~~~~~~~~~~~~~~~~~~~~~~~~~~~~~~~~~~~~~~~~~~~~~~\frac{\partial \tau}{\partial x'} + \frac{v}{c^2-v^2}\frac{\partial \tau}{\partial x} = 0.~~~~~~~~~~~~~~~~~~~~~~~~~~~~~~~({\rm E}3.5) \]
      In view of the linearity of the equations, ({\rm E}3.5) is satisfied providing that:
 \[~~~~~~~~~~~~~~~~~~~~~~~~~~~~~~~~~~~~~~~~~~~~~~~~\tau = a \left(t- \frac{v}{c^2-v^2}x'\right).~~~~~~~~~~~~~~~~~~~~~~~~~~~~~~~~({\rm E}3.7) \]
      For the clock at A, where $x' = 0$, the time transformation is therefore 
    \begin{equation}
    \tau_A = a t_A.
     \end{equation}
      In agreement with (6.1). However $a = a(v)$ remains undetermined.
        
      Since the clock at B has $x'_B = L$, it follows from ({\rm E}3.7) that:
      \begin{equation}
  \tau_B = a \left(t_B- \frac{v}{c^2-v^2}L\right).
  \end{equation}
    Eq.~({\rm E}3.7) or (6.7), together with (6.6), already implies a `relativity of simultaneity' effect. If $\tau_A = \tau_B$ then $t_A \ne t_B$
    and vice versa. Comparing ({\rm E}3.5) and (6.5) this can be seen to be the consequence of two errors in
     Einstein's calculation:
   \begin{itemize}
   \item[(i)] Neglect of the essential $x$-dependence of $\tau$.
   \item[(ii)] Subsitution of incorrect corresponding events in the frame S in the LSCSP relation.
   \end{itemize}
    When the correct events in S are substituted into Eq.~(6.2),(as in Eq.~(6.4)), and the necessary
     $x$-dependence of the transformation is taken into account, the spurious relativity of simultaneity
    effect of Eq.~(6.7) (the dependence of $\tau_B$ on $L$) is removed by the condition (6.5),
    equivalent to the substitutions $x' \rightarrow x' - L$ and $x \rightarrow x - L$ in the LT of Eqs.~(2.5)
    and (2.6) which describe a synchronised clock with $x' = L = 0$.
    \par Einstein now considers a light signal emitted in the frame S' at $\tau = 0$ along the
      $\xi$ axis with the equation of motion:
\[~~~~~~~~~~~~~~~~~~~~~~~~~~~~~~~~~~~~~~~~~~~~~~~~~~~~~\xi = c \tau.~~~~~~~~~~~~~~~~~~~~~~~~~~~~~~~~~~~~~~~~~~~({\rm E}3.8) \] 
  
  The time in S of the reflection event at M' is then assumed (incorrectly, see Table 1 and Figs.6b,6c) to be
 \[~~~~~~~~~~~~~~~~~~~~~~~~~~~~~~~~~~~~~~~~~~~~~~~t_B = \frac{x'_B}{c-v}.~~~~~~~~~~~~~~~~~~~~~~~~~~~~~~~~~~~~~~~~~~~({\rm E}3.9) \]
   Substituting this value of $t_B$ in ({\rm E}3.7) and using ({\rm E}3.8) then gives the space-coordinate transformation
   equation:
  \[~~~~~~~~~~~~~~~~~~~~~~~~~~~~~~~~~~~~~~~~~~~~\xi_B = a\frac{c^2}{c^2-v^2} x'_B.~~~~~~~~~~~~~~~~~~~~~~~~~~~~~~~~~~~~~~~~({\rm E}3.10) \]
     Considering light signals moving along the $\eta$ and $\zeta$ axes in S' and assuming that the apparent 
    speed of light in S is $c$ leads to the relations:
\[~~~~~~~~~~~~~~~~~~~~~~~~~~~~~~~~~~~~~~~\eta = a\frac{c}{\sqrt{c^2-v^2}} y \equiv \phi(v) y,~~~~~~~~~~~~~~~~~~~~~~~~~~~~~~~~~({\rm E}3.16) \]
 \[~~~~~~~~~~~~~~~~~~~~~~~~~~~~~~~~~~~~~~~\zeta = a\frac{c}{\sqrt{c^2-v^2}} z = \phi(v) z.~~~~~~~~~~~~~~~~~~~~~~~~~~~~~~~~~({\rm E}3.17) \]
     Note that the $y$, $z$ LT equations (2.3) and (2.4) (({\rm E}3.29) and ({\rm E}3.30) in Einstein's notation)
     together with
     the TD relation (6.1) predict, in accordance with Einstein's assumption in deriving ({\rm E}3.16) and ({\rm E}3.17)
    that the apparent speed, in S, of light signals propagating perpendicular to the $\xi$ axis
    in S' is $c$. If $r$ is the path length of the signal in S, then the apparent speed of light in S
     of a light signal propagating parallel to the $\eta$ axis is:
  \begin{equation}
     c_{\eta}^{app} = \frac{r}{t} = \frac{\sqrt{y^2+v^2t^2}}{t} = \sqrt{(\frac{c\tau}{t})^2 + v^2} = c       
   \end{equation}
     where the time dilation relation (6.1), $y = \eta = c \tau$ and the definition of $\gamma $ have been used.
     This is to be contrasted with the apparent speed of light in S of signals moving parallel
    to the $\xi$ axis. The entries of Table 1 give the following apparent velocities for the
    photons following the paths AB and BA in S':
   \begin{eqnarray}
    c_{\xi,AB}^{app} & = & c\left(\frac{1}{\gamma} + \beta\right), \\
  c_{\xi,BA}^{app} & = & c\left(\frac{1}{\gamma} - \beta\right).
   \end{eqnarray}
   \par Subsituting $x'$ from Eq.~({\rm E}3.1) in ({\rm E}3.7) and ({\rm E}3.10) gives, respectively:
    \begin{eqnarray}
 ~~~~~~~~~~~~~~~~~~~~~~~~~~~~~~~~\tau & = & \phi(v) \gamma (t-\frac{v x}{c^2}),~~~~~~~~~~~~~~~~~~~~~~~~~~~~~~~~~~~~~~~({\rm E}3.14) \nonumber  \\
 ~~~~~~~~~~~~~~~~~~~~~~~~~~~~~~~~ \xi & = & \phi(v) \gamma (x-vt)~~~~~~~~~~~~~~~~~~~~~~~~~~~~~~~~~~~~~~~~({\rm E}3.15) \nonumber 
 \end{eqnarray}
      where the subscript $B$ is dropped. 
    Einstein having now arrived at the space-time LT up to a multiplicative constant in ({\rm E}3.14)-({\rm E}3.17), 
   the use of the inverse transformation (tacitly assuming here the Reciprocity Principle~\cite{BG,JHFLT1})
    and spatial isotropy, to derive the relations:
 \[~~~~~~~~~~~~~~~~~~~~~~~~~~~~~~~~~~~~~~~~~\phi(v)\phi(-v) = 1,~~~~~~~~~~~~~~~~~~~~~~~~~~~~~~~~~~~~~~~~({\rm E}3.25) \] 
\[~~~~~~~~~~~~~~~~~~~~~~~~~~~~~~~~~~~~~~~~~~\phi(v) = \phi(-v)~~~~~~~~~~~~~~~~~~~~~~~~~~~~~~~~~~~~~~~~~({\rm E}3.26) \] 
      and hence $\phi(v) = \phi(-v) = 1$, is unexceptionable, giving finally the LT as:
   \begin{eqnarray}
  ~~~~~~~~~~~~~~~~~~~~~~~~~~~~~~~~~~~~\tau & = & \gamma(t-\frac{v x}{c^2}),~~~~~~~~~~~~~~~~~~~~~~~~~~~~~~~~~~~~~({\rm E}3.27) \nonumber  \\
  ~~~~~~~~~~~~~~~~~~~~~~~~~~~~~~~~~~~~~~~~~~\xi & = & \gamma(x-vt),~~~~~~~~~~~~~~~~~~~~~~~~~~~~~~~~~~~~~~({\rm E}3.28) \nonumber  \\
  ~~~~~~~~~~~~~~~~~~~~~~~~~~~~~~~~~~~~~~~~~~\eta & = & y,~~~~~~~~~~~~~~~~~~~~~~~~~~~~~~~~~~~~~~~~~~~~~~~~~({\rm E}3.29) \nonumber  \\
  ~~~~~~~~~~~~~~~~~~~~~~~~~~~~~~~~~~~~~~~~~~\zeta & = & z.~~~~~~~~~~~~~~~~~~~~~~~~~~~~~~~~~~~~~~~~~~~~~~~~~({\rm E}3.30) \nonumber 
    \end{eqnarray}  

    \par The correct LT ({\rm E}3.27)-({\rm E}3.30) for a synchronised clock at A has now been derived by misapplication
    of Eqs.~({\rm E}3.14),({\rm E}3.15) to the clock at B, for which, lacking the required $x'$ (or $L$) dependence of (2.15)
    and (2.16), they are not valid, and, in parallel, misinterpreting the meaning of
    Einstein's postulate 2, by assigning
    incorrect events in S in correspondence with the LSCSP events in S' (see Fig.6). It is interesting to
     note here that Einstein knew that the LT must depend explicitly on $x'$ both because of the
     initial ansatz $\tau = \tau(x',y,z,t)$ and the explicit statement~\cite{Ein1A} concerning the necessary
      additional additive constants on the
      right sides of ({\rm E}3.14)-({\rm E}3.17), noted in Section 2 above. However, by omitting the $x$-dependence
     of $\tau$ in the initial ansatz, the LT for the clock at B (with $x' = L$) was found to be the
     one that is actually valid for the clock at A (where $x' = 0$) and so does not correctly describe
   a synchronised 
     clock at B. Such synchronised clocks actually exhibit the universal TD effect of Eq.~(2.18), that is, there
 is no `relativity of simultaneity' as implied if Eqs.~(6.6) and (6.7). Afterwards,
      in the
    applications of the LT in Ref.~\cite{Ein1},
    the missing $x'$ dependence of the final LT equations ({\rm E}3.27)-({\rm E}3.30), for any synchronised clock that
     does not have $\xi = 0$,
    is forgotten\footnote{It was universally forgotten also throughout the 20th Century},
    leading to the prediction,
      as will be seen below,  
    of a spurious `length contraction' effect in \S 4 of
    Ref.~\cite{Ein1}.       

 \SECTION{\bf{\S 4. Physical Meaning of the Equations Obtained In Respect to Moving Rigid Bodies and Moving
   Clocks}}
  Einstein first considers a rigid sphere of radius $R$ at rest in S' with its centre at the origin. The equation
  of this sphere is 
 \[~~~~~~~~~~~~~~~~~~~~~~~~~~~~~~~~~~~~~~~~~~~\xi^2+\eta^2+ \zeta^2 = R^2.~~~~~~~~~~~~~~~~~~~~~~~~~~~~~~~~~~~~~~~~~({\rm E}4.1) \]
    Setting $t=0$ in the LT ({\rm E}3.27)-({\rm E}3.30) and using it to eliminate $\xi$, $\eta$ and $\zeta$ in favour of
    $x$, $y$ and $z$ in Eq.~(4.1) gives the equation:
 \[~~~~~~~~~~~~~~~~~~~~~~~~~~~~~~~~~~~~~~~\frac{x^2}{1-(v/c)^2} + y^2 +z^2 = R^2~~~~~~~~~~~~~~~~~~~~~~~~~~~~~~~~~~~({\rm E}4.2) \]
   which represents an ellipsoid of revolution. This however does not correctly describe the moving sphere
   as seen from S, since the LT ({\rm E}3.27)-({\rm E}3.30) appropriate for $\xi = 0$ (i.e. points
     on the intersection of the sphere with the $\eta$-$\zeta$ plane) has been
   also used to transform other points on the surface of the sphere where it is no longer applicable).  
  \par Introducing spherical coordinates $\theta$, $\phi$ the coordinates of an arbitary point P on the sphere
  are:
  \begin{eqnarray}
  \xi & = & R\sin \theta \cos \phi, \\
  \eta & = & R\sin \theta \sin \phi, \\
   \zeta & = & R \cos \theta.
  \end{eqnarray}
  Assume that a clock, situated at the point P, is synchronised with a clock in S so that $\tau = t = 0$ when the center
   of the sphere is at the origin of coordinates in S. The LT giving the corresponding space-time coordinates
   in S is given, in Einstein's notation by Eqs.~(2.15), (2.16), (2.3) and(2.4) as:
 \begin{eqnarray}
\tau & = & \gamma(t-\beta(x - R \sin \theta \cos \phi)/c], \\ 
 \xi - R \sin \theta \cos \phi  & = & \gamma[x - R \sin \theta \cos \phi -v t] = 0, \\
  \eta & = & R\sin \theta \sin \phi = y, \\
   \zeta & = & R \cos \theta = z.
  \end{eqnarray}
  Combining ({\rm E}4.1), (7.5), (7.6) and (7.7) the surface observed in S at time $t$ is therefore:
 \begin{equation}
   (x-vt)^2 + y^2 +z^2 = R^2
 \end{equation}
  which is a sphere centered at $x = v t$. There is no therefore no shortening of the $x$-dimension
  of the sphere when the LT correctly describing a synchronised clock at any point on its
  surface, (7.4)-(7.7), is applied.
   \par Einstein's error is to use the LT ({\rm E}3.27)-({\rm E}3.30), that correctly describes a synchronised clock
     (i.e. one for which $\tau = 0$ when $t = 0$) at the origin of S', that is the center of the sphere,
     incorrectly to describe a synchronised clock at the point P on the surface of the sphere at which $\xi \ne 0$.
 At $t = 0$
 Eq.~({\rm E}3.27) gives
     \begin{equation}
 \tau_P = -\frac{ \gamma \beta}{c} R \sin \theta \cos \phi.
    \end{equation}
   The time in S' varies with the position of P so that clearly ({\rm E}3.27) does not describe a synchronised clock
   at P. The operation of clock synchronisation is a purely mechanical or electronic one, entirely under
    the control of the experimenter and independent of space-time physics. For example, clocks at different positions
   on the sphere can be synchronised by signals sent along equal-length cables connected to the centre of
   the sphere. When the latter coincides with the origin in S a triggering system sends synchronous signals
    along each cable to start the clocks, which each have an initial time offset to compensate
    for the signal delays~\cite{JHFST1}. Since the LT ({\rm E}3.27)-({\rm E}3.30) does not describe such a synchronised
   clock when $\xi \ne 0$, the length contraction effect of Eq.~({\rm E}4.2) is spurious. For further discussion of this
   point See Refs.~\cite{JHFST1,JHFUMC,JHFCRCS,JHFLLT}.
     \par The most important, and correct, conclusion of Ref~\cite{Ein1} concerning
      space-time physics is the time dilation
       formula ({\rm E}4.5) which is derived in an identical manner to Eq.~(2.18) in Section 2 above. Note that, unlike
      for the discussion of length contraction, the derivation of the TD formula is not dependent on the 
      value assigned to Einstein's parameter $x' = x(t=0)$ ($L$ in Eqs.~(2.15) and (2.16)). It is indeed 
      precisely assigning  incorrect values of $x'$ to clocks on the surface of the sphere that results in the
      spurious prediction of length contraction.
    \par After deriving the TD formula (6.1) Einstein discusses observations of times recorded by stationary
         or moving clocks. The case when a moving clock starts at the same position as the stationary one,
          moves along a curve ({\rm E}instein considers this curve as the limit of a polygonal line) and returns
        to its starting position, is now
           referred to as the `Twin Paradox'~\cite{Langevin} ---less time has elapsed during the round trip according
      to observations of the moving clock as compared to the stationary one. 
      \par  Einstein concludes \S 4 with the statement:
    \par {\tt  Thence we conclude that a balance-clock (Not a pendulum clock which is \newline physically a system
      to which the Earth belongs. This case had to be \newline excluded.) at the equator must go more slowly, by a small
      amount, than a \newline precisely similar clock situated at one of the poles under otherwise \newline
      identical conditions.}
      \par The correctness of the phrase `go more slowly'  depends on the frame from which the clock
      is being observed. Greater clarity would be obtained by replacing  `go more slowly'  with
        `appears to go more slowly when observed from the frame F than when it is observed in its own
          rest frame'. The question  then arises: what are the frames F for which this statement is true?
         Einstein specifies an observer at a pole of the Earth, that is one that does not
       partake of the rotational motion of the Earth and a clock at the equator that does. Thus, to a very
     good approximation, the observer's frame F is an inertial one co-moving with the centroid of the 
      Earth, whereas the equator clock is in an accelerated frame. 
   \par Only special relativistic effects were considered by Einstein in Ref.~\cite{Ein1}. In fact, due
     to the rotation of the Earth, its surface bulges outward at the equator, which  places an observer 
      there at a higher gravitational potential\footnote{The gravitational potential is defined to be
        negative and to vanish at infinity} than one at the pole. It turns out~\cite{HSPT,BH,Cocke} that,
    for example,
    for an atomic clock, the increase of frequency of equator clock photons observed at the pole due to
    gravitational
     blue shift almost exactly compensates TD (the transverse Doppler effect) for an observer at a pole.
     Indeed, as shown in Refs.~\cite{BH,Cocke}, this cancellation between TD and gravitational
    frequency shifts  occurs for any pair of clocks situated on the Earth's geoid ---a surface at sea-level
    perpendicular to the direction of a local plumb line which is a equipotential
        surface of the combined gravitational and centrifigal potential.
     \par Einstein's exclusion from consideration of pendulum clocks in the English translation 
    of Ref.~\cite{Ein1} means that he must have later considered the importance of gravitational
     effects, if not already gravitational frequency shifts, at least the dependence of the period of
     a pendulum on the size of the ambient gravitational field.

\SECTION{\bf{\S 5. The Composition of Velocities}}
   Einstein derived the composition laws for longitudinal and transverse velocity components by simple
  substitution of the relations:
  \[~~~~~~~~~~~~~~~~~~~~~~~~~~~~~~~~~\xi = \omega_{\xi}\tau,~~~~ \eta =  \omega_{\eta}\tau,~~~~\zeta = 0~~~~~~~~~~~~~~~~~~~~~~~~~~~~~~~~~~~~({\rm E}5.1) \] 
 in the LT equations ({\rm E}3.27)-({\rm E}3.30). This leads immediately to the relations:
   \begin{eqnarray}
 ~~~~~~~~~~~~~~~~~~~~~~~~~~~~~~~~~~~~x & = & \frac{(\omega_{\xi}+v)t}{1+\frac{v \omega_{\xi}}{c^2}},~~~~~~~~~~~~~~~~~~~~~~~~~~~~~~~~~~~~~~~~~~~~~~~~~~~({\rm E}5.2) \nonumber  \\
  ~~~~~~~~~~~~~~~~~~~~~~~~~~~~~~~~~~~~y & = & \omega_{\eta}t \frac{\sqrt{1-(v/c)^2}}{1+\frac{v \omega_{\xi}}{c^2}}.~~~~~~~~~~~~~~~~~~~~~~~~~~~~~~~~~~~~~~~~~~~({\rm E}5.3) \nonumber
 \end{eqnarray}
  Specialising to the case of motion in S' parallel to the $\xi$-axis (i.e. $\omega_{\eta} = 0$)
  and setting $x = Vt$ ({\rm E}5.2) yields the parallel velocity addition relation (PVAR)
 \[~~~~~~~~~~~~~~~~~~~~~~~~~~~~~~~~~~~~~~~  V = \frac{v + \omega}{1+\frac{v  \omega}{c^2}}~~~~~~~~~~~~~~~~~~~~~~~~~~~~~~~~~~~~~~~~~~~~~~~~~~~({\rm E}5.9) \]
   where
  \[~~~~~~~~~~~~~~~~~~~~~~~~~~~~~~~~~~~~~~~\omega^2 = \omega_{\xi}^2 + \omega_{\eta}^2.~~~~~~~~~~~~~~~~~~~~~~~~~~~~~~~~~~~~~~~~~~~~~~~~~~({\rm E}5.6) \]
  \par Einstein derived the PVAR directly from the LT, but this relation may also be obtained assuming
   only Einstein's second postulate~\cite{Mermin,Peres} or, alternatively, by assuming that $V$ is a
    single-valued function of $v$ and $\omega$ and using the Reciprocity Principle~\cite{JHFLT1}.
     \par The PVAR really involves three inertial frames S, S' and S$_0$ where the last is the
     proper frame of the moving object. Even so, Einstein's derivation uses only the LT
  ({\rm E}3.27)-({\rm E}3.30) connecting the frames S and S'. To understand the operational meaning of the 
   symbols used in this derivation it is convenient to revert to the notation af Section 2.
    The observer in the thought experiment corresponding to the derivation of the PVAR is at rest in 
    the frame S and the PVAR gives the velocity of the object relative to this observer, $V$, in the
    case that that the velocity of the object relative to an observer at rest in the frame S' is
    $\omega$. The appropriate LT is therefore (2.15) and (2.16) with $L = 0$ and
     $x' = \omega  \tau'(\Cp) = \omega t'({\rm S})$,
      $x = V \tau$:
     \begin{eqnarray}
    \omega t'({\rm S}) & = & \gamma [V - v] \tau, \\
     t'({\rm S}) & = & \gamma [1 - \frac{v V}{c^2}]\tau
    \end{eqnarray}
        where the equivalence of Eq.~(2.31) between the apparent time interval t'({\rm S}) and
       a proper time interval $\tau'(\Cp)$ has been used.
    In these equations, whereas $\tau$ is a proper time recorded by a clock
  at rest in S, $t'({\rm S})$ is the apparent time of a clock at rest in S' as viewed from S. 
   The ratio of (8.1) to (8.2) is:
   \begin{equation}
     \omega = \frac{V-v}{1 - \frac{v V}{c^2}}
   \end{equation} 
    which gives, after rearrangement, ({\rm E}5.9).
    \par With the definitions: $\beta_V \equiv V/c$,  $\beta_\omega \equiv \omega/c$
      (8.3) may be written as:
    \begin{equation}
     \beta_\omega = \frac{\beta_V-\beta}{1 -\beta_V\beta}.
   \end{equation}
   By algebraic manipulation, this equation may, equivalently, be written in either of the forms:
        \begin{eqnarray}
        \gamma_{\omega} \beta_{\omega} &  = & \gamma[ \gamma_V \beta_V - \beta \gamma_V ], \\
        \gamma_{\omega} &  = & \gamma[ \gamma_V - \beta  \gamma_V \beta_V ]
    \end{eqnarray}  
     where $\gamma_V \equiv 1/\sqrt{1-\beta_V^2}$ and $\gamma_{\omega} \equiv 1/\sqrt{1-\beta_{\omega}^2}$.
     These equations are the LT between the frames S and S' of the spatial and temporal components,
     respectively, of the dimensionless velocity 4-vector ($\gamma_V$;$\gamma_V \beta_V$,0,0).
     These formulae will be found useful in discussing the transformation properties of electric
     and magnetic fields in the following section.
     \par A more transparent method to derive the PVAR, mentioned by Einstein at the end of
      \S 5, is by successive application of the LT ({\rm E}3.27)-({\rm E}3.30), that is by invoking the group 
       property of the LT. However, no such calculation was presented.

\SECTION{\bf{\S 6. Transformation of the Maxwell-Hertz Equations for Empty Space. On the Nature
    of the Electromotive Forces in a Magnetic Field During Motion}}

   In the first part of this section Einstein derives the transformation laws for
   electric and magnetic fields in `empty space' from the assumed covariance of Maxwell's equations
 obtaining the well-known results, in modern notation and Gaussian units:
    \begin{eqnarray}
 ~~~~~~~~~~~~~~~~~~~~~~~~~~~~~~ E_{\xi} & = & E_x,~~~~ B_{\xi} = B_x,~~~~~~~~~~~~~~~~~~~~~ ~~~~~~~~~~~~~~~~~~~~~~~~~({\rm E}6.13) \nonumber  \\
    ~~~~~~~~~~~~~~~~~~~~~~~~~~~~~~E_{\eta} & = & \gamma [E_y-\beta B_z],~~~~B_{\eta} = \gamma [B_y+\beta E_z],~~~~~~~~~~~~~~~~~~~~~({\rm E}6.14) \nonumber  \\
  ~~~~~~~~~~~~~~~~~~~~~~~~~~~~~~ E_{\zeta} & = & \gamma [E_z+\beta B_y],~~~~B_{\zeta} = \gamma [B_z-\beta E_y].~~~~~~~~~~~~~~~~~~~~~({\rm E}6.15) \nonumber
  \end{eqnarray}
   The fields in these equations are `force fields', that, when substituted into the Lorentz
  force equation
   \begin{equation}
   \frac{d \vec{p}}{d t} = \vec{F} = q(\vec{E}+\vec{\beta} \times \vec{B})
    \end{equation}
      where $\vec{p} \equiv  \gamma \vec{\beta} m c$ is the relativistic 3-momentum,
     give the force $\vec{F}$ on a test charge $q$
    at the field point under consideration. The fields of \S 8 of the paper although represented by identical
    symbols are actually `radiation fields' that provide the classical description of the creation,
    propagation and destruction of real photons. However in \S 9 and  \S 10 force fields are again considered.
     \par As will be seen shortly, a certain tacit asumption concerning the nature of electric and
      magnetic fields underlies the transformation equations ({\rm E}6.13)-({\rm E}6.15). To understand this assumption
        it will be found convenient to write the transformation equations in tensor notation.
        To render the equations more transparent the  notation of Section 2 for space-time
     coordinates in the frame S' ($x'$,$y'$,$z'$,$t'$) instead of Einstein's ($\xi$,$\eta$,$\zeta$,$\tau$) is
     used. The electromagnetic field tensor $F^{\alpha \beta}$ is defined by the 
    equations:
    \begin{eqnarray}
    -E^i \equiv F^{0i} \equiv \partial^0 A^i-\partial^i A^0, \\
    -B^i \equiv \epsilon_{ijk} F^{jk} \equiv  \epsilon_{ijk}(\partial^j A^k-\partial^k A^j)
     \end{eqnarray}
      where $A^{\alpha}$ is the 4-vector electromagnetic potential.
       The  alternating tensor, $\epsilon_{ijk}$, is equal to $+1(-1)$ for even(odd) permutations of $ijk$.
      Greek indices $\alpha$,$\beta$,... take
     the values 0,1,2,3 and latin indices $i$,$j$,... take the values 1,2,3. Repeated upper and lower
         indices in (9.3) are summed over 1,2 and 3. The contravariant space-time
      4-vector $x^{\alpha}$ is defined according to the relations:
     \begin{equation}
         (x^0=x_0,x^1,x^2,x^3) \equiv (ct,x,y,z).
    \end{equation}
   The following notation for partial derivatives is also used:
    \begin{eqnarray}
    \partial^0 & \equiv & \frac{\partial~}{\partial x^0} = \frac{1}{c}\frac{\partial~}{\partial t}.  \\
      \partial^i & \equiv & -\frac{\partial~}{\partial x^i} \equiv -\nabla_i.
    \end{eqnarray}
  With this notation the transformation laws ({\rm E}6.13)-({\rm E}6.15) are all subsumed in the single tensor equation:
     \begin{equation}
         F'^{\alpha \beta} =\frac{\partial x'^{\alpha}}{\partial x^{\gamma}}
        \frac{\partial x'^{\beta}}{\partial x^{\delta}} F^{\gamma \delta}
    \end{equation}
  where repeated upper and lower indices are summed over 0,1,2 and 3. In (9.7) the values of the
  partial derivatives are derived from the infinitesimal LT between the frames S and S':
   \begin{eqnarray}
 dx' & = & \gamma(dx-\beta dx^0), \\ 
  dx'^0 & = & \gamma(dx^0-\beta dx), \\
   dy'& = & dy,~~~dz' = dz. 
    \end{eqnarray}
    The values of the partial derivatives in (9.7) that give the transformation laws
    ({\rm E}6.13)-({\rm E}6.15) require a certain assumption, concerning the space-time functional
   dependence of the fields, that will now be discussed. This assumption is that all components of
   the electric and magnetic fields are `classical' ones in the sense that all space-time
   coordinates may be considered to be independent variables. If this is the case, then the chain 
    rule of differential calculus is valid for for any generic field component $F$~\cite{Rosser}:
    \begin{eqnarray}
       dF & = &   \frac{\partial F}{\partial x^0}dx^0 + \frac{\partial F}{\partial x} dx
       + \frac{\partial F}{\partial y}dy + \frac{\partial F}{\partial z} dz \nonumber \\
       & = &  \frac{\partial F}{\partial x'^0}dx'^0 + \frac{\partial F}{\partial x'} dx'
       + \frac{\partial F}{\partial y'}dy' + \frac{\partial F}{\partial z'}dz'.
   \end{eqnarray}
     The transformation law between $\partial^{\alpha}$ and $\partial'^{\alpha}$ is found
    by using the inverse equations of (9.8)-(9.10) to eliminate $d\vec{x}$ and $dx^0$ in favour
    of $d\vec{x}'$ and $dx'^0$ in Eq.~(9.11). Collecting terms and equating the coefficients 
     of the three components of  $d\vec{x}'$ and of $dx'^0$ in the second and third members
   of (9.11) gives, since the field $F$ is arbitary, the transformation laws of the partial
    differential operators as:
    \begin{eqnarray}
       \frac{\partial ~ }{\partial x'} & = & \gamma \left(\frac{\partial ~ }{\partial x}+ \beta
      \frac{\partial ~ }{\partial x^0}\right),    \\
      \frac{\partial ~ }{\partial x'^0} & = & \gamma \left(\frac{\partial ~ }{\partial x^0}+ \beta
      \frac{\partial ~ }{\partial x} \right),   \\ 
    \frac{\partial ~ }{\partial y'} & = &   \frac{\partial ~ }{\partial y},~~~
      \frac{\partial ~ }{\partial z'}  =   \frac{\partial ~ }{\partial z}.
    \end{eqnarray}
      These equations show that the partial derivatives transform as a 4-vector under the LT. In virtue
     of the definitions (9.2) and (9.3) of $F^{\alpha \beta}$ and the 4-vector character of $A^{\alpha}$,
     the electric and magnetic fields then transform as a second rank tensor according to Eq.~(9.7).
     The transformation coefficients in (9.7) then have the values, obtained from (9.8)-(9.10):
            \begin{eqnarray}
     \frac{\partial x'}{\partial x} & = & \gamma,~~~~\frac{\partial x'}{\partial y} = \frac{\partial x'}{\partial z}=0,~~~~
      \frac{\partial x'}{\partial x_0} = -\gamma \beta, \\
    \frac{\partial x'_0}{\partial x} & = & -\gamma \beta,~~~~\frac{\partial x_0'}{\partial y}
      =  \frac{\partial x'_0}{\partial z}=0,~~~~\frac{\partial x'_0}{\partial x_0}  =  \gamma, \\
   \frac{\partial y'}{\partial x} & = & 0,~~~~\frac{\partial y'}{\partial y} = 1,~~~~\frac{\partial y'}{\partial z}  =  0,~~~~
   \frac{\partial y'}{\partial x_0} = 0, \\
   \frac{\partial z'}{\partial x} & = & 0,~~~~\frac{\partial z'}{\partial y} = 0,~~~~\frac{\partial z'}{\partial z}  =  1,~~~~
   \frac{\partial z'}{\partial x_0} = 0.
   \end{eqnarray}
    These coefficients, when subsituted into (9.7) give exactly Einstein's  transformation
    laws ({\rm E}6.13)-({\rm E}6.15). It is important to stress again that, in order derive (9.15)-(9.18) 
     the field components $F$ are required to
     respect the chain rule relation (9.11). This requires, in particular, that the x-coordinate 
     and the time must be independent variables. 
     \par In reality, however the force fields in `empty space' considered by Einstein do not
       exist. Since they are defined by the force they exert on a test charge they can be non-zero
        only if, in the problem, there is at least one source charge to produce the fields. 
     The space in the region of the test charge therefore cannot be `empty'. The simplest possible
      electrodynamic system described by force fields therefore contains two electric
      charges, one of which may be considered as the `source' of the fields that act on the
     other 'test' charge, although it is clear, from the symmetry of the system, that these roles
     may be interchanged. Here, the simplest case of a single source charge in uniform
     motion will be considered. The corresponding electric and magnetic fields of pre-relativistic
     classical electromagnetism (CEM) are given by the equations~\cite{Heaviside}:
           \begin{eqnarray}
\vec{E}({\rm CEM}) & = & \frac{Q \vec{r}}{r^3 \gamma_{u}^2(1- \beta_{u}^2 \sin^2 \psi)^{\frac{3}{2}}}, \\
\vec{B}({\rm CEM}) & = & \frac{\vec{u} \times \vec{E}}{c}
    \end{eqnarray}
 whereas, in RCED,  they are given by~\cite{JHFRCED}:
   \begin{eqnarray}
  \vec{E}({\rm RCED}) & = & \frac{Q}{r^2}\left[ \frac{\hat{\imath} \cos \psi}{\gamma_{u}} +
   \hat{\jmath}\gamma_{u}\sin \psi \right], \\
\vec{B}({\rm RCED}) & = & \frac{\vec{u} \times \vec{E}}{c}.
    \end{eqnarray}
  In these equations the velocity of the source charge $Q$ is $\vec{u} = \hat{\imath} \beta_u c$
 and the radius vector is $\vec{r} = r(\hat{\imath} \cos \psi + \hat{\jmath} \sin \psi)$ where
   $\hat{\imath}$, $\hat{\jmath}$ and  $\hat{k}$ are unit vectors directed along the $x$, $y$, $z$ axes. 
    In (9.19)-(9.22) $\vec{r}$ is specified at the instant at which the fields are defined, although
    the Heaviside formula (9.19) is commonly derived~\cite{PP} 
  from retarded Li\'{e}nard-Wiechert potentials~\cite{LW}.
    As mentioned in Section 2 above,
    the pre-relativistic Heaviside formulae (9.19), (19.20) are actually incorrect, but what
    is important for the present discussion is only the functional space-time dependence of the
    fields in (9.19)-(9.22) and of the corresponding 4-vector potential $A^{\alpha}$ in the defining
    equations of the fields (9.2) and (9.3). In all of these equations, the field point, $\vec{x}$,
    always appears in combination with the vector, $\vec{x}_Q$, specifying the instantaneous
    position of the source charge, to define the radius vector, $\vec{r}$, giving the relative 
     positions of the source and test charges:
     \begin{equation}
       \vec{r} \equiv \vec{x} - \vec{x}_Q.
    \end{equation}
      When the source charge is in motion, the functional time dependence of the fields is implicit in that
       of the source charge position $\vec{x}_Q = \vec{x}_Q(t)$. Because $A^{\alpha}$ and all the fields
      are functions of $r =  |\vec{x} - \vec{x}_Q(t)|$ it follows, for the case of Eqs.~(9.19)-(9.22) 
      where the source charge is in motion parallel to the $x$-axis, that $x$ and $t$ are no longer
       independent variables, as assumed in Eq.~(9.11), but instead their partial derivatives
       satisfy the relation~\cite{JHFSTF}:
     \begin{equation}
 \left. \frac{\partial ~}{\partial t}\right|_x = -u  \left. \frac{\partial ~}{\partial x}\right|_t.
   \end{equation}
      Also, as demonstrated in Ref.~\cite{JHFSTF} $r$ is a Lorentz-invariant quantity so that
    in virtue of Eqs.~(9.14), and in contrast to Eq.~(9.12), the transformation of the $x$ partial
     derivative is, like the $y$ and $z$ partial derivatives, time independent:
     \begin{equation}
       \frac{\partial ~}{\partial x'} =  \frac{\partial ~}{\partial x}. 
    \end{equation}
     A consequence of (9.24) and (9.25) is that the $x$ and $t$ partial derivatives do not 
     transform as components of a 4-vector according to Eqs.~(9.12) and (9.13) so that the
     electric and magnetic fields produced by a uniformly moving charge do not transform,
     according to (9.7), as components of a second-rank tensor.
     \par The actual transformation laws of the RCED fields of (9.21) and (9.22) are easily
     read off from these equations. Denoting by $\vec{E}^{\ast}$ the electric field in the
     rest frame S$^{\ast}$ of the source charge $Q$, (9.21) gives:     
  \begin{eqnarray}
     E_x & = & \frac{E_x^{\ast}}{\gamma_u},~~~~  E_y = \gamma_u E_y^{\ast}, \\
     B_z & = & \beta_u E_y = \beta_u \gamma_u E_y^{\ast}. 
  \end{eqnarray}
   For comparison with Einstein's transformation equations ({\rm E}6.13)-({\rm E}6.15), suppose that
 the velocity of the source charge in the frame S' is $w$. Then (9.21) and (9.22) give, 
  for the fields in the frame S': 
   \begin{eqnarray}
  \vec{E}' & = & \frac{Q}{r^2}\left[ \frac{\hat{\imath} \cos \psi}{\gamma_{w}} +
   \hat{\jmath}\gamma_{w}\sin \psi \right], \\
\vec{B'} & = & \frac{\vec{w} \times \vec{E}'}{c}.
    \end{eqnarray}
   Considering the transformations similar to (9.26) and (9.27) between  S$^{\ast}$ 
    and S', and eliminating the S$^{\ast}$ fields between these equations and (9.26) and
    (9.27) gives, for the field transformations from S to S':
     \begin{equation}
     E_x' = \frac{\gamma_u}{\gamma_w} E_x,~~~~ E_y' = \frac{\gamma_w}{\gamma_u} E_y,~~~~
      B_z' = \frac{\beta_w \gamma_w}{\beta_u \gamma_u} B_z
     \end{equation}
     Since, from Eq.~(9.27), $E_y = B_z/\beta_u$ the second of Eqs.~(9.30) may be written as
       \begin{equation}
       E_y' = \frac{\gamma_w}{\beta_u \gamma_u} B_z.   
      \end{equation}
       Because the velocities $w$, $u$ and $v$ satisfy the PVAR:
     \begin{equation}
     \beta_w =\frac{\beta_u-\beta}{1-\beta_u \beta}
      \end{equation}
  the temporal component of the dimensionless 4-vector velocity $U=$($\gamma_u$;$\gamma_u \beta_u$,0,0)
   transforms between S and S' (see Eq.~(8.6)) as:
        \begin{equation}
          \gamma_w = \gamma(\gamma_u-\beta\gamma_u \beta_u).
        \end{equation}     
   Combining (9.31) and (9.33) then gives
          \begin{equation}    
     E_y' = \frac{\gamma}{\beta_u}(B_z - \beta \beta_u B_z) = \gamma ( \frac{B_z}{\beta_u}-\beta B_z) 
          =  \gamma (E_y-\beta B_z).
   \end{equation}
   The Lorentz transformation of the spatial component of $U$
  between S and S' is (see Eq.~(8.5)):
       \begin{equation}
      \gamma_w \beta_w = \gamma(\gamma_u \beta_u-\beta\gamma_u).
      \end{equation}    
 Combining (9.35) with the last equation in (9.30) gives:
         \begin{equation}    
     B_z' = \frac{\gamma}{\beta_u}(\beta_u B_z - \beta B_z) = \gamma ( B_z-\frac{\beta  B_z}{\beta_u}) 
          =  \gamma (B_z-\beta E_y).
    \end{equation}    
     Eqs.~(9.34) and (9.36) for the transformation of the transverse electric field and the
    magnetic field agree with the corresponding equations in ({\rm E}6.14) and ({\rm E}6.15).
    Since the temporal derivatives, and the $y$- and $z$-component spatial derivatives, that appear
     in the defining equations (9.2) and (9.3) for these components, are independent, the chain rule (9.11)
     and hence the tensorial transformation equation (9.7) is respected in this case.
    Since the transformation coefficients of (9.34) and (9.36)
    depend only on the relative velocity of the frames S and S', and are independent
    of the velocity of the source charge, these equations are covariant.
    It is otherwise with the transformation law of the longitudinal electric field in 
    the first of Eqs.~(9.30) which does not agree with Einstein's covariant transformation
     law in ({\rm E}6.13), $E_x' = E_x$. In this case the transformation coefficient does depend
   on the velocity of the source charge. There is therefore a preferred frame (the frame
    in which the source charge is at rest) in the problem. Because of this, relativistic
     covariance is broken. Fundamentally, the tacit assumption underlying Einstein's
     analysis, introduced into theoretical electromagnetism by Faraday and Maxwell, that electric
    and magnetic forces may be correctly described by {\it local classical fields}
     is  not true of the actual force fields of a uniformly moving charge. This
   applies both to the  pre-relativity fields of Eqs.~(9.19) and (9.20)
    and the RCED fields of (9.21) and (9.22). The actual force fields are produced by the 
    source charge and the motion of this charge affects, instantaneously, the force experienced
    by the test charge. Note that that this remains true for the `present time' Heaviside
     formulae (9.19) and (9.20) even though the corresponding interaction is retarded. 
     All inertial frames are not equivalent, so that special relativistic invariance
    breaks down in the description of inter-charge forces.
    \par Another consequence of this breakdown of covariance is the frame dependence of some
     of the Maxwell equations~\cite{JHFSTF}. For example, the Amp\`{e}re law equation in the
     first of Eqs.~({\rm E}6.1):
      \begin{equation}
      \frac{1}{c}\frac{\partial E_x}{\partial t} = \frac{\partial B_z}{\partial y}
       -  \frac{\partial B_y}{\partial z}
     \end{equation}
      is replaced by the non-covariant equation
       \begin{equation}
      \frac{1}{c}\frac{\partial E_x}{\partial t} + \frac{Q \beta_u}{r^3}\left(\gamma_u-\frac{1}{\gamma_u}\right)
       (2-3\sin^2 \psi)
 = \frac{\partial B_z}{\partial y}  -  \frac{\partial B_y}{\partial z}.
     \end{equation}
    This is demonstrated by direct calculation of the partial derivatives of the fields
    of (9.21) and (9.22), taking into account the relation (9.24) connecting the $x$ and $t$ 
    partial derivatives. A similar calculation of the $y$- and $z$-components of  Amp\`{e}re's law
    show that the first equations of ({\rm E}6.11) and ({\rm E}6.12) remain valid. In contrast 
    to the Amp\`{e}re law,  all components of the Faraday-Lenz law in the second equations in ({\rm E}6.10)-({\rm E}6.12) 
    remain valid for the RCED force fields of Eqs.~(9.21) and (9.22) and the CEM fields of (9.19)
      and (9.20). This follows directly from 
    the defining equations of the fields in terms of $A^{\alpha}$, (9.2) and (9.3), written in
     3-vector notation as:
          \begin{equation}
         \vec{B} = \vec{\nabla} \times \vec{A},~~~~ \vec{E} = 
           - \vec{\nabla} A^0 -\frac{1}{c}\frac{\partial \vec{A}}{\partial t}.          
  \end{equation}
     Taking the time derivative of the first of these equations and substituting the value
   of $\partial \vec{A}/\partial t$ from the second equation gives:
   \begin{equation}
   \frac{1}{c}\frac{\partial \vec{B}}{\partial t}= -\vec{\nabla} \times  (\vec{\nabla} A^0)
     - \vec{\nabla} \times \vec{E} = - \vec{\nabla} \times \vec{E}
   \end{equation}
     in virtue of the 3-vector identity $\vec{\nabla} \times  (\vec{\nabla \phi)} = 0$, for arbitary $\phi$. 
     Thus the Faraday-Lenz law is valid in any inertial frame (is covariant) independently
     of the space-time functional dependence of  $A^{\alpha}$ or the force fields.
     \par Another consequence of the non-covariant character of the electric force fields is the 
      breakdown of the electric Gauss law if the source charge is in motion~\cite{JHFRSKO,JHFSTF}:
   \begin{equation}
  \vec{\nabla} \cdot \vec{E} =  \frac{Q}{r^3}\left(\gamma_u-\frac{1}{\gamma_u}\right)
       (2-3\sin^2 \psi).
   \end{equation}
     The appearence of a similar covariance-breaking term in the Gauss law (9.41) and the  Amp\`{e}re law
      (3.38) follows from Eq.~(9.24) and the first member of Eq.~(9.27). Since the electric field
       in Eq.~(9.19) or (9.21) is confined to the $x$-$y$ plane, and only the $z$-component
      of the magnetic field is non-vanishing:
 \begin{eqnarray}
  \vec{\nabla} \cdot \vec{E} & = & \frac{\partial E_x}{\partial x}+ \frac{\partial E_y}{\partial y}
     = \frac{1}{\beta_u}\left[\beta_u \frac{\partial E_x}{\partial x}+
       \frac{\partial (\beta_u  E_y)}{\partial y}\right] \nonumber \\
      & = & \frac{1}{\beta_u}\left[-\frac{1}{c}\frac{\partial E_x}{\partial t}
      + \frac{\partial B_z}{\partial y}\right] 
 \end{eqnarray}
    or 
   \begin{equation}
  -\frac{1}{c}\frac{\partial E_x}{\partial t} + \frac{\partial B_z}{\partial y}
 = \beta_u  \vec{\nabla} \cdot \vec{E}.
    \end{equation}  
   So that if Gauss law is valid $\vec{\nabla} \cdot \vec{E} = 0$ then so is the $x$-component of the
    Amp\`{e}re law (the first of Eqs.~({\rm E}6.1) with $B_y = 0$). In the case of the field in  (9.41)
    where the Gauss law is not respected (9.38) follows directly from (9.41) and (9.43) on setting
    $B_y$ to zero.
   \par As shown on Ref~\cite{JHFSTF} the wave equations of all components of the electric
    and magnetic fields derived from Eqs.~(9.21) and (9.22) are also modified by the addition 
    of $\beta_u$ dependent covariance-breaking terms, so that Maxwell's derivation of
     `electromagnetic waves' associated with these fields also breaks down. The leading 
      covariance breaking terms in Gauss' law and the electric field wave equations are of O($\beta_u^4$), and
     for the Amp\`{e}re law and the magnetic field wave equations of O($\beta_u^5$), so that
     no practical applications of  Maxwell's equations in the non-relativistic domain
      are affected significantly by these modifications. However it is clear that the heretofore fundamental
     physical status of Maxwell's equations is profoundly modified by the the lack of covariance
     of the Amp\`{e}re law and the Gauss' law for electric fields.
      \par In the second part of \S 6 Einstein returns to the general problem
      of the relationship between electric and magnetic forces that was discussed for the special case
      of electromagnetic induction in the Introduction. Einstein's remarks here are important,
      both for what is said, and for what is left unsaid:
 
      \par {\tt 1. If a unit electric point charge is in motion in an electromagnetic \newline
         field there acts upon it, in addition to the electric force, an \newline ``electromative force''
       which, if we neglect the terms multiplied by the \newline second and higher powers of $v/c$,
       is equal to the vector-product of the \newline velocity of the charge and the magnetic force
     divided by the velocity of \newline light. (Old manner of expression).
      \par 2. If a unit electric point charge is in motion in an electromagnetic \newline field, the force
      acting upon it is equal to the electric force which is \newline present at the locality of the
      charge, and which we ascertain by \newline transformation of the field to a system of co-ordinates
     at rest relatively to the electrical charge (New manner of expression).
       \par The analogy holds for ``electromotive forces''. We see that electromotive force plays in the
      developed theory merely the part of an auxiliary concept which owes its introduction to the
      circumstance that electric and magnet \newline forces do not exist independently of the state of 
      motion of the system of  \newline coordinates.
      \par Furthermore it is clear that the asymmetry mentioned in the introduction as arising 
       when we consider the currents produced by the relative motion of a magnet and a conductor
      now disappears. Moreover, questions as to the  \newline ``seat'' of electrodynamic electromotive
      forces (unipolar machines) now have no point.}
       \par Einstein clearly states here that the essential physical phenomenon is the
       force on the test charge and that the electric and magnetic `fields' serve only
      as labels for different terms in the force equation, the relative contributions of which
      vary according to the inertial frame in which the force is evaluated. It is pointed out that
       the``electromotive force'' (Lorentz force due to motion in a magnetic field) that acts
       on the test charge may be annuled by transformation into the rest frame of the test
       charge, in which only the force associated with the electric field acts. Einstein later
      uses this argument in \S 10 to derive, correctly to lowest order in $\beta$,
       the Lorentz force from the transformation laws
      of electric and magnetic fields. Because it can be `transformed away' in this manner
      the Lorentz force (and by implication the `magnetic field' itself) is assigned the status
     of an `auxiliary concept'. But by the same token the electric field is also
     an `auxiliary concept' in that it, like the magnetic field, is completly specified by the 
     configuration of electric charges that constitute its source. Throughout Ref~\cite{Ein1}
      Einstein discusses only fields and either their interactions with test charges or the description
      they provide of electromagnetic radiation. In what manner these fields are produced by
     their sources, which is essential for a complete physical description, is not even 
     mentioned. The covariance of Maxwell's equations with sources is discussed in \S 9
       but not the physical interpretation of the fields and sources, which, in their
         most important physical application, the description of radiation, is quite different
       to that of the force fields discussed in \S 6, \S 9 and \S 10. 
      \par In fact, since all mechanical effects on a test charge are described by Eq.~(3.4)
       (or by its generalistation to the case of multiple source charges) in which only the
       physical parameters specifying the spatial and kinematical configuration of
      the source charges appear, both the electric and magnetic `fields', that serve only
     as labels for the different terms on the right side of (3.4), are inessential `auxiliary concepts', or 
     alternatively, `mathematical abstractions'. The same is true of the 4-vector
     potential, $A^{\alpha}$, in terms of which, electric and magnetic fields are defined
     by Eqs.~(9.2) and (9.3). This, in turn, is unambiguously specified~\footnote{A particular gauge,
       Lorenz gauge~\cite{JO}, is a necessary consequence of the postulates from which Eq.~(3.4) is derived.
      Thus there is no freedom of the choice of gauge used to define $A^{\alpha}$ in RCED.} by the physical
     parameters that describe the configuration of source charges~\cite{JHFRCED}. This is shown in 
      Fig.1. All force predictions are contained in the `Intercharge Force; box inside
    the dot-dashed line. They are obtained without the necessity to introduce any `field' concept
     whatever.  
     \par At the end of \S 6, Einstein returns to the problem of induction, as analysed in 
     different inertial frames discussed in the Introduction and considers it solved
     by the comments in the paragraphs 1. and 2. cited above. If he had actually calculated
    the electric field in the rest frame of the test charge for the configuration shown in
     Fig.1b, using the then-available Heaviside equation with a radial electric 
    field he would have found, as explained in Section 2 above, that the induction force is not the 
    same as the Lorentz
     force in Fig5a, but actually vanishes! As described in ~\cite{JHFIND} consistent
     results in the two frames are given by the RCED formulae (9.21) and (9.22). The
     calculation shows that the induction force is, however, not exactly equal in the
     frames S and S', as stated by Einstein, but differs by terms of O($\beta_u^4$). The origin of this difference
     is the source velocity dependence of the electric field in Eq.~(9.21), which breaks
     covariance for the longitudinal component of the electric field.
     \par The closing remark of \S 6 concerns the absence, in electromagnetism, of magnetic monopoles.
          Since the Gauss law for the magnetic field $\vec{\nabla}\cdot\vec{B} = 0$
        follows from the definition of the magnetic field in terms of $\vec{A}$ in first of
        equations (9.39) and the three-vector identity $\vec{a} \cdot ( \vec{a} \times \vec{b}) = 0$
         this comment of Einstein's is certainly well-founded.

\SECTION{\bf{\S 7. Theory of Doppler's Principle and Aberration}}
   Einstein first writes down in Eqs.~({\rm E}7.1)-({\rm E}7.3) the field components, in the frame S,
 of an electromagnetic wave propagating in the direction defined by the direction cosines $l$, $m$, $n$ relative
 to the $x$, $y$, $z$ axes. The phase of the wave in S is written as:

\[~~~~~~~~~~~~~~~~~~~~~~~~~~~~~~~~~\Phi = \omega \{t-\frac{1}{c}(l x +m y + n z)\}.~~~~~~~~~~~~~~~~~~~~~~~~~~~~~~~~~~~({\rm E}7.4) \]
   Similarly the phase of the wave in S' is written as:
\[~~~~~~~~~~~~~~~~~~~~~~~~~~~~~~~~~\Phi' = \omega'\{\tau-\frac{1}{c}(l' \xi +m' \eta + n' \zeta)\}.~~~~~~~~~~~~~~~~~~~~~~~~~~~~~~~~~({\rm E}7.8) \]
  At this point there is a logical hiatus in the discussion. Einstein immediately writes 
  down the transformation equations giving the quantities $\omega'$, $l'$, $m'$ and $n'$ in terms of
  $\omega$, $l$, $m$ and $n$:
   \begin{eqnarray}
~~~~~~~~~~~~~~~~~~~~~~~~~~~~~~~~~~~~~~~~~~\omega' & = & \omega \gamma(1-\beta l),~~~~~~~~~~~~~~~~~~~~~~~~~~~~~~~~~~~~~~({\rm E}7.9) \nonumber  \\
~~~~~~~~~~~~~~~~~~~~~~~~~~~~~~~~~~~~~~~~~~ l'& = & \frac{l-\beta}{1-\beta l},~~~~~~~~~~~~~~~~~~~~~~~~~~~~~~~~~~~~~~~~~~~({\rm E}7.10) \nonumber  \\
~~~~~~~~~~~~~~~~~~~~~~~~~~~~~~~~~~~~~~~~~~~~m'& = & \frac{m}{\gamma(1-\beta l)},~~~~~~~~~~~~~~~~~~~~~~~~~~~~~~~~~~~~~({\rm E}7.11) \nonumber   \\ 
~~~~~~~~~~~~~~~~~~~~~~~~~~~~~~~~~~~~~~~~~~~~ n'& = & \frac{n}{\gamma(1-\beta l)}.~~~~~~~~~~~~~~~~~~~~~~~~~~~~~~~~~~~~~({\rm E}7.12) \nonumber  
     \end{eqnarray}
    The field components are also transformed into the frame S' in Eq.~({\rm E}7.5)-({\rm E}7.7), but the
     actual values of the field components are irrelevant for the discussion of the Doppler effect 
    and aberration in this section. 
    \par In order to derive ({\rm E}7.9)-({\rm E}7.12) another important hypothesis, not mentioned by
     Einstein, is needed: $\Phi' = \Phi$. That is, it must be assumed that the phase of the
    wave is a Lorentz invariant quantity. Presumably Einstein did this, used the LT ({\rm E}3.27)-({\rm E}3.30)
 to replace $\tau$, $\xi$, $\eta$ and $\zeta$ in ({\rm E}7.8) by $t$, $x$, $y$ and $z$, and equating
   coefficients of these variables in the thus-transformed Eq.~({\rm E}7.8) with those in Eq.~({\rm E}7.4) arrived
  (after some algebraic manipulation) at Eqs.~({\rm E}7.9)-({\rm E}7.12). This method does yield the 
   correct Doppler effect and aberration formulae, but, as discussed below, poses a certain 
  problem of mathematical logic.
    \par From a modern viewpoint The postulate of the invariance of $\Phi$ and the fact that 
     $X \equiv (ct = x^0; x,y,z)$ is a 4-vector, implies that $\Omega \equiv (\omega/c;\omega l/c,\omega m/c,
     ,\omega n/c)$ must also be a 4-vector so that ({\rm E}7.4) may be written as:
       \begin{equation}
        \Phi = \Omega \cdot X.
       \end{equation}
      Since $\Omega$ is a unidimensional 4-vector its LT between the frames S and S' is 
     given by the space-time symmetric LT~\cite{JHFLT2}:
     \begin{eqnarray}
 \Omega'_0 & = &\gamma(\Omega_0-\beta \Omega_x), \\
 \Omega'_x & = &\gamma(\Omega_x-\beta \Omega_0), \\
\Omega'_y & = & \Omega_y,~~~\Omega'_z = \Omega_z
 \end{eqnarray}
 or
     \begin{eqnarray}
       \omega' & = & \omega \gamma(1-\beta l), \\
  \omega'l' & = & \omega  \gamma(l-\beta ), \\
     \omega'm' & = & \omega m,   \\
      \omega'n' & = & \omega n  
   \end{eqnarray}
    where factors $1/c$ on both sides of (10.5)-(10.8) have been divided out. Eq.~(10.5) is identical to 
    ({\rm E}7.9) and yields ({\rm E}7.11) and ({\rm E}7.12) on eliminating  $\omega/\omega'$ from (10.7) and (10.8).
      Eq.~({\rm E}7.10) follows on eliminating $\omega/\omega'$ between (10.5) and (10.6).
    \par The problem of mathematical logic in Einstein's surmised proof of Eqs.~({\rm E}7.9)-({\rm E}7.12) arises
    from the factor in curly brackets in ({\rm E}7.4). The quantity $l x + m y + n z$ is a rotational
    invariant that has the value:
    \begin{equation}
      s = \sqrt{x^2 + y^2 + z^2}.
    \end{equation}
     The distance $s$ is the displacement of the wave front (i.e. a surface of constant phase)
     in its direction of propagation
     during the time $t$. Thus, ({\rm E}7.4) may be written as:
   \begin{equation}
   \Phi = \omega(t -\frac{s}{c}).
   \end{equation}  
    Since the wave front moves at speed $c$, $s = c t$
     and (10.10) gives $\Phi = 0$. 
     Since the field components in Eqs.~({\rm E}7.1)-({\rm E}7.3) are all proportional to $\sin \Phi$ they 
     all vanish identically. So Einstein's electromagnetic wave (as defined) does not exist!
      In a similar way, ({\rm E}7.8) may be written as:
    \begin{equation}
   \Phi' = \omega(\tau -\frac{s'}{c}).
   \end{equation}
    Since, by the second postulate, the wave front also moves at speed $c$ in S', $s' = c \tau$
    and  $\Phi' = 0$. Thus (10.10) and (10.11) and the Lorentz invariance of the phase $\Phi = \Phi'$
   reduce to the equation:
      \begin{equation}
       \Phi = \omega \times 0 = \omega' \times 0 = \Phi'
  \end{equation}
   which is true for any finite values of $ \omega$ and $\omega'$. This would seem to indicate
   that Einstein's presumed derivation of ({\rm E}7.9)-({\rm E}7.12) by (tacitly) setting $\Phi$ equal to $\Phi'$,
    and Lorentz transforming one set of space-time coordinates,
   was a mathematically hazardous enterprise that finally `fell on its feet'!
   \par At the end of \S 7 Einstein writes down, without derivation, two equations for the transformation of
    the `intensity' of an electromagnetic wave:
   \begin{eqnarray}
    ~~~~~~~~~~~~~~~~~~~~~~~~~~~~~~~~~~~~~~(A')^2 & = & \frac{A^2(1-\beta \cos \phi)^2}{1-\beta^2},~~~~~~~~~~~~~~~~~~~~~~~~~~~~~~({\rm E}7.17)
 \nonumber \\
   ~~~~~~~~~~~~~~~~~~~~~~~~~~~~~~~~~~~~~~~~~(A')^2 & = & \frac{A^2(1-\beta)}{1+\beta}.~~~~~~~~~~~~~~~~~~~~~~~~~~~~~~~~~~~~~~({\rm E}7.18) \nonumber
  \end{eqnarray}
    The quantities $A$ and $A'$ are called `the amplitudes of the electric or
     magnetic force' which shows some confusion over the physical meaning of the
    field components in ({\rm E}7.1)-({\rm E}7.3) and ({\rm E}7.5)-({\rm E}7.7). These are radiation fields,
    not force fields. The squares of these radiation fields give the energy density $\rho_E$ in the 
     electromagnetic wave according to the well-known formula:
        \begin{equation}
        \rho_E = \frac{1}{8 \pi}(\vec{E}^2+\vec{B}^2).
      \end{equation}
    As pointed out above, as defined, Eqs.~({\rm E}7.1)-({\rm E}7.3) and  Eqs.~({\rm E}7.5)-({\rm E}7.7) imply
     $A = A' = 0$. This problem is removed by making the permissible substitutions
      $\sin \Phi \rightarrow \cos\Phi$ and  $\sin \Phi' \rightarrow \cos\Phi'$ in these equations.
      However, ({\rm E}7.17) and ({\rm E}7.18) derived from  the transformation of force fields
      in Eqs.~({\rm E}7.6) and ({\rm E}7.7) still make no sense physically. This can be seen by
      noting that a `plane electromagnetic wave' is in fact a beam of monochromatic
     photons~\cite{JHFEJP}. The transformation of the energy density of the wave may therefore
     be derived from the relativistic kinematics of photons. Since the energy density
      $\rho_E \simeq A^2$ is the product
     of the photon energy $E_{\gamma}$ and the photon number density $n_{\gamma}$, which is 
      Lorentz invariant, the transformation law of $\rho_E$ may be derived from that of the
      photon energy. This gives, instead of ({\rm E}7.17) and ({\rm E}7.18):
        \begin{eqnarray}
       \rho'_E  & = &  \rho_E \frac{1-\beta \cos \phi}{\sqrt{1-\beta^2}},  \\
        \rho'_E  & = &  \rho_E \sqrt{\frac{1-\beta}{1+\beta}}. 
       \end{eqnarray}
       Notice that the transformation law of the the photon energy, and hence of the energy
       density of the beam, is the same as that of the  photon frequency in Eqs.~({\rm E}7.13) and ({\rm E}7.14)
       not that of the square of the force fields in Eqs.~({\rm E}7.6) and ({\rm E}7.7). This is a consequence
       of the Planck-Einstein relation, $E_{\gamma} = h \nu_{\gamma}$, to be further discussed in the 
       following section. Einstein's error in deriving Eqs.~({\rm E}7.17) and ({\rm E}7.18) is the misidentification
      of force fields, describing the effects of space-like virtual photon exchange between
      electric charges, with the radiation fields appropriate to the description of the monochromatic beams of 
      real photons, which manifest the Doppler effect and aberration. The distinction between radiation
      fields and force fields is further discussed in Section 12 below.
      \par Einstein's final statement in \S7 :
       \par {\tt It follows from these results that to an observer approaching the source of
         light with velocity c, this source of light must appear of infinite \newline intensity.}
       \par remains true according to (10.15). although the approach to `infinite intensity' is
        less rapid than in Eq.~({\rm E}7.18). 
       
 \SECTION{\bf{\S 8. Transformation of the Energy of Light Rays. Theory of the
      Pressure of Radiation Exerted on Perfect Reflectors}}
   In this section Einstein discusses, firstly, the transformation law of the energy density
  of an electromagnetic wave and, secondly, the radiation pressure due to a plane electromagnetic
   wave at oblique incidence on a uniformly moving plane mirror. Since plane electromagnetic
    waves may be identified with a parallel beam of monochromatic photons, all the results 
   of this section may, alternatively, be obtained in a simple and transparent manner by
   invoking the relativistic kinematics of real photons.
    \par To obtain the energy transformation law of a `light ray' Einstein considers a 
    spherical surface in the frame S with the
     equation:
 \[~~~~~~~~~~~~~~~~~~~~~~~~~~~~~~~~(x-l c t)^2 + (y-m c t)^2+(z-n c t)^2 = R^2.~~~~~~~~~~~~~~~~~~~~~~~~~~~~({\rm E}8.1) \]
      This equation represents a sphere of radius $R$ centered at $x = l c t$, $y = m c t$, 
       $z = n c t$. The center moves in the same direction and at the same speed as the plane
     wave considered in \S 7. Even though the plane wave is of infinite extent in all
    three dimensions, Einstein considers only the energy of the  `light complex'
    (Lichtcomplexe) inside the spherical surface. The spurious `length contraction' in the direction
    of motion
     of the spherical surface in S as viewed from S' according to ({\rm E}4.2) is now invoked 
     to claim that the surface ({\rm E}8.1), as  observed in S' at $\tau = 0$ is:
\[~~~~~~~~~~~~~~~~~~~~~~~~~~~~~~~~\xi^2\frac{(1-\beta l)^2}{1-\beta^2}+(\eta- m \gamma \beta \xi)^2 + (\zeta - n \gamma \beta \xi)^2
    = R^2.~~~~~~~~~~~~~~~~~~~~~~~~~~~~({\rm E}8.2) \] 
  The ratio of the volume, $V$, inside the sphere ({\rm E}8.1) to that, $V'$, inside the ellipsoid of
  revolution ({\rm E}8.2) is:
 \[~~~~~~~~~~~~~~~~~~~~~~~~~~~~~~~~~~~~~~~~~~~~~~\frac{V'}{V} = \frac{\sqrt{1-\beta^2}}{1- \beta \cos \phi}~~~~~~~~~~~~~~~~~~~~~~~~~~~~~~~~~~~~~~~~~~({\rm E}8.3) \]
    Setting the total wave energy inside ({\rm E}8.1) equal to $A^2 V /8 \pi$, that inside
    ({\rm E}8.2) to  $(A')^2 V' /8 \pi$ and using ({\rm E}7.17) for the ratio  $(A')^2/A^2$ the transformation
   law of the energy of the `light complex' is found to be:
\[~~~~~~~~~~~~~~~~~~~~~~~~~~~~~~~~~~~~~\frac{E'}{E} = \frac{(A')^2 V'}{A^2 V} = \frac{1-\beta \cos \phi}{\sqrt{1-\beta^2}}.~~~~~ ~~~~~~~~~~~~~~~~~~~~~~~~~~~~({\rm E}8.4) \]
    This equation, which does correctly describe the transformation law of the
   energy density of an electromagnetic wave, as in Eq.~(10.14) above, has been obtained by combining
  the incorrect formula for the transformation of the energy density ({\rm E}7.17) with the spurious
   length contraction effect in ({\rm E}8.2). A factor $\gamma(1-\beta \cos \phi)$ cancels to give 
   the correct transformation law for a `light ray' identified with a `light complex'. 
    \par The above calculation has defined a `light complex' as a spherical region inside
      an infinite plane electromagnetic wave in the frame S, which is then transformed into
     the frame S'. However, from the Reciprocity Principle of special relativity~\cite{JHFLT1}
     such a spherical  `light complex'  can equally well be defined for the electromagnetic
     wave in the frame S'. Viewed from the frame S, by the same argument leading from ({\rm E}8.1)
     to ({\rm E}8.2), it will appear as an ellipsoid of revolution with volume $V$ related to the volume
     of the sphere in S', $V'$ by Eq.~(8.3) with $V$ and $V'$ interchanged. The transformtion
      law of the energy of this  `light complex' is then different to ({\rm E}8.4) and inconsistent
     with that, (10.14), obtained from photon kinematics. The logical footing of this manifestly
     incorrect calculation is identical to that used by Einstein to derive ({\rm E}8.4). That the
     latter describes correctly the transformation of the energy density of an electromagnetic
     wave is purely fortuitous.
    \par Since the photon number density, $n_{\gamma}$, is Lorentz invariant, the relation
    ({\rm E}8.4) is most simply obtained by transformation of unidimensional photon energy-momentum
     4-vector:
    \begin{equation}
  p_{\gamma} \equiv ({\rm E}_{\gamma}; ({\rm E}_{\gamma}/c)\cos \phi, ({\rm E}_{\gamma}/c)\sin \phi, 0)
    \end{equation}
    according to the space-time symmetric LT (10.3). After the derivation, by a fallacious
    argument, of the correct Eq.~({\rm E}8.4), Einstein makes the following important remark:
     \par{\tt It is remarkable that the energy and the frequency of a light complex \newline vary with
       the state of motion of the observer in accordance with the same\newline law.}
      \par In a paper written earlier in 1905~\cite{Ein2}, Einstein had introduced into physics the concept
   of the `light quantum' ---light as a particle, later to be called the photon. However, at
    the time of writing the special relativity paper Einstein was not yet aware of the
    formalism of relativistic kinematics, to be developed, somewhat later, by Planck~\cite{PlanckRK}, 
     and so could not have been known of the simple kinematical derivation of ({\rm E}8.4) provided
     by the photon concept. Had he done so, he may have realised a deep and important connection
    between the light quantum and special relativity papers, that is, ultimately, one between
    quantum mechanics and special relativity. In fact, in the spirit of Einsten's remark just quoted,
     the ratio of the frequency transformation equation:
     \[~~~~~~~~~~~~~~~~~~~~~~~~~~~~~~~~~~~~~~~~~~~~~~\nu' =  \nu \frac{1-\beta \cos \phi}{\sqrt{1-\beta^2}} ~~~~~~~~~~~~~~~~~~~~~~~~~~~~~~~~~~~~~~~~~~({\rm E}7.13) \]
    to the energy transformation equation ({\rm E}8.4) implies the Lorentz invariance of $E/\nu$, or that
    ~\cite{JHFEJP}:
    \begin{equation}
       \frac{E'}{\nu'}= \frac{E}{\nu}= {\rm constant} \equiv    h.
    \end{equation}
       This is the Planck-Einstein relation. If Einstein had had more confidence, in 1905, in
    the reality of photons the equations ({\rm E}7.13) and ({\rm E}8.4) of Ref.~\cite{Ein1} might have lead to an independent
    discovery of Planck's constant and the associated fundamental quantum mechanical
    equation (11.2).

    \par  As discussed in Ref.~\cite{JHFEJP}, a more detailed comparison
    of plane electromagnetic waves with the equivalent photonic description enables other important
    quantum mechanical formulae and concepts to be understood as consistency conditions
    on the wave and particle representations of the same phenomenon.

     \par The photonic description enables a simple and direct calculation of radiation
   pressure. The photon flux per unit area per unit time, $f_{\gamma}$, of a beam with angle
   of incidence $\phi$ on a mirror moving away from the beam with velocity $v$ in the
    direction normal to its surface is:
    \begin{equation}
    f_{\gamma} =n_{\gamma} (c \cos \phi - v)
     \end{equation}
    where $n_{\gamma}$ is the photon number density. The momentum transfer, per photon,
    perpendicular to the surface of the mirror is:
   \begin{equation}
   \Delta p_{\gamma} = \frac{E_{\gamma}}{c} \cos \phi - \frac{E'''_{\gamma}}{c} \cos \phi'''.
    \end{equation}
     Use of (11.2) and ({\rm E}8.14) enables this equation to be written as:
    \begin{equation}
   \Delta p_{\gamma} = \frac{2 E_{\gamma}}{c} \frac{(\cos \phi - \beta)}{1-\beta^2}.
   \end{equation}
    By Newton's Second Law the force per unit area perpendicular to the surface
   of the mirror, which is, by definition, the radiation pressure, is then:
  \begin{equation}
    P = f_{\gamma} \Delta p_{\gamma} 
     =  \frac{2 E_{\gamma} n_{\gamma}(\cos \phi - \beta)^2}{1-\beta^2}
     =   \frac{2 \rho_{E}(\cos \phi - \beta)^2}{1-\beta^2}. 
    \end{equation}
     With the replacement $\rho_E \rightarrow  A^2/8 \pi$ this agrees with Einstein's
     formula ({\rm E}8.17). 
       \par Eq.~(11.6) can be written in terms of the energy densities $\rho_E$ and $\rho_{E'''}$
       of the incident and reflected wave, respectively, as:
 \begin{equation}
 P = (\rho_E \cos \phi-\rho_{E'''}\cos \phi''')(\cos \phi - \beta).  
   \end{equation}
  Setting $\rho_E = A^2/8 \pi$ and $\rho_{E'''} = (A''')^2/8 \pi$ and using ({\rm E}8.12) for
  $A'''$ gives, for the radiation pressure:
  \begin{equation}
     P = \rho_E \frac{[2(1-\beta^2)-(1-2 \beta\cos\phi+\beta^2)(2\beta-(1+\beta^2)\cos \phi)]}
                {(1-\beta^2)^2}(\cos \phi - \beta)
    \end{equation}
  in evident contradiction with (11.6).

\begin{figure}[htbc]
\begin{center}\hspace*{-0.5cm}\mbox{
\epsfysize15.0cm\epsffile{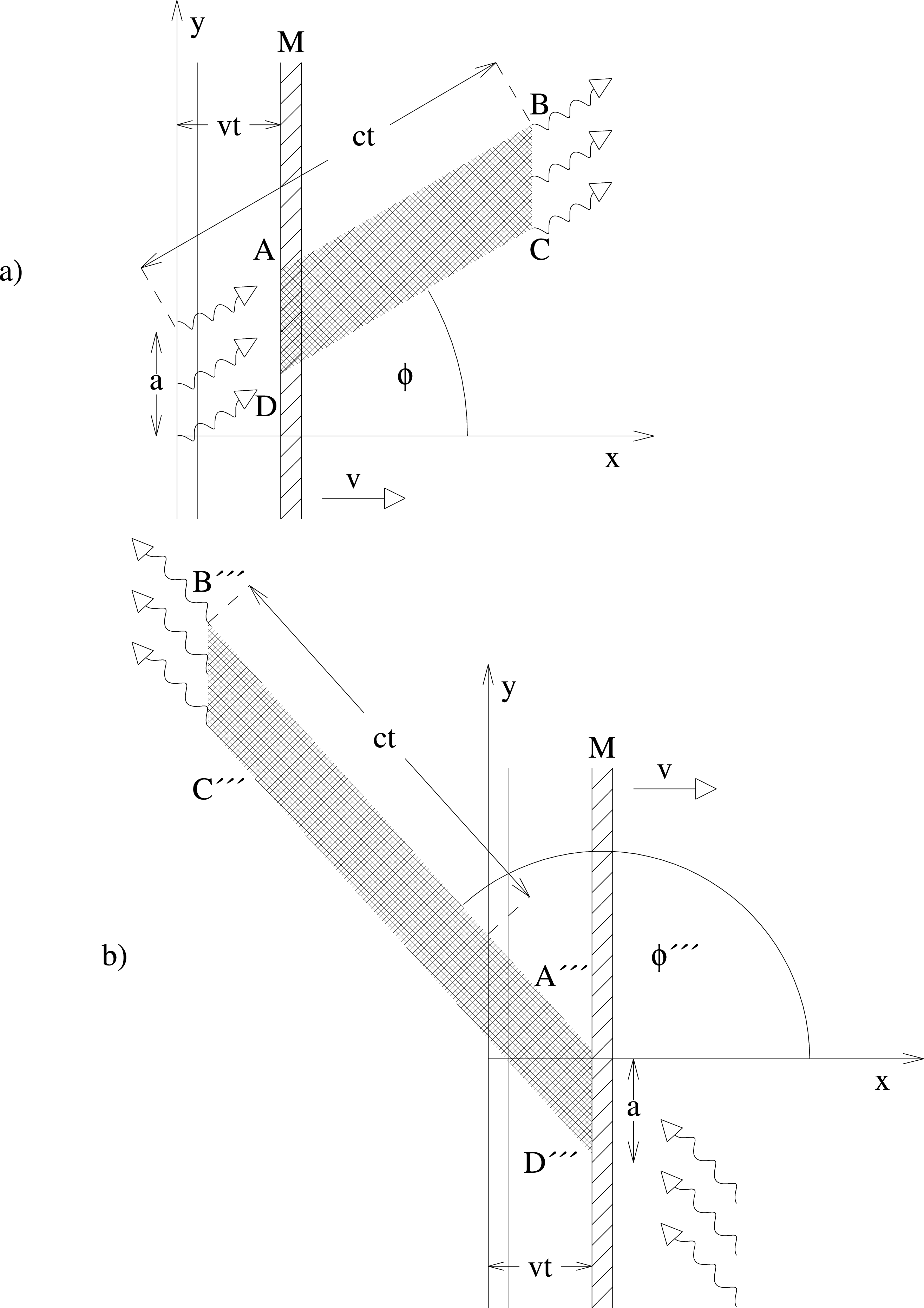}}
\caption{ {\em  a) a plane electromagnetic wave (a parallel beam of monochromatic photons)
   in the $x$-$y$ plane is incident at an angle $\phi$ on a plane mirror, M, with surface perpendicular to
   the $x$-axis moving with uniform velocity $v$ parallel to the latter. The number of photons in the beam
   of $y$- and $z$-dimensions $a$ striking the front surface of the mirror in time $t$ is proportional to the area ABCD.
   b) a parallel beam of monochromatic photons in the $x$-$y$ plane is incident at an angle $\pi-\phi'''$ on
      the \underline{back surface}! of the mirror M. The number of photons in the beam
   of $y$- and $z$-dimensions $a$ striking the \underline{back surface} of the mirror is proportional to the area
  A'''B'''C'''D'''.  Einstein incorrectly assumed that this was the number of photons reflected from
   the front surface of the M in time $t$. In this figure, $\phi = 30^{\circ}$, $\phi''' = 132^{\circ}$ and
    $v = 0.25c$.}}
\label{fig-fig7}
\end{center}
\end{figure}

  \par To understand how Einstein nevertheless obtained the correct result (11.6) by using the incorrect
    energy density of the reflected wave, given by ({\rm E}8.12), it is necessary to examine in more detail
    the calculation presented. The energy flux per unit area per unit time, $f_E$, incident on the mirror
    is:
\[~~~~~~~~~~~~~~~~~~~~~~~~~~~~~~~~~~~~~~~~~~~~~~f_E = \frac{A^2(c \cos \phi - v)}{8 \pi}.~~~~~~~~~~~~~~~~~~~~~~~~~~~~~~~~~~~~~~~~~~({\rm E}8.15) \]
  The derivation of this formula is demonstrated in Fig.7a. A portion of the plane wave
   (or, equivalently, of a monochromatic, parallel, photon beam) of dimensions $\Delta y = \Delta z = a$
 is incident on the plane mirror moving with velocity $v$ along the $x$-axis. It can be seen from
   the geometry of this figure that the number of photons crossing the mirror during the time
   interval $t$ is:
  \begin{equation}
    N_{\gamma}^{IN} = n_{\gamma} a {\rm Area(ABCD)} =  n_{\gamma} a^2 (c \cos \phi-v) t.
   \end{equation}
 The incoming energy flux is therefore:
  \begin{equation}
 f_E = \frac{ N_{\gamma}^{IN} E_{\gamma}}{a^2 t}
  = n_{\gamma} E_{\gamma} (c \cos \phi-v) =  \frac{A^2(c \cos \phi - v)}{8 \pi}   
   \end{equation}
  since $ n_{\gamma} E_{\gamma} = \rho_E = A^2/8 \pi$, in agreement with ({\rm E}8.15).
   The outgoing energy flux is given by Einstein as:
\[~~~~~~~~~~~~~~~~~~~~~~~~~~~~~~~~~~~~~~~~~~~~~~f_{E'''} = \frac{(A''')^2(-c \cos \phi''' + v)}{8 \pi}.~~~~~~~~~~~~~~~~~~~~~~~~~~~~~~~~~~~~~~~~~~({\rm E}8.16) \]
    This formula is derived from the geometry of Fig.7b. However, the photon flux
    considered in this figure is evidently not that of the reflected beam (the number of photons
    reflected must be equal to the number of photons incident on the mirror) but rather  
    is that of a beam of the same size and in the same direction as the reflected beam
      but {\it incident \underline{behind} the
    mirror!} Indeed, it is found from the geometry of Fig.7b that:
  \begin{equation}
       N_{\gamma}^{OUT} =  n_{\gamma}  a {\rm Area(A'''B'''C'''D''')} =  n_{\gamma} a^2 (-c \cos \phi''' + v)t
      \end{equation}
   which yields ({\rm E}8.16) for the energy flux. The photon number density must be the
   same for the incident and reflected beams (the reflection process changes only the
    kinematical parameters of the photons, not their number). However, it is clear from (11.9) and (11.11)
 that in Einstein's calculation, $ N_{\gamma}^{OUT} \ne   N_{\gamma}^{IN}$. In Fig.3, $\phi = 30^{\circ}$, and $v = 0.25 c$ so that 
    $\phi''' = 132^{\circ}$. This gives:
  \begin{equation}
   \frac{ N_{\gamma}^{OUT}}{ N_{\gamma}^{IN}} = \frac{-\cos \phi'''+\beta}{\cos \phi -\beta} = 1.49.
   \end{equation}
   Therefore Einstein's reflected energy flux formula ({\rm E}8.16) requires creation of photons
   in the reflection process, an evident absurdity. 
   \par Repeating Einstein's calculation of the radiation pressure by equating the net energy flow
      on to the mirror to the work done by the radiation pressure, but using the correct
     incident and reflected energy fluxes:
    \begin{eqnarray}
      f_E^{IN} & = &  n_{\gamma} E_{\gamma}(c \cos \phi - v), \\
       f_E^{OUT} & = &  n_{\gamma} E'''_{\gamma}(c \cos \phi - v)
      = \frac{ n_{\gamma} E_{\gamma}(1-2 \beta \cos \phi+ \beta^2)}{1-\beta^2}(c \cos \phi - v)
      \end{eqnarray}
   gives
   \begin{equation}
   P = \frac{c n_{\gamma} E_{\gamma}}{v}\left[1- \frac{(1-2 \beta \cos \phi+ \beta^2)}{1-\beta^2}\right]
   (\cos \phi - \beta) = 2 \left(\frac{A^2}{8 \pi}\right)\frac{(\cos \phi - \beta)^2}{1-\beta^2}
  \end{equation}
    in agreement with ({\rm E}8.17) and (11.5). 
    \par Finally, it must be concluded that that Einstein obtained fortuitously the correct radiation
    pressure formula ({\rm E}8.17) in spite of using the incorrect energy density formula ({\rm E}7.17),
    and the absurd geometrical analysis of the reflected wave shown in (Fig.7b).

   \par It remains that Einstein's final remark in this section is of great importance for the
    effective application of special relativity to physical problems.
     \par{\tt All problems in the optics of moving bodies can be solved by the method here employed.
    What is essential is, that the electric and magnetic force of light which is influenced
    by a moving body be transformed into the system of co-ordinates at rest relatively to the body.
    By this means all problems in the optics of moving bodies will be reduced to a series of problems
    in \newline the optics of stationary bodies.}
    \par The essential idea here, that physical problems are solved in the most elegant way by choosing
      the reference frame in which the description is simplest, and then using relativistic transformations
     to obtain predictions in the frame of interest, has applications extending far beyond
     `the optics of moving bodies'. A related idea is to express physical equations, in any conveniently
     chosen frame, in terms of Lorentz-invariant quantities, to obtain an equation valid in all inertial reference
     frames~\cite{BMT,JHFRCED}.
      
 \SECTION{\bf{\S 9. Transformation of the Maxwell-Hertz Equations when Convection-Currents are
     Taken into Account}}
   In this section the notation of Section 9 will be adopted, i.e. ($x'$,$y'$,$z'$,$t'$) will be used for
   space-time coordinates in S' rather than Einstein's ($\xi$,$\eta$,$\zeta$,$\tau$). Also a field point in
  the xy-plane is considered where the source charge moves parallel to the $x$-axis. This choice of
  coordinates entails no loss of generality and much simplifies the equations since
  $\vec{E}_z  = \vec{B}_x   = \vec{B}_y = 0$. Also the substitution $\rho = 4 \pi \rho_Q$ is made in 
    Einstein's equations to obtain the conventional charge density  $\rho_Q$ in Gaussian units.
  In this case ({\rm E}9.1)-({\rm E}9.11) simplify to:
   \begin{eqnarray}
  ~~~~~~~~~~~~~~~~~~~~~~~~~~~~~~ \frac{1}{c}\frac{\partial E_x}{\partial t} + 4 \pi \beta_u \rho_Q & = & \frac{\partial B_z}{\partial y},~~~~~~~~~~ ~~~~~~~~~~~~~~~~~~~~~~~~~~~~~~~({\rm E}9.1A) \nonumber  \\
 ~~~~~~~~~~~~~~~~~~~~~~~~~~~~~~\frac{1}{c}\frac{\partial E_y}{\partial t} & = & -\frac{\partial B_z}{\partial x},~~~~~~~ ~~~~~~~~~~~~~~~~~~~~~~~~~~~~~~~({\rm E}9.2A) \nonumber  \\
 ~~~~~~~~~~~~~~~~~~~~~~~~~~~~~~\frac{1}{c}\frac{\partial B_z}{\partial t} & = & \frac{\partial E_x}{\partial y}
   -\frac{\partial E_y}{\partial x},~~~~~~~~~~~~~~~~~~~~~~~~~~~~~~~({\rm E}9.3A) \nonumber  \\
~~~~~~~~~~~~~~~~~~~~~~~~~~~~~~  4 \pi \rho_Q & = & \frac{\partial E_x}{\partial x}+  \frac{\partial E_y}{\partial y},~~~~~~~~~~~~~~~~~~~~~~~~~~~~~~~({\rm E}9.4A) \nonumber  \\
 ~~~~~~~~~~~~~~~~~~~~~~~~~~~~~~ \frac{1}{c}\frac{\partial E'_{x'}}{\partial t'} + 4 \pi \beta_{u'} \rho'_Q & = & \frac{\partial B'_{z'}}{\partial y'},~~~~~~~  ~~~~~~~~~~~~~~~~~~~~~~~~~~~~~~~~~({\rm E}9.5A) \nonumber  \\
 ~~~~~~~~~~~~~~~~~~~~~~~~~~~~~~~~~~~ \frac{1}{c}\frac{\partial E'_{y'}}{\partial t'} & = & -\frac{\partial B'_{z'}}{\partial x'},~~~~~~~~~~~~~~~~~~~~~~~~~~~~~~~~~~~~~~({\rm E}9.6A) \nonumber  \\
~~~~~~~~~~~~~~~~~~~~~~~~~~~~~~~~~~~~\frac{1}{c}\frac{\partial B'_{z'}}{\partial t'} & = & \frac{\partial E'_{x'}}{\partial y'}
   -\frac{\partial E'_{y'}}{\partial
 x'},~~~~~~~~~~~~~~~~~~~~~~~~~~~~~~({\rm E}9.7A) \nonumber  
 \end{eqnarray}
  \begin{eqnarray}
~~~~~~~~~~~~~~~~~~~~~~~~~~~~~~~~~~~~~~~~~u' & = & \frac{u-v}{1-uv/c^2},~~~~~~~~~~~~~~~~~~~~~~~~~~~~~~~~~~~~~~~~~({\rm E}9.8A) \nonumber  \\
 ~~~~~~~~~~~~~~~~~~~~~~~~~~~~~~~~~ 4 \pi \rho'_Q & = & \frac{\partial E'_{x'}}{\partial x'}+  \frac{\partial E'_{y'}}{\partial y'}
       = 4 \pi \rho_Q \gamma(1-\beta_u \beta).~~~~~~~~~~~({\rm E}9.11A) \nonumber  
  \end{eqnarray}    
   In these equations the force fields are specified at the space point ($x$,$y$,0). In general the 
   source charge density  $\rho_Q$ is an arbitary function of the coordinate ($x_S$,$y_S$,$z_S$) specifying
   an element of the source. For definitiness, this is specialised to the case of a single point-like source charge
   in uniform motion, where the charge density  $\rho^{\ast}_Q$ in the rest frame S$^{\ast}$
    of the charge is:
    \begin{equation}
    \rho^{\ast}_Q(\vec{x}_S) = Q\delta(\vec{x}_S-\vec{x}_Q)
    \end{equation}
     where $\vec{x}_Q$ is the position vector of the source charge $Q$. In RCED and QED the force fields and the
     charge density are specified at the same time, so that the intercharge force is instantaneous,
     not retarded~\cite{JHFRCED}.
      \par Since the position of the field point and that of the source charge are always distinct
     there is in fact no difference between the `empty space' case discussed in \S 6 and the
     `convection current included' case of  \S 9. Without proximate source charges or currents all
     force fields must vanish. The considerations of Section 9 above then imply that while the
     Faraday-Lenz law ({\rm E}9.3A) and ({\rm E}9.7A) and Amp\`{e}re's law for the transverse (to the direction of
      motion of the source charge) electric field, ({\rm E}9.2A) and ({\rm E}9.6A) are correct, 
     the Amp\`{e}re law equations for the longitudinal electric field, ({\rm E}9.1A) and ({\rm E}9.5A),
     and the Gauss law for the electric field, ({\rm E}9.4A) and ({\rm E}9.11A), are modified by covariance-breaking
     terms to:
   \begin{eqnarray}
  \frac{1}{c}\frac{\partial E_x}{\partial t} + 4 \pi J_x +\frac{Q\beta_u}{r^3}\left(\gamma_u -\frac{1}{\gamma_u}\right) 
   (2-3\sin^2 \psi)   & = & \frac{\partial B_z}{\partial y},     \\
  4 \pi J^0 +\frac{Q}{r^3}\left(\gamma_u -\frac{1}{\gamma_u}\right) 
   (2-3\sin^2 \psi) & = & \frac{\partial E_x}{\partial x}+  \frac{\partial E_y}{\partial y}, \\
  \frac{1}{c}\frac{\partial E'_{x'}}{\partial t'} + 4 \pi J'_{x'}+\frac{Q\beta_{u'}}{r^3}\left(\gamma_{u'}
    -\frac{1}{\gamma_{u'}}\right) (2-3\sin^2 \psi)   & = & \frac{\partial B'_{z'}}{\partial y'},     \\
  4 \pi (J')^{0} + \frac{Q}{r^3}\left(\gamma_{u'} -\frac{1}{\gamma_{u'}}\right) 
   (2-3\sin^2 \psi)  & = & \frac{\partial E'_{x'}}{\partial x'}+  \frac{\partial E'_{y'}}{\partial y'} 
  \end{eqnarray}  
    where the transformation law of the charge density in the last member of ({\rm E}9.11A) as applied between the frames
    S$^{\ast}$  and S has been used to introduce the 4-vector current density:
   \begin{equation}
    (J^0;\vec{J}) \equiv ( \rho_Q; \beta_u \rho_Q,0,0) = (\gamma_u \rho^{\ast}_Q;\gamma_u \beta_u \rho^{\ast}_Q,0,0).
    \end{equation}
     \par Given that the `empty space' Maxwell equations of \S 6 already describe the electric and magnetic
     force fields at `source free' field points in an identical manner to the `with source' equations
     of the present section it is perhaps worthwhile to ask what additional physics information
     is provided by the latter set of equations? There are two answers to this question, the first
     first concerning predictions for the force fields and the second the description of
      of radiative processes where real photons are created. 
      \par The operational meaning of the force fields is provided by the Lorentz equation (9.1) in terms
      of the force on a test charge. The Maxwell equations contain spatial and temporal partial
      derivatives of these fields. What additional information is provided by these first-order
     partial differential equations? The answer, insofar as the force fields are concerned, is
     not at all clear. In fact predictions of electromagnetic forces are typically extracted from
     Maxwell's equations by invoking the corresponding integral relations, which,
    in a certain sense, annul the partial derivatives:
    \par Gauss' Law for the electric field
     \begin{equation}
       \int_S \vec{E} \cdot d \vec{S} = 4 \pi \int_V \rho_Q^{\ast}dV. 
      \end{equation}
    \par The magnetostatic Amp\'{e}re law        
  \begin{equation}
       \int_s \vec{B} \cdot d \vec{s} = 4 \pi \int_S \vec{J} \cdot d \vec{S}. 
      \end{equation}
 \par The Faraday-Lenz law   
  \begin{equation}
       \int_s \vec{E} \cdot d \vec{s} = -\frac{d~}{dt}\int_S \vec{B} \cdot d \vec{S}. 
      \end{equation}
  \par In highly symmetrical configurations these integral relations provide a simple
    manner to calculate the fields and hence, by use of the Lorentz equation, the physical
    prediction, which is the force on a test charge. For example Coulomb's inverse
    square force law for a point charge may be derived from (12.7)~\cite{LLCF} and the magnetic field
     due to an infinite straight conductor from (12.8). Eq.~(12.9) gives directly 
     the induction force on conduction electrons in a circular coil. However, all
    the intercharge forces of electrodynamics are described by the single formula (3.4)
    in which no `fields' appear and no Lorentz equation must be invoked.
    Maxwell's equations, including their covariance-breaking relativistic
    corrections, are all found {\it en route} to the derivation of Eq.~(3.4)~\cite{JHFRCED,JHFSTF} (see Fig.1). 
    Electric and magnetic force fields, although of tremendous historical
    importance and phenomenological utility, do not provide the most economical
    and fundamental description of electromagnetic forces. This is given by 
    quantum electrodynamics, in which the underlying quantum mechanical mechanism
    is the exchange between electric charges of space-like virtual photons~\cite{JHFRCED}.
     \par The second application of Maxwell's equations with sources concerns an entirely different
      process: the production and propagation of real photons, associated, in classical electrodynamics, 
     with accelerated electric charges. In common with essentially all text book authors
     Einstein did not distingish between force fields (effects of virtual photon exchange)
     and radiation fields (creation propagation and destruction of real photons) using identical
     symbols for the different types of fields and referring in the text to the `force'
     of the radiation fields discussed in \S 7 and \S 8. This is not the place to describe
     in detail how Maxwell's equations with sources are used to calculate radiation
     fluxes. This is done in all advanced text books on classical electromagnetism. Only the essential
     steps are reviewed, and in view of the incorrect transformation law of the energy  density
     of an electromagnetic wave ({\rm E}7.17) given by Einstein the (lack of) covariance of the
     `radiation fields' will be commented on
       \par  The electric and magnetic fields in the Gauss and Amp\`{e}re laws ({\rm E}9.4A) and ({\rm E}9.1A)
            are replaced by the derivatives of the 4-vector potential $A^{\alpha}$ according
   to the defining equations (9.2) and (9.3) to yield second order partial differential
     equations in components of  $A^{\alpha}$. The Lorenz condition:
     \begin{equation} 
      \vec{\nabla} \cdot \vec{A} + \frac{1}{c}\frac{\partial A^0}{\partial t} = 0
     \end{equation} 
    is then used to obtain separate second-order partial differential equations (d'Alembert Equations)
   for $\vec{A}$ and $A^0$. Solving these equations by the Green's function technique
      and calculating the correponding retarded electric and magnetic fields, using (9.2) and (9.3),
     terms with both $1/r^2$ and $1/r$ dependence are obtained. Terms of the second type,
    associated with source charge acceleration, are identified as radiation fields.
    Making the non-relativisic approximation that the maximum source-charge velocity
    is much less than $c$, and using the Poynting vector to calculate the energy flux, the rate of real photon 
    radiation in, for example, dipole radiation is predicted, in good agreement both with experiment
    and, in the appropriate limit with the corresponding quantum mechanical calculation~\cite{JHFSTF}.
    At very large distances from the source, the radiation field, denoted by Einstein
     as $A$, has the same operational meaning as the
     probability amplitude to observe a photon at the corresponding spatial position~\cite{JHFSTF,JHFEJP}.
    The physical interpretation of the radiation field is therefore quite different from that of force
    fields which are quite unrelated to the real photon degrees of freedom. 
    \par It is interesting to note that the Lorenz condition (12.10) is an identity for the instantaneous
     force-field 4-vector potential, being a consequence of the relation (9.24) connecting temporal 
    and spatial partial derivatives~\cite{JHFRCED}. Nevertheless its introduction is essential for the
     derivation of the retarded potential associated with the radiation fields,
     \par It was found above in Section 10 that the force fields of Eqs.~({\rm E}7.1)-({\rm E}7.3) and ({\rm E}7.5)-({\rm E}7.7)
      naively identified as radiation fields gave the incorrect transformation laws ({\rm E}7.17),({\rm E}7.18) for the 
      energy density of a plane electromagnetic wave. Another example of the inadequacy of classical
     electromagnetic fields to give a correct relativistic description of real photon processes
      is discussed in Ref.~\cite{JHFSTF}. Although radiation fields derived from Maxwell's
      equations, as outlined above, give a correct description of dipole radiation
      in the non-relativistic limit of source charge motion, this is no longer
      true if the fields are boosted from the frame in which the average velocity
    of the source charge vanishes, into a frame with $\beta_u \simeq 1$.
    The energy flow predicted
    by the Poynting vector of the transformed radiation fields does then not agree with that obtained by
    direct Lorentz transformation of the energy-momentum 4-vectors of the radiated photons.
    \par Einstein closes \S 9 with the following remark:
     \par{\tt In addition I may briefly remark that the following important law may be easily 
     deduced from the developed equations: If an electrically charged body is in motion
     anywhere in space without altering its charge when regarded from a system of co-ordinates
     moving with the body, its charge also remains --when regarded from the ``stationary'' system S--
      constant.}
     \par This statement is true insofar as the value of the charge $Q$ in the 4-vector electromagnetic
      current (12.6) is a Lorentz-invariant quantity. However the `effective charge', proportional
     to the force on a test charge at a fixed relative position, does vary with the motion of
     the source. Inspection of (9.21) shows that the  longitudinal electric field is reduced by
      the fraction $\beta_u^2/2$ at lowest order in $\beta_u$, while the transverse field is increased 
     by the same fraction. On averaging over an isotropic distribution of directions for a source of
     fixed velocity the radial electric field is increased, at leading order in $\beta_u$,
     by the fraction $\beta_u^2/6$.
     \par A consequence of this is that a test charge close to a magnetostatic system experiences
     an electric field $\propto I^2$ where $I$ is the magnet current. This is because the effective charge
    of moving conduction electrons is different to that of the static positive ions of the conductor
    material. The static transverse field $\propto I^2$ of a superconducting magnet was
     observed and published~\cite{Edwards} in 1976, in agreement with the prediction of the RCED 
      formula (9.21). The same transverse electric field is predicted by the CEM formula
       (9.19) when $\psi = \pi/2$.
     \par It is important to notice that this `electrostatic field in magnetostatics' has no effect
     on the experiments which have tested, to very high precision, the neutrality of matter by subjecting 
    it to strong external electric fields and searching for the effect of electric forces~\cite{King,ZCH}.
    This is because, unlike the dependence of electric field produced by a charge on the
     motion of the charge, the force on a test charge due to an external electric field
     does not depend on the motion of the charge. See the Lorentz force equation (9.1).

 \SECTION{\bf{\S 10. Dynamics of the Slowly Accelerated Electron}}
    In this section, the motion in different reference frames, of a particle of charge $e$ under the
   influence of electric and magnetic fields is considered. The particle is assumed to be 
   instantaeously at rest at the origin of S' at the instant at which the origins of S and S'
  coincide. Only the electric force acts in the frame S' to give the equations of motion
  (Newton's Second Law):
   \begin{eqnarray}
~~~~~~~~~~~~~~~~~~~~~~~~~~~~~~~~~~~~~~~~~~~m\frac{d^2\xi}{d \tau^2} & = & e E'_{\xi},~~~~~~~~~~~~~~~~~~~~~~~~~~~~~~~~~~~~~~~~({\rm E}10.4) \nonumber  \\
~~~~~~~~~~~~~~~~~~~~~~~~~~~~~~~~~~~~~~~~~~~ m\frac{d^2\eta}{d \tau^2} & = & e E'_{\eta},~~~~~~~~~~~~~~~~~~~~~~~~~~~~~~~~~~~~~~~~({\rm E}10.5) \nonumber  \\
~~~~~~~~~~~~~~~~~~~~~~~~~~~~~~~~~~~~~~~~~~~ m\frac{d^2\zeta}{d \tau^2} & = & e E'_{\zeta}.~~~~~~~~~~~~~~~~~~~~~~~~~~~~~~~~~~~~~~~~({\rm E}10.6) \nonumber
 \end{eqnarray}
   Using the transformation of space-time coordinates and field
   components:
    \[~~~~~~~~~~~~ \xi = \gamma(x-vt),~~~\eta = y,~~~\zeta = z,~~~\tau = \gamma(t-vx/c^2,)~~~~~~~~~~~~~~~~~~~~~~~({\rm E}10.7) \]
\[~~~~~~~~~~~~ E'_{\xi} = E_x,~~~ E'_{\eta} = \gamma( E_y-\beta B_z),~~~E'_{\zeta} = \gamma(E_z+\beta B_y)~~~~~~~~~~~~~~~~~~~~~~~({\rm E}10.8) \]
 the following equations of motion in the frame S are obtained:
   \begin{eqnarray}
~~~~~~~~~~~~~~~~~~~~~~~~~~~~~~~m\gamma^3\frac{d^2 x}{d t^2} & = & e E_x = e E'_{\xi},~~~~~~~~~~~~~~~~~~~~~~~~~~~~~~~~~~~~~~~~({\rm E}10.12) \nonumber  \\
~~~~~~~~~~~~~~~~~~~~~~~~~~~~~~~m\gamma^2\frac{d^2 y}{d t^2} & = & e \gamma(E_y-\beta B_z) = e E'_{\eta},~~~~~~~~~~~~~~~~~~~~~~~~~~~({\rm E}10.13) \nonumber  \\
~~~~~~~~~~~~~~~~~~~~~~~~~~~~~~~m\gamma^2\frac{d^2 z}{d t^2} & = & e \gamma( E_z+\beta B_y) = e E'_{\zeta}.~~~~~~~~~~~~~~~~~~~~~~~~~~~({\rm E}10.14) \nonumber 
 \end{eqnarray}
  The derivation of ({\rm E}10.12)-({\rm E}10.14) from ({\rm E}10.4)-({\rm E}10.8) is discussed below, but in order to
  compare these predictions with those of the RCED formulae, a specific source of the fields,
  a charge $Q$ moving with fixed velocity $u$ along the $x$-axis in S, as in Sections 9 and 12, will be
   considered. The same notation for space time-coordinates and field components as in these
   sections will also now be employed. Since, in this case, $E_z = B_x = B_y = 0$ Eqs.~({\rm E}10.6) and ({\rm E}10.14) 
   are no longer relevant. Eqs.~({\rm E}10.4) and ({\rm E}10.5) are written as:
     \begin{eqnarray}
~~~~~~~~~~~~~~~~~~~~~~~~~~~~~~~~~~~~~~~~~~~m\frac{d^2 x'}{d t'^2} & = & e E'_{x'}, ~~~~~~~~~~~~~~~~~~~~~~~~~~~~~~~~~~~~~~~~({\rm E}10.4A) \nonumber  \\
~~~~~~~~~~~~~~~~~~~~~~~~~~~~~~~~~~~~~~~~~~~m\frac{d^2 y'}{d t'^2} & =
       & e E'_{y'} ~~~~~~~~~~~~~~~~~~~~~~~~~~~~~~~~~~~~~~~~({\rm E}10.5A)
       \nonumber  
 \end{eqnarray}
  and Eqs.~({\rm E}10.12) and ({\rm E}10.13) as:
 \begin{eqnarray}
~~~~~~~~~~~~~~~~~~~~~~~~~~~~~~~m\gamma^3\frac{d^2 x}{d t^2} & = & e E_x = e E'_{x'}, ~~~~~~~~~~~~~~~~~~~~~~~~~~~~~~~~~~~~~~~~({\rm E}10.12A) \nonumber  \\
~~~~~~~~~~~~~~~~~~~~~~~~~~~~~~~ m\gamma^2\frac{d^2 y}{d t^2} & = & e \gamma(E_y-\beta B_z) = e E'_{y'}.~~~~~~~~~~~~~~~~~~~~~~~~~~~({\rm E}10.13A) \nonumber
 \end{eqnarray}
 Using the RCED force fields of (9.21) and (9.22), the Lorentz force equation (9.1), 
 and the field transformation laws discussed in Section 9, the following equations
 of motion are found in the frame S:
  \begin{eqnarray}
  \frac{d p_x}{dt} & = & \frac{d(\gamma \beta_x m c) }{dt}  = m\gamma^3\frac{d^2 x}{d t^2}  =  e E_x, 
   = e \frac{\gamma_w  E'_{x'}}{\gamma_u}  \\  
 \frac{d p_y}{dt} & = & \frac{d(\gamma \beta_y m c) }{dt}  = m\gamma \frac{d^2 y}{d t^2} 
   =  e ( E_y-\beta B_z) = e \frac{E'_{y'}}{\gamma} 
 \end{eqnarray}
 where the source charge moves with velocity $w$ along the $x'$-axis in S'. To derive the third member
  of Eq.~(13.2) from the second the condition $\beta_y = 0$ has been used.
 Agreement is found between the first member of Eq.~({\rm E}10.12A) and the third member of (13.1), however the last
  member of ({\rm E}10.12A) is at variance with that of (13.1) and, although ({\rm E}10.13A) and (13.2) are algebraically
  equivalent, the forces defined by these equations differ by a multiplicative factor $\gamma$.
 \par As will be shown below the relation:
  \begin{equation}
  \frac{{\rm transverse~force~in~test~charge~rest~frame}}
     {{\rm transverse~force~in~frame~with~relative}~x{\rm-velocity~}v} = \gamma
  \end{equation} 
   satisfied by (13.2) but not by ({\rm E}10.13A), is of general validity, following
    from the invariance of transverse coordinates with respect to the LT.
   \par Since the methods used to derive, on the one hand ({\rm E}10.12A) and({\rm E}10.13A), and on the
   other (13.1) and (13.2), are different it is perhaps not surprising that different results are obtained.
   Einstein introduces as a {\it hypothesis} the equality of the electromagnetic forces in
   the frames S' and S, and simply rewrites the S' frame fields in terms of the S frame ones according
   to the transformation equations ({\rm E}10.8). In the case of the RCED equations (13.1) and (13.2), the
   fields are directly
   calculated from the source currents using (9.21) and (9.22) for the frame S and (9.28) and (9.29)
   for the frame S' and substituted into the Lorentz force equation (9.1). The left-hand sides of
    (13.1) and (13.2) are just an expression of the relativistic version of Newton's Second Law.
    The forces calculated in this way are found not to be equal in the frames S' and S, in contradiction
   to Einstein's assumption in Eqs.~({\rm E}10.12A) and ({\rm E}10.13A).
    \par In order to derive the left sides of ({\rm E}10.12A) and ({\rm E}10.13A) ,the coordinates $x'$, $y'$ and $t'$ 
    of ({\rm E}10.4A)
    and ({\rm E}10.5A) are transformed using the LT of ({\rm E}10.7) with the replacements $\xi \rightarrow x'$,
     $\eta \rightarrow y'$ and  $\tau\rightarrow t'$. No details were provided of the calculation,
     so it must be conjectured how it was performed. A calculation based on a certain assumption will
     now be shown to yield the left sides of ({\rm E}10.12A) and ({\rm E}10.13A) from those of
     ({\rm E}10.4A) and ({\rm E}10.5A). The assumption is that $v$ and $\gamma$ in the LT in ({\rm E}10.7) may be considered
     constant under the conditions of the problem. This is clearly in the spirit of the title
     of \S 10: `Dynamics of the \underline{Slowly} Acclerated Electron' and the supposed `slow' motion
     of the electron, the latter meaning, presumably, $v \ll c$. However $v \equiv$ $dx$/$dt$
      (and hence $v$ and $\gamma$)
     cannot be constant, as assumed, as this implies that $d^2x/dt^2 = 0$. Making, in any case, this assumption,
      and considering infinitesimal space and time intervals, the ratio of the first to the fourth equation
      in ({\rm E}10.7) gives:
       \begin{equation}
       \frac{d x'}{d t'} =\frac{\frac{d x}{d t} -v}{1-\frac{v}{c^2}\frac{d x}{d t}}.
    \end{equation}
      The ratio of the second to the fourth equations of ({\rm E}10.7) gives
     \begin{equation}
       \frac{d y'}{d t'} =\frac{\frac{d y}{d t}}{\gamma(1-\frac{v}{c^2}\frac{d x}{d t})}.
    \end{equation}
     The fourth equation of ({\rm E}10.7), written with infinitesimal intervals, yields the differential
     operator relation:
    \begin{equation}
        \frac{d ~}{d t'} =\frac{1}{\gamma(1-\frac{v}{c^2}\frac{d x}{d t})}\frac{d ~}{d t}.
    \end{equation}
     Using (13.6) in combination with (13.4) or (13.5) and assuming that $v$ (but not $dx/dt$) is constant
     yields the equations:
     \begin{eqnarray}
       \frac{d^2 x'}{d t'^2} & = & \frac{1}{\gamma(1-\frac{v}{c^2}\frac{d x}{d t})}
  \left[\frac{1}{1-\frac{v}{c^2}\frac{d x}{d t}}+\frac{v(\frac{d x}{d t} -v)}
    {c^2(1-\frac{v}{c^2}\frac{d x}{d t})^2}\right]
     \frac{d^2 x}{d t^2}, \\  
      \frac{d^2 y'}{d t'^2} & = &\frac{1}{\gamma^2(1-\frac{v}{c^2}\frac{d x}{d t})}
  \left[\frac{1}{1-\frac{v}{c^2}\frac{d x}{d t}}+\frac{v\frac{d y}{d t}}
    {c^2(1-\frac{v}{c^2}\frac{d x}{d t})^2}\right]
     \frac{d^2 y}{d t^2}.
      \end{eqnarray}
      Setting finally $dx/dt = v$ and $dy/dt = 0$ (the test charge moves along the $x$-axis in S)
      in (13.7) and (13.8)  give:
     \begin{equation}
        \frac{d^2 x'}{d t'^2} = \gamma^3 \frac{d^2 x}{d t^2},
        ~~~ \frac{d^2 y'}{d t'^2} = \gamma^2 \frac{d^2 x}{d t^2}
      \end{equation} 
   relating the left sides of ({\rm E}10.12A),({\rm E}10.13A) to those ({\rm E}10.4A),({\rm E}10.5A). 
    \par Two comments are in order:
    \begin{itemize}
    \item[(i)] The velocity $v$  is not constant under the conditions of the problem.
    \item[(ii)] Since the electron is being accelerated, S' is not an inertial frame, and the
      LT (E10.7) and (E10.8), with constant $v$ is therefore not applicable to the problem.
    \end{itemize}
   Since $dx/dt = v$ the differential operator relation (13.6) may be written as:
         \begin{equation}
        \frac{d ~}{d t'} =  \gamma(t)\frac{d ~}{d t}.
    \end{equation} 
     As discussed in Section 9, space intervals are Lorentz invariant, so that $dx = dx'$ and (13.13)
     gives:
   \begin{equation}
      \frac{d^2 x'}{d t'^2} =  \gamma(t)\frac{d ~}{d t}\left[\gamma(t)\frac{d x}{d t}\right] =\gamma(t)^4 \frac{d^2 x}{d t^2}.
     \end{equation} 
     Since $dy = dy'$, it follows simililarly that;
    \begin{equation}
      \frac{d^2 y'}{d t'^2} =\gamma(t)^4 \frac{d^2 y}{d t^2}.
     \end{equation}     
     These equations should be contrasted with Einstein's results in (13.9). Correcting ({\rm E}10.12A) by replacing
     the factor $\gamma^3$ on the left by $\gamma^4$ it can be seen that this equation is no longer
     in agreement with Eq.~(13.1). Also the factor $\gamma^2$ on the left side of ({\rm E}10.13A) should be replaced
     by $\gamma^4$. This equation of transverse motion will be discussed further below. 
      \par The evident difference between the corrected equations ({\rm E}10.12A) and ({\rm E}10.13A) and the RCED equations
     of motion (13.1) and (13.2) should come as no surprise. The physical assumptions underlying
     these equations are quite different. In  ({\rm E}10.12A) and ({\rm E}10.13A) the forces in S and S' are,
      by hypothesis, equal. Simply, the electric fields in S' have been transformed
     into the fields in S according to the transformation laws in ({\rm E}10.8). Similarly the derivatives on the
     left sides are transformed (actually incorrectly), according to the space-time transformations of ({\rm E}10.7).
     But there is no physics in such a purely mathematical transformation. In the case of (13.1) and (13.2)
      the electromagentic forces are calculated using the appropriate formula in each  frame,
      and are found not be equal. The left
       sides of the equations contain no transformations, but are expressions of the relativistic
      generalisation of Newton's Second Law ---a `force', in any frame, may be operationally  
     defined as the time derivative of
      of the momentum in that frame. The agreement between ({\rm E}10.12A) and (13.1) as a consequence of the incorrect
     second derivative transformation equation in (13.9) must then be regarded as purely fortuitous.

    \par The general relation (13.3), not respected by ({\rm E}10.13A), will now be proved. It follows from the 
   invariance of the transverse coordinate interval under the LT:
       \begin{equation}
         dy = dy'
       \end{equation}
        and the time dilation relation, $dt  =\gamma dt'$, that Eq.~(13.13) implies conservation of relativistic transverse
    momentum:
       \begin{equation}
        p'_T = m \frac{d y'}{d t'} = m \frac{d y}{d t'} = \gamma m \frac{d y}{d t} = p_T.
      \end{equation}
       Since the transverse force is defined, in any frame, as $d p_T/d t$ and (13.14) states that
      transverse momentum
     is conserved, $d p_T = d p'_T$,
      \begin{equation}
         F_T \equiv \frac{d p_T}{d t}=  \frac{d p'_T}{\gamma d t'} \equiv \frac{F'_T}{\gamma}
      \end{equation}
       which implies
    \begin{equation}
      \frac{F'_T}{F_T} = \gamma
    \end{equation}
      which is Eq.~(13.3)
\par It is important to remark that if Einstein had consistently applied the condition
    that the electron is `slow' in the frame S, i.e. $v \ll c$, as well a similar condition
    on the velocity of the source charge  $u \ll c$, so that terms of O($\beta^2$), O($\beta_u^2$)
      and  O($\beta \beta_u$) and higher orders are neglected, Eqs.~(13.1), ({\rm E}10.12A) and (13.2),({\rm E}10.13A) are
    identical to first order in $\beta$:
    \begin{eqnarray}
     m \frac{d^2 x}{d t^2} &  =  &   e E_x =  e E_{x'} + O(\beta^2,\beta_u^2), \\
     m \frac{d^2 y}{d t^2} &  =  &   e( E_y -\beta B_z) = e E'_{y'} + O(\beta^2).
      \end{eqnarray}
    So, at this order, Einstein has indeed derived the Lorentz force equation as a consequence of the transformation
    of the electric field in the rest frame of the test charge, as previously described in \S 6. In this context
  it is interesting to recall Einstein's remark in \S 6, concerning the equality of the forces on a test 
   charge in frames in which it is either in motion or at rest. The important caveat there mentioned:
      `which, if we neglect the terms multiplied by the second and higher powers of $v/c$', is forgotten
      in the last members of ({\rm E}10.12A) and ({\rm E}10.13A) where the forces in the frames S and S' are instead assumed
     to be exactly equal, in equations where some terms of `the second and higher powers of $v/c$' have been 
     retained
   \par The RCED equations (13.1) and (13.2) are valid for all velocities, not only in the small $\beta$
    limit. It is crucial for Einstein's subsequent arguments that the first member of ({\rm E}10.13A), although
     derived on the basis of the false assumptions pointed out in the remarks (i) and (ii) above, is
    actually identical to the third member of (13.1) and so is relativistically correct, that is, valid
    to all orders in $\beta$. This is essential for the derivation of the seminal equation ({\rm E}10.17) where
    Einstein first introduces the `$E = m c^2$' concept. 
     \par Invoking the Newtonian definition: Force $=$ mass $\times$ acceleration, the concepts of
        `longitudinal' and `transverse' masses are suggested by the left sides of Eqs.~({\rm E}10.12A)
      and ({\rm E}10.13A) respectively:
   \[~~~~~~~~~~~~~~~~~~~~~~~~~~~~~~~~{\rm Longitudinal~mass~} = \frac {m}{(\sqrt{1-\beta^2})^3}.~~~~~~~~~~~~~~~~~~~~~~~~~~~~({\rm E}10.15) \]

     \[~~~~~~~~~~~~~~~~~~~~~~~~~~~~~~~~~~~~~~~~{\rm Transverse~mass~} = \frac {m}{1-\beta^2}.~~~~~~~ ~~~~~~~~~~~~~~~~~~~~~~~~~~~~({\rm E}10.16) \]
     Although the definition of `longitudinal mass' is the same as that given by Lorentz in the previous
     year~\cite{Lor} the `transverse mass' is $\propto \gamma^2$ to be compared with Lorentz'
     definition  where it is  $\propto \gamma$. Thus Lorentz' definitions of both masses
      are consistent with the RCED formulae (13.1) and (13.2). The difference between the Lorentz
      and Einstein definitions of transverse mass has been described in the literature as a
      different `choice of convention in the expression for force and mass in the dynamics of
      charged particles.'~\cite{Holton}. This is misleading. The `convention' happens to be
      correct for the longitudinal mass but not for the transverse one. Stated more bluntly,
       Eqs.~({\rm E}10.13A) and ({\rm E}10.16) are both wrong since the mathematically illicit calculation
      by which they are derived does not, unlike in the case of ({\rm E}10.12A) and ({\rm E}10.15), give, 
      fortuitously, the correct result.

      In the English translation of Ref.~\cite{Ein1} a footnote mentioning the work of Planck
      ~\cite{PlanckRK} states that the Newtonian definition of force as mass times
        acceleration is perhaps not the one best
      adapted to special relativity, but rather one in which `the laws momentum and energy
      assume the simplest form'. This prompts another `What if?' speculation similar to
      the one in Section 11 above, concerning the Planck-Einstein relation (11.2). 
       What if Einstein had adopted instead the definition of a force (suggested by Newton's
        First Law), as the time derivative of the momentum, as in Eqs.~(13.1) and (13.2)? 
        Having obtained the correct equation in the first member of ({\rm E}10.12A) this would suggest
        the following  definition for the momentum of the electron:
      \begin{eqnarray}
  p & = & \int e E_x dt = m \int \gamma^3 \frac{d^2 x}{d t^2} dt = m c\int \gamma^3 \frac{d \beta}{d t}dt
    \nonumber \\
    & = & m c  \int_0^{\beta} \frac{d \beta}{(1-\beta^2)^{\frac{3}{2}}} = m \gamma \beta c. 
       \end{eqnarray}
     This is just the relativistic momentum, introduced later by Planck~\cite{PlanckRK}.
      \par After obtaining the equations of motion in the frame S, ({\rm E}10.12)-({\rm E}10.14), and introducing
          Longitudinal and Transverse masses, the following comment is made:
       \par {\tt We remark that these results as to the mass are also valid for ponderable material points,
             because a ponderable material point can be made into an \newline electron  (in our sense of the
       word) by the addition of an electric charge}{ \it no matter how small.}
       \par (Italics in the original) This is interesting, because, from a modern viewpoint, it is understood
      that the mass variation is a purely kinematical effect, valid for both charged and neutral
      particles independently of any dynamics. Einstein derived
     the formulae for charged particles in the context of classical electrodynamics and it is not clear from
      this statement that dynamical and purely kinematical consequences of special relativity were clearly separated
      in his mind at this point.  
       \par The next paragraph concerns the most important result of Ref.~\cite{Ein1}, both as a revolutionary
     physical concept and for its practical consequences. The kinetic energy, $W$, of the electron is calculated
     by the space integral of the $x$-component of the force in the frame S using the first member of
      Eq.~({\rm E}10.12) or ({\rm E}10.12A):
   \[~~~~~~~~~~~~~~~~~~~~~~~~~~~~W = \int e E_x d x = m \int_0^v \gamma^3 v dv = mc^2\left\{\frac{1}{\sqrt{1-\beta^2}}-1\right\}.~~~~~~~~~~~~~~~~~~~~~~~~({\rm E}10.17) \] 
     There follows the statement:
     \par {\tt Thus when $v = c$, W becomes infinite. Velocities greater than that of \newline light have
       --as in our previous results-- no possiblity of existence.}
     \par This has become a paradigm of 20th Century physics, often termed `causality'.
          It is true of all physical objects with a time-like or light-like energy-momentum 
          4-vector such that
        \begin{equation}
         E^2 - p^2 c^2 \ge 0.
         \end{equation}
        It is not, however, true for virtual photons that, according to QED, are responsible for inter-charge
          electromagnetic forces. These have a space-like energy-momentum  4-vector with
          \begin{equation}
         E^2 - p^2 c^2 < 0.
         \end{equation}
         The speed of such particles is always greater than that of light and is predicted by QED to be 
         infinite in the overall centre-of-mass frame of the interacting charges~\cite{JHFRCED}.
          \par Einstein did not state in Ref.~\cite{Ein1} the equivalence of mass and energy that is implicit
         in ({\rm E}10.17). Later in 1905 he published another paper with the title `Does the Inertia of a Body
         Depend upon its Energy Content'~\cite{Ein3} in which an ingenious thought experiment 
            involving (in modern terms) the emission of a pair of photons, as viewed from two different
         inertial frames, was analysed, to draw the conclusion:
          \par { \tt If a body gives off energy $L$ in the form of radiation its mass \newline 
                  diminishes by $L/c^2$.}
          \par The analysis of the thought experiment was based on the previously found and correct
           (but incorrectly  derived) formula ({\rm E}10.4) for the transformation of the energy of a plane
            electromagnetic wave. At this stage light quanta were still `heuristic' and the equations
            of relativistic particle kinematics, later to be written down by Planck, were not available.
           \par However, it is already clear from ({\rm E}10.17) that the term $m c^2$ has the dimensions of
            energy and is associated with the mass of the particle. This equation may be rewritten as
       \begin{equation}
         E = W +m c^2
         \end{equation}
       where
       \begin{equation}
         E \equiv \frac{m c^2}{\sqrt{1-\beta^2}} = \gamma m c^2. 
         \end{equation}
          If Einstein had taken the time integral of ({\rm E}10.12) to obtain the momentum,
          Eq.~(13.19) would have been obtained. This equation, (13.23) and the 
          identity $\gamma^2 \equiv \gamma^2 \beta^2+1$
           gives
        \begin{equation}
         E^2 = p^2 c^2 + m^2 c^4
         \end{equation}
        while the ratio of (13.19) to (13.23) gives
          \begin{equation}
         v = \frac{p c^2}{E}.
         \end{equation}
      Combining (13.24) and  (13.25):
        \begin{equation}
         E(v=0) \equiv E_0 = m c^2.
         \end{equation}  
       Thus the equations of relativistic kinematics (13.22)-(13.26) were already
      implicit in the formulae given in Ref.~\cite{Ein1}, so that no further thought experiment
       was required to obtain Eq.~(13.26).
       \par Finally, after noting that, by energy conservation, a change in potential energy
        equal to the kinetic energy $W$ in ({\rm E}10.12) is obtained ({\rm E}qn({\rm E}10.19)), the motion of
      an electron in a constant
      magnetic field was considered, making use of ({\rm E}10.13). The magnetic field is 
      parallel to the $z$-axis, but as throughout the paper, the source
       remains unspecified. The equation of motion obtained from
      ({\rm E}10.13) is:
    \[~~~~~~~~~~~~~~~~~~~~~~~~~~~~~~~~~~~~ -\frac{d^2 y}{d x^2} = \frac{v^2}{R} = \frac{e}{m}\beta B_z \sqrt{1-\beta^2}~~~~~~~~~~~~~~~~~~~~~~~~~~~~~~~~({\rm E}10.20) \]
         which yields, for the radius of curvature, $R$, of the
         circular orbit:
 \[~~~~~~~~~~~~~~~~~~~~~~~~~~~~~~~~~~~~R = \frac{m c^2}{e}\frac{\beta}{\sqrt{1-\beta^2}} \frac{1}{B_z}.~~~~~~~~~~~~~~~~~~~~~~~~~~~~~~~~~~~~~~~({\rm E}10.21) \]
          It is informative to compare these results with the prediction of Eq.~(13.2).
         Placing a source charge of opposite sign, at rest in S, at the same position as the
        moving source charge, will produce no additional magnetic field, and according to
        Eq.~(9.21) (setting $\psi = \pi/2$) will reduce the value of the electric field in (13.2) from $E_y$ to
        $(1-1/\gamma_u)E_y$. This opposite charge corresponds, in the present problem,
         to the effect of the positive charge of the ions of the bulk matter in a
         conductor used to produced a magnetic field. The non-vanishing electric field is
          just the `electrostatics in magnetostatics' effect discussed in the previous section.
           Taking it into account gives, for the instantaneous radius of curvature, ${\cal R}$,
         of the electron orbit:
        \begin{equation}
       {\cal R} = \frac{\gamma m c^2}{e} \frac{\beta^2}{[\beta B_z-E_y(1-1/\gamma_u)]}.
         \end{equation}     
         The right side of ({\rm E}10.21) is recovered in the limit $\beta_u \rightarrow 0$.
 
\SECTION{\bf{Summary: Einstein's Mistakes}}
  As stated in the Introduction, and implicit in the title of Ref.~\cite{Ein1}, the purpose
  of Einstein's formulation of Special Relativity Theory (SRT)\footnote{For the reader's
  convenience, some acronyms introduced earlier in the paper are redefined in the present
  section}, on the basis of the Special Relativity Principle and the constancy of the speed
  of light, was:
  \par {\tt ...the attainment of a simple and consistent theory of the \newline
          electrodynamics
      of moving bodies based on Maxwell's theory for stationary  \newline bodies.}
  \par This reliance on the phenomena of Classical Electromagnetism (CEM) in 
    constructing the theory is manifest in the `Kinematical Part' of Ref.~\cite{Ein1},
    through the adoption of the Light Signal Clock Synchronisation Procedure (LSCSP)
 to {\it define} simultaneous and spatially separated events. Since SRT describes
    the physics of (flat) space-time that underlies {\it all} dynamical physical
   phenomena (with the possible exception of gravity), Einstein's approach is,
   {\it a priori}, a perfectly valid one. However,
   applying Occam's razor, it is not the simplest and most economical one (i.e. that
   with the smallest number of, and the simplest, necessary initial postulates) to obtain
   SRT. This was aleady realised as early as 1910 by Ignatowsky~\cite{Ignatowsky}.
     The space-time Lorentz Transformation (LT), from which all purely kinematical
     consequences of SRT may be obtained, does not require, for its derivation,
      the consideration of CEM or of any other dynamical theory~\cite{Pauli,JHFST1,JHFLT1,JHFLT2}.
     The derivations, of the latter type, of the LT, in Refs.~\cite{JHFLT1,JHFLT2}, shown
     in Fig.4, may be compared with Einstein's (putative) derivation shown in Fig.3.
 \par After establising the LSCSP in \S 1, Einstein discusses in \S 2 a thought experiment
      which purports to demonstrate, without invoking the LT,
     `relativity if simultaneity' (RS) by applying the LSCSP of \S 2 to two moving clocks 
      at different positions in the same inertial frame. The analysis presented of this
     thought experiment is Einstein's first major mistake. The LSCSP is defined for
     clocks at rest in some inertial frame; it makes no sense to apply the same equations
     to moving clocks, the velocities of which, relative to light signals, in the
     frame under consideration, are not equal to $c$. The physical inportance
     of {\it relative velocities} of light signals greater than, or less than, $c$ 
     as clearly demonstrated by the existence of ths Sagnac effect~\cite{Sagnac,Post} discovered
      in 1913. The LSCSP is clearly based
     on the assumption of a constant velocity of light signals relative to the
     to-be-synchronised clocks, which are at rest in the frame in which the
     procedure is applied. As shown in Fig.5 and Table 1, the events considered
        by Einstein in the `stationary' frame S are not the Lorentz-transformed
      LSCSP events in the `moving' frame S'. The explicit calculation presented
      in Section 5 shows that the clocks introduced by Einstein at A and B,
      which have an apparent (time dilated) rate equal to that of clocks at rest in 
       S, are judged to be synchronous by observers in both S and S', in contradiction
      to Einstein's conclusion.
   \par In the derivation of the LT in \S 3, the false correspondence of the events in
      S in Fig.5c and 5e with those in S' in Fig.5b and 5d respectively, assumed in 
      \S 2, is maintained. As described in detail in Section 6 above, that the correct LT
      for a synchronisaed clock at the origin of S' ({\rm E}3.27)-({\rm E}3.30) is obtained
       from the incorrect assignment of events in S and S' just mentioned, is
       due to other, compensating, errors in the derivation. The first is the neglect
       of the $x$-dependence of the LT in the initial ansatz for the time 
       transformation equation: $\tau = \tau(x',y,z,t)$ instead of
      $\tau = \tau(x',x,y,z,t)$ where $x' = L = x(t=0)$ is a  constant depending 
      on the choice of spatial coordinate system in the frame S.
     . The second is the neglect of this constant in the LT finally 
      obtained. In the derivation, the space time coordinates in the final equations
      obtained are actually those of events at the point B, whereas the LT derived is 
       appropriate only to a clock at A (the origin in S'). Einstein's `derivation'
      of the LT is therefore fallacious, both with respect to the initial assumptions
       concerning the transformed events, and with respect to the correct physical
       interpretation of the space-time coordinates appearing in the final
        result. 
        \par The second major mistake in Ref.~\cite{Ein1} is the `length contraction'
        (LC) effect of Eq.~({\rm E}4.2) resulting from the failure to take into account
         the appropriate value of $x'$ (or $L$) for the clocks concerned. The LT
         for a synchronised clock at the center of the considered sphere,
        at rest in S', is applied also to points on the surface of the sphere,
        where it does not correctly describe synchronised clocks. Use of the 
          correct LT for an arbitary point on the surface of the sphere shows ({\rm E}qn(7.8))
         that there is no apparent distortion of the sphere due to the LT
          when it is viewed from S. Interestingly enough, the standard RS effect
           (7.9) associated with the spurious LC effect of ({\rm E}4.2), as described
        in all text books on SRT, is not mentioned in Ref.~\cite{Ein1}. The RS
        effect of \S 2, obtained without invoking the LT, although equally
        spurious, is related to different events in S and S', than those
        giving (7.9).   
      \par Einstein's discussion of asymmetric aging, by considering observations
        of a stationary clock and a moving one, after the correct derivation
        of the Time Dilation (TD) formula ({\rm E}4.4), is incomplete, since only
         an observer in the `stationary' frame S is considered. Similar
        asymmetric aging (but in the opposite sense) will be seen by an observer
        at rest in the `moving' frame S' when clocks at rest in S are compared
        with those at rest in S' in the experiment reciprocal to the one considered.
         Einstein did not comment on this apparently paradoxical
          situation in Ref.~\cite{Ein1}.
      \par As pointed out in Refs.~\cite{HSPT,BH,Cocke}, Einstein's discussion of the relative rates of clocks at a
         pole of the Earth and the equator did not take into account gravitational (general relativistic) effects
         that, to a high degree of accuracy, render equal the rates of clocks at sea level at any position
          on the Earth's surface.
     \par Turning now to the `Electrodynamic Part' of Ref.~\cite{Ein1}, it should be pointed
     out that the transformation laws of electric and magnetic fields, ({\rm E}9.13)-({\rm E}6.15),
  obtained by imposing the covariance of Maxwell's equations in `free space' under the LT,
    are based on the 
   tacit assumption that electric and magnetic fields are local `classical' ones for which
   the space and time coordinates may be taken as independent variables in the differential
   calculus. Recalling the operational meaning of electromagnetic fields as proportional
    to the force on a test charge at the corresponding space-time position, it is clear
    that such fields cannot exist in `free space' ---there must be at least one source
     charge in the proximity of the field point to produce the fields. Taking this fact into
    account, the simplest non-trivial electrodynamical system, where mechanical forces act,
    must consist of a least two charges, one the `test charge' and the other the `source
    charge'. When such a system is considered it is clear from the formulae giving
     the fields of a uniformly moving charge, in either CEM or the recently developed
    relativistic theory (RCED)~\cite{JHFRCED}, that, in the case that the source
   charge moves along the $x$-axis, the derivatives with respect to time and the
   $x$-coordinate are not independent, as required for a local classical field,
    but are related by Eq.~(9.24). This formula has been previously given in text books
    on CEM ~\cite{PP1,LL1} but its implication for the transformation laws
    of electric and magnetic fields has only recently been worked out~\cite{JHFSTF}. 
     Because time and the $x$-coordinate are not independent variables for the force
       fields of both CEM and RCED, these fields do not necessarily transform as a
    second-rank tensor according to the equations (9.7) and (9.15)-(9.18) or
     Eqs.~({\rm E}6.13)-({\rm E}6.15). Explicit calculation of the transformation laws of the RCED fields
       between the frames S and S' using Eqs.~(9.21)-(9.22) and (9.28)-(9.29) shows agreement with
      the tensor transformation law for the transverse electric and magnetic fields
      given by Eqs.~({\rm E}6.14) and ({\rm E}6.15), but not for the longitudinal component 
     of the electric field. The first equation in ({\rm E}6.13):
     \begin{equation}
        E'_{x'} = E_x
    \end{equation}
    is replaced by the first formula in (9.30):
     \begin{equation}
        E'_{x'} = \frac{\gamma_u}{\gamma_w} E_x
    \end{equation}
   where the source charge moves with velocities $u$,$w$ along the $x$-axis in the frames S and S'
    respectively. Thus the longitudinal electric field, unlike the transverse electric and magnetic
    fields, does not transform in a covariant manner ---its value depends on the source charge
    velocities in both S and S'. In fact there is, in the problem, a `preferred frame' that breaks
    special relativistic covariance ---it is the frame, S$^{\ast}$, where the source charge is at 
     rest.
    \par It is also shown by direct calculation of spatial and temporal derivatives using the
      RCED fields
    (9.21) and (9.22), and the relation (9.24), that Amp\`{e}re's Law is a necessary 
     consequence of the electric field Gauss Law, thus deriving Maxwell's `Displacement Current'
      term in the former, and that both the  $x$-component
     of  Amp\`{e}re's Law and the electric field Gauss Law are modified by covariance-breaking
     terms that depend on the velocity of the source charge. Einstein's initial hypotheis
     for the derivation of Eqs({\rm E}6.13)-({\rm E}6.15) ---the covariance of Maxwell's Equations--- is thererefore
     invalid for the RCED force fields. However, the $y$- and $z$-components of Amp\`{e}re's Law,
        the Faraday-Lenz Law and the magnetic Gauss Law remain covariant for the RCED force fields.
    \par At the end of \S 6, Einstein re-discusses the problem, introduced in the Introduction,
      of electromagnetic induction in different reference frames, in the light of the field
       transformation laws ({\rm E}6.13)-({\rm E}6.15). Already in the Introduction, Einstein indicates
       that the `luminiferous aether' of the 19th Century will become superfluous; here it
      is further suggested that the `electromotive force' (the magnetic Lorentz force) and,
      by implication, the magnetic field itself, is only an `auxiliary concept'. By the same
       token it could be argued that the electric field, also defined by the force on a test
      charge, is a similar `auxiliary concept', but Einstein does not take this further
     step. Somewhat inconsistently, Einstein is prepared to throw away the aether, but not
     the `electromagnetic waves' supposed throughout the 19th Century and beyond, to propagate
     in it. It is tantamount to throwing away the ocean but not the waves on the shore.
     What was needed to complete Einstein's revolution was the concept of light as a
     particle, that replaces, in an ontological sense, the `electromagnetic waves'. In
       spite, however, of having won the Nobel Prize for the discovery of the photon it is not
     clear that Einstein,
      even to the end of his life\footnote{In a letter to Besso in 1951, Einstein made the comment
      on the subject of the light quantum `Heute glaubt zwar jeder Lump, er wisse es, aber er tr\"{a}ucht
   sich'\cite{AEMBC}.}, ever embraced the concept, trivially accepted
      by all practitioners of experimental High Energy Physics, that the photon
       is a elementary particle like any other such. From a modern viewpoint (see Ref.~\cite{JHFRCED})
        both electric and magnetic fields are second level mathematical abstractions. The
       true seat of the fundamental underlying physics is to be found in QED processes.
       For the mechanical effects described, classically, by electric and magnetic force fields, it
       is the exchange, in a symmetric manner, of space-like virtual photons between the
       source and test charges.
        \par Einstein's failure to obtain the noncovariant transformation law (14.2)
           of the longitudinal electric field results from the introduction of such
          fields as the {\it a priori} physical concepts in terms of which the theory is
          framed, without properly taking into account the nature of the physical phenomenon
         (inter-charge forces) that is to be described with the aid of the fields.
          When this is done it is quite clear that the fields are not classical
        and local but have an essential functional dependence on the physical
          parameters of the source charge (or charges) both for the case of retarded and instantaneous
        inter-charge interactions. Einstein, also, in common with
          the authors of many text books on CEM makes no distinction (using
          identical mathematical symbols) between the `force fields'
         considered in \S 4, \S 9 and \S 10 and the `radiation fields' 
          considered in \S 7 and \S 8, although the operational meaning of the fields
         is quite different in each case.
        \par In \S 7, devoted to the Doppler effect for electromagnetic waves and the
      aberration of light, Einstein presumably derives the associated formulae by assuming
       equality of the phases defined by Eqs.~({\rm E}7.4) and ({\rm E}7.8), Lorentz transforming
        the space-time coordinates in ({\rm E}7.8) into the S frame using the LT  ({\rm E}3.27)-({\rm E}3.30)
         and equatiing the coefficients of the S frame coordinates to obtain
        the correct relations ({\rm E}7.9)-({\rm E}7.15). 
         However, as explained in Section 10, such a procedure is of somewhat dubious
         mathematical validity. Both phases actually vanish, due to the vanishing
         of the factor in curly brackets, in each case. The phases are therefore equal
         (and zero) for arbitary finite values of $\omega$ and $\omega'$ as well as those
         satisfying Eq.~({\rm E}7.9). As shown in Section 10, a less dubious derivation
         of ({\rm E}7.9)-({\rm E}7.15) is provided by exploiting the manifestly Lorentz-scalar property
          of the phase.
         \par At the end of \S 7, the formulae ({\rm E}7.17) and ({\rm E}7.18) for the transformation
           law of the square of the amplitude of what Einstein calls `the amplitude of the
          electric of magnetic force' are written down, without any derivation. In view of the similarity
           between the numerators and denominators of the right sides of ({\rm E}7.13) and ({\rm E}7.17)
           it may be conjectured that Einstein is confusing the concepts of frequency and 
          field amplitude (or incorrectly assuming that they are proportional) at this point. 
          At the beginning of the following \S 8 it is stated that:
           \par{\tt Since $A^2/8 \pi$ equals the energy of light per unit volume
          we have to regard $(A')^2/8 \pi$, by the principle of relativity, as the energy
          of light} (presumably also per unit volume) {\tt in the moving system.}
         \par This statement correctly identifies $A$ and $A'$ as the field amplitudes
         of a plane electromagnetic wave in the frames S and S' respectively. They are
          not  `the amplitude of the electric or magnetic force' as stated in the previous
          section. Indeed this last definition is physically meaningless, since  $A$ and $A'$
          correspond to radiation fields that describe, classically, a parallel beam of
           monochromatic photons~\cite{JHFEJP}, not intercharge forces. This demonstrates
           Einstein's
           confusion concerning the different physical interpretations that must be 
           assigned to force fields and radiation fields. Since $A^2/8 \pi = \rho_E$ 
           where $\rho_E$ is the energy density of the plane electromagnetic wave, Eqs.~({\rm E}7.17)
             and ({\rm E}7.18) give, according to Einstein, the transformation
            law of this energy density. They are incorrect. This is the third major
            error in Ref.~\cite{Ein1}. The correct transformation law for $\rho_E$, Eq.~(10.11),  
             is actually given by the later equation ({\rm E}8.4), on making the
             replacements $E \rightarrow \rho_E$, $E' \rightarrow \rho'_E$. The 
             quantities $E$ and $E'$ in Eq.~({\rm E}8.4) are the total energies of what Einstein calls a
            `light complex'. In order to derive the correct transformation for
         $\rho_E$ from the incorrect transformation law for $A^2$, ({\rm E}7.17), the 
            `light complex' is defined as an arbitary spherical region, in the frame S,
            of an (infinite) plane electromagetic wave, and the spurious LC effect
           derived in \S 4 is applied. The volume of the resulting ellipsoid of revolution,
           supposedly observed in the frame S', to that of the sphere in S
            is the reciprocal of the ratio of $A'$ to $A$ given by Eq.~({\rm E}7.17).
             Consideration of the total energy of the contracted `light complex'
               in S' then leads to the formula ({\rm E}8.4), which must actually replace
    the incorrect formula ({\rm E}7.17) as the transformation law for $\rho_E$ or $A^2$.
        As previously, in the derivation of the LT, incorrect formulae, ({\rm E}7.17) and
        ({\rm E}8.2) are combined in such a way as to give a correct result, here the transformation
        law ({\rm E}8.4), on replacing the spurious quantity, E: `energy of the light complex', by
         $A^2/8 \pi$ or $\rho_E$. Indeed, in view of the frequency transformation
          formula ({\rm E}7.13), the Planck-Einstein relation and the Lorentz invariance of the
           photon number density, the correct transformation law of $\rho_E$,
           Eq.~(10.11), is an immediate consequence of ({\rm E}7.13).
      \par In the discussion of the radiation pressure due to reflection of a plane
         electromagnetic wave at oblique incidence on a uniformly moving plane mirror
       in the second part of \S 8, the incorrect formula ({\rm E}7.17) for the transformation
        of the amplitude of an electomagnetic wave is used, yielding the incorrect
        formulae ({\rm E}8.6) and ({\rm E}8.12) used in the analysis of the reflection problem.
        The formula finally obtained for the light pressure ({\rm E}8.17) is correct, but in
         order to derive it using the wrong amplitude transformation equations
         ({\rm E}8.6) and ({\rm E}8.12) Einstein commits the fourth (and most flagrant) major
          mistake in Ref.~\cite{Ein1}. The formula ({\rm E}8.16), that is purported to
         be the energy flux of the reflected wave is actually the energy flux
         of a wave of the same dimensions and direction as the reflected wave
         but {\it incident on the back surface of the mirror!} (see Fig.7b).
          The writer of the present paper finds in very difficult to understand
          why, to his best knowledge, this obvious blunder has not been pointed 
           out in the century since Ref.~\cite{Ein1} was written. Combining this 
          absurd flux calculation with the incorrect formula ({\rm E}8.12) for
          $A'''$ , gives, fortuitously, the correct result ({\rm E}8.17) for the 
          radiation pressure. In fact the flux of reflected photons is equal 
          to the flux of incident photons (in the reflection process exactly
           one photon is created for each photon that is destroyed). The different
          energy flux then results solely from the transformation of photon energy
          by reflection. The correct result for the radiation pressure is then obtained
            by using ({\rm E}8.4) or (10.11) to transform $\rho_E$ and assuming equal fluxes
          of incident and reflected photons. This calculation is in Eqs.~(11.3)-(11.6).
          Alternatively an  `energy flow' analysis, as done by Einstein, can be performed, usng the correct 
         photon fluxes and energy densities to obtain the same result. This calculation yields
         Eqs.~(11.13)-(11.15).
        \par The fifth and sixth major mistakes in Ref~\cite{Ein1} occur in the concluding 
        section \S 10, `Dynamics of the Slowly Accelerated Electron', but as in previous cases
        discussed above, the effects of the errors cancel in the final, and in this case crucial, result.
        Thus the formula ({\rm E}10.12), that is integrated to obtain the seminal relation ({\rm E}10.7)
        demonstrating the equivalence of mass and energy, is correct.
         \par The fifth error is a conceptual one. In order to derive the differential equation
        of motion of the `Slowly Accelerated Electron' in the frame S from that in the frame S'
         (where it is instantaneously at rest), the coordinates and fields in the former
         equation are simply transformed into those of the frame S' using the equations
          ({\rm E}10.7)-({\rm E}10.8). Such a procedure correctly predicts the Lorentz Force Equation in
          the frame S to the lowest order in $\beta$
          (i.e. neglecting corrections of O($\beta^2$)). However, since only a change of variables
           is performed, it is assumed that the force components are equal in the frames S and S'.
           Explicit caclulation of the forces for the case of a uniformly moving source
           charge using the RCED equations (9.21) and (9.22) shows that the forces are not
           equal in the two frames when  O($\beta^2$) and higher order corrections are
           included. Einstein's initial hypothesis in deriving Eqs.~({\rm E}10.12)-({\rm E}10.14) is
           therefore not correct in the relativistic theory.
         \par The sixth error occurs in the calculation of the Jacobian relating
           $d^2\xi/d \tau^2$ in Eq.~(10.4) to  $d^2x/d t^2$ in Eq.~(10.12). This calculation
           assumes that $v$ and $\gamma$ are constant in Eq.~(10.7), but that $dx/dt$ is not.
           Since $v \equiv dx/dt$, this is impossible. Calculating correctly
          the Jacobian, allowing for the time dependence of  $v$ and $\gamma$,
          and the Lorentz invariance of spatial intervals $d \xi = d x$, results in the replacement
          $\gamma^3 \rightarrow \gamma^4$ in the left side of (10.12) and 
          $\gamma^2 \rightarrow \gamma^4$ in the left sides of ({\rm E}10.13) and ({\rm E}10.14). 
           The spatial integration of the so-modified  Eq.~({\rm E}10.12) then no longer yields
           the mass-energy equivalence relation ({\rm E}10.17). In fact the $\gamma^3$ factor 
           on the left side of ({\rm E}10.12) is correct, but it orginates not from the
           Jacobian of a coordinate transformation but from the {\it definition of force as the
          time derivative of momentum} and the definition, (13.32), of relativistic momentum
          (see Eq.~(13.1).
         \par To conclude this section a list of the incorrect formulae from Ref.~\cite{Ein1} is 
              given, together with, in each case, the corresponding correct relativistic 
              formula.
       \begin{itemize}
          \item[(i)]\underline{ Space-time Lorentz Transformation} 
         \begin{eqnarray}
 ~~~~~~~~~~~~~~~~~~~~~~~~~~~~~~\tau & = & \gamma(t-\frac{v x}{c^2}),~~~~~~ ~~~~~~ ~~~~~~~~~~~~~~~~~~~~~~~~~~~~~~~({\rm E}3.27) \nonumber  \\
 ~~~~~~~~~~~~~~~~~~~~~~~~~~~~~~\xi & = & \gamma(x-vt),~~~~~~~~~~~~~~~~~~~~~~~~~~~~~~~~~~~~~~~~~~~~({\rm E}3.28) \nonumber  \\
 ~~~~~~~~~~~~~~~~~~~~~~~~~~~~~~~~~  \eta & = & y,~~~~~~~~~~~~~~~~~~~~~~~~~~~~~~~~~~~~~~~~~~~~~~~~~~~~~~({\rm E}3.29) \nonumber  \\
 ~~~~~~~~~~~~~~~~~~~~~~~~~~~~~~~~~  \zeta & = & z.~~~~~~~~~~~~~~~~~~~~~~~~~~~~~~~~~~~~~~~~~~~~~~~~~~~~~~({\rm E}3.30) \nonumber
    \end{eqnarray}  
        The equations ({\rm E}3.27) and ({\rm E}3.28) are correct only for a synchronised
     clock at $\xi = 0$. For such a clock
        at  $\xi = L$ they are replaced by:
         \begin{eqnarray}
    \tau & = & \gamma[t-\frac{v(x-L)}{c^2}],  \\
 \xi-L & = & \gamma[x-L-vt] = 0
    \end{eqnarray}  
   or, equivalently:
          \begin{eqnarray}
    t & = & \gamma \tau,  \\
    x & = & L+v t.
    \end{eqnarray}   

  \item[(ii)] \underline{Spurious Length Contraction Effect}

 \[~~~~~~~~~~~~~~~~~~~~~~~~~~~~~~~~~~~~~~~\frac{x^2}{1-(v/c)^2} + y^2 +z^2 = R^2 ~~~~~~~~~~~~~~~~~~~~~~~~~~~~~~~~~~~({\rm E}4.2) \]
         is replaced by
   \begin{equation}
     x^2 + y^2 +z^2 = R^2.
  \end{equation}
 \item[(iii)]\underline{ Transformation of the Energy Density of Plane {\it EM} Wave}
  \begin{eqnarray}
~~~~~~~~~~~~~~~~~~~~~~~~~~~~~~(A')^2 & = & \frac{A^2(1-\beta \cos \phi)^2}{1-\beta^2},~~~~~~~~~~~~~~~~~~~~~~~~~~~~~~~~({\rm E}7.17) \nonumber  \\
~~~~~~~~~~~~~~~~~~~~~~~~~~~~~~(A')^2 & = & \frac{A^2(1-\beta)}{1+\beta}~~~~~~~~~~~~~~~~~~~~~~~~~~~~~~~~~~~~~~~~~~({\rm E}7.18) \nonumber 
     \end{eqnarray}
    are replaced by:
   \begin{eqnarray}
     (A')^2 & = & \frac{A^2(1-\beta \cos \phi)}{\sqrt{1-\beta^2}}, \\
    (A')^2 & = & A^2\sqrt{\frac{1-\beta}{1+\beta}}. 
     \end{eqnarray}
    Also E'/E in Eqs.~({\rm E}8.4), ({\rm E}8.5) is replaced by  $(A')^2/A^2$.       
   \begin{eqnarray}
~~~~~~~~~~~~~~~~~~~~~~~~~~~~~~ A' & = &  A \frac{1-\beta \cos \phi}{\sqrt{1-\beta^2}},~~~~~~~~~~~~~~~~~~~~~~~~~~~~~~~~~~~~~~~~~~~({\rm E}8.6) \nonumber  \\
~~~~~~~~~~~~~~~~~~~ A''' & = &   A'' \frac{1+\beta \cos\phi''}{\sqrt{1-\beta^2}} =  A \frac{(1-2\beta \cos \phi+ \beta^2)}{1-\beta^2}~~~~~~~~({\rm E}8.12) \nonumber 
   \end{eqnarray}
      are replaced by:     
   \begin{eqnarray}
       (A')^2 & = &  A^2 \frac{1-\beta \cos \phi}{\sqrt{1-\beta^2}}, \\
       (A''')^2 & = &   (A'')^2 \frac{1+\beta \cos \phi''}{\sqrt{1-\beta^2}}
           =  A^2 \frac{(1-2\beta \cos \phi+ \beta^2)}{1-\beta^2}.
   \end{eqnarray}
   The reflected energy flux:
 \[~~~~~~~~~~~~~~~~~~~~~~~~~~~~~~~~~~~~~~~f_{E'''} = \frac{(A''')^2(-c \cos \phi''' + v)}{8 \pi}, ~~~~~~~~~~~~~~~~~~~~~~~~~~~~~~~~~~~({\rm E}8.16) \]
      is replaced by:
   \begin{equation}
     f_{E'''} = \frac{(A''')^2(c \cos \phi - v)}{8 \pi} =  f_{E}(A''')^2/A^2
   \end{equation}
     where $ f_{E}$ is the incident energy flux.

  \item[(iv)] \underline{Amp\`{e}re's Law and the Gauss Law for the Electric Field}

  \begin{eqnarray}
  ~~~~~~~~~~~~~~~~~~~~~~~~~~~~~~ \frac{1}{c}\frac{\partial E_x}{\partial t} + 4 \pi \beta_u \rho_Q & = & \frac{\partial B_z}{\partial y},~~~~~~~~~~ ~~~~~~~~~~~~~~~~~~~~~~~~~~~~~~~({\rm E}9.1A) \nonumber  \\
~~~~~~~~~~~~~~~~~~~~~~~~~~~~~~  4 \pi \rho_Q & = & \frac{\partial E_x}{\partial x}+  \frac{\partial E_y}{\partial y},~~~~~~~~~~~~~~~~~~~~~~~~~~~~~~~({\rm E}9.4A) \nonumber  \\
 ~~~~~~~~~~~~~~~~~~~~~~~~~~~~~~ \frac{1}{c}\frac{\partial E'_{x'}}{\partial t'} + 4 \pi \beta_{u'} \rho'_Q & = & \frac{\partial B'_{z'}}{\partial y'},~~~~~~~  ~~~~~~~~~~~~~~~~~~~~~~~~~~~~~~~~~({\rm E}9.5A) \nonumber 
 \end{eqnarray}
  \begin{eqnarray}
~~~~~~~~~~~~~~~~~~~~~~~~~~~~~~~~ 4 \pi \rho'_Q & = & \frac{\partial E'_{x'}}{\partial x'}+  \frac{\partial E'_{y'}}{\partial y'}
 ~~~~~~~~~~~~~~~~~~~~~~~~~~~~~~~~~~~~~~~~~~({\rm E}9.11A) \nonumber  
  \end{eqnarray}
     are replaced by
    \begin{eqnarray}
  \frac{1}{c}\frac{\partial E_x}{\partial t} + 4 \pi  \beta_u \rho_Q +\frac{Q\beta_u}{r^3}\left(\gamma_u -\frac{1}{\gamma_u}\right) 
   (2-3\sin^2 \psi)   & = & \frac{\partial B_z}{\partial y},~~~~~~~~~      \\
  4 \pi \rho_Q  +\frac{Q}{r^3}\left(\gamma_u -\frac{1}{\gamma_u}\right) 
   (2-3\sin^2 \psi) & = & \frac{\partial E_x}{\partial x}+  \frac{\partial E_y}{\partial y},~~~~~~~~~ \\
  \frac{1}{c}\frac{\partial E'_{x'}}{\partial t'} + 4 \pi \beta_{u'} \rho'_Q +\frac{Q\beta_{u'}}{r^3}\left(\gamma_{u'}
    -\frac{1}{\gamma_{u'}}\right) (2-3\sin^2 \psi)   & = & \frac{\partial B'_{z'}}{\partial y'}, ~~~~~~~~~   \\
  4 \pi \rho'_Q + \frac{Q}{r^3}\left(\gamma_{u'} -\frac{1}{\gamma_{u'}}\right) 
   (2-3\sin^2 \psi)  & = & \frac{\partial E'_{x'}}{\partial x'}+  \frac{\partial E'_{y'}}{\partial y'}.~~~~~~~~~
  \end{eqnarray}     
  \item[(v)]\underline{Differential Equation of Motion of Acclerated Electron}
   \begin{eqnarray}
~~~~~~~~~~~~~~~~~~~~~~~~~~~~~~~m\gamma^3\frac{d^2 x}{d t^2} & = & e E_x = e E'_{\xi},~~~~~~~~~~~~~~~~~~~~~~~~~~~~~~~~~~~~~~~~({\rm E}10.12) \nonumber  \\
~~~~~~~~~~~~~~~~~~~~~~~~~~~~~~~m\gamma^2\frac{d^2 y}{d t^2} & = & e \gamma({\rm E}_y-\beta B_z) = e E'_{\eta},~~~~~~~~~~~~~~~~~~~~~~~~~~~({\rm E}10.13) \nonumber  \\
~~~~~~~~~~~~~~~~~~~~~~~~~~~~~~~m\gamma^2\frac{d^2 z}{d t^2} & = & e \gamma({\rm E}_z+\beta B_y) = e E'_{\zeta}~~~~~~~~~~~~~~~~~~~~~~~~~~~({\rm E}10.14) \nonumber 
 \end{eqnarray} 
     are replaced by  
  \begin{eqnarray}
  \frac{d p_x}{dt} & = &  \frac{d(\gamma \beta_x m c) }{dt}  = m\gamma^3\frac{d^2 x}{d t^2}  =  e E_x 
   = e \frac{\gamma_w  E'_{x'}}{\gamma_u}, \\ 
 \frac{d p_y}{dt} & = & \frac{d(\gamma \beta_y m c) }{dt} =   m\gamma\frac{d^2 y}{d t^2} = 
  e ({\rm E}_y-\beta B_z) = e \frac{E'_{y'}}{\gamma}, \\
 \frac{d p_z}{dt} & = & \frac{d(\gamma \beta_z m c) }{dt} =   m\gamma\frac{d^2 z}{d t^2} = 
  e ({\rm E}_z+\beta B_y) = e \frac{E'_{z'}}{\gamma}.
 \end{eqnarray} 
     \[~~~~~~~~~~~~~~~~~~~~~~~~~~~~~~~~~~~~~~~~{\rm Transverse~mass~} = \frac {m}{1-\beta^2}~~~~~~~ ~~~~~~~~~~~~~~~~~~~~~~~~~~~~({\rm E}10.16) \]
     is replaced by
 \begin{equation}
{\rm~~~~~ Transverse~mass~} = \frac {m}{\sqrt{1-\beta^2}}.
 \end{equation}

 \end{itemize}
\SECTION{\bf{Conclusions
. The Enduring Legacy of Einstein's 1905 Special Relativity Paper}}
   To complement the list of wrong formulae presented at the end of the previous section
   the present one starts with a list of the correct formulae from Ref.~\cite{Ein1}, leaving aside
   the question whether the derivations given were physically valid or, as discussed above, in 
    some cases, fallacious.
   In conclusion, the importance and orginality of the relativistic predictions of Ref.~\cite{Ein1}
   will be discussed. In all equations in this section, the notation for times and spatial coordinates
   introduced in Section 2 will be employed.
  \begin{itemize}
 \item[(a)] \underline{The Lorentz transformation for a clock at the origin of S' and
     the time dilation effect}
   \begin{eqnarray}
    t' & = & \gamma[\tau-\frac{v x}{c^2}], \\
    x' & = & \gamma[x -v \tau] = 0, \\
    y' & = & y, \\
    z' & = & z.
   \end{eqnarray}
      (15.1) and (15.2) may also be written as:
   \begin{eqnarray}
      \tau & = & \gamma t', \\
        x & = & v \tau
   \end{eqnarray}
   where (15.5) describes the time dilation effect, giving the apparent time, $t'$, of a clock
      at rest in S', as viewed from S, and $\tau$ is the corresponding time registered by a 
      similar and sychronised clock at rest in S. 

  \item[(b)] \underline{Velocity addition formulae}
     \begin{equation}
   V =\frac{\sqrt{v^2+\omega^2 + 2 v\omega \cos \alpha -[(v \omega \sin \alpha)/c^2]^2}}{
         1+ \frac{v \omega}{c^2} \cos \alpha}
   \end{equation}
    or, when $\alpha = 0$ (parallel velocity addition)
      \begin{equation}
   V =\frac{v+\omega}{1+ \frac{v \omega}{c^2}}. 
   \end{equation}   
   See also the formulae for $x$- $y$- and $z$-components in (h) below.

    \item[(c)] \underline{Transformation laws for transverse electric fields and magnetic
        fields}
       \begin{eqnarray}
         E'_{y'} & = & \gamma[E_y-\beta B_z], \\
       E'_{z'} & = & \gamma[E_z+\beta B_y], \\
       B'_{y'} & = & \gamma[B_y+\beta E_z], \\
       B'_{z'} & = & \gamma[B_z-\beta E_y].
  \end{eqnarray}
    \item[(d)] \underline{Doppler shift and aberration of light}
         \begin{equation}
         \nu ' = \nu \gamma(1- \beta \cos \phi)
   \end{equation}
    or, when $\phi = 0$
           \begin{equation}
         \nu ' = \nu \sqrt{\frac{1-\beta}{1+\beta}},
  \end{equation}
   \begin{equation}
     \cos \phi' = \frac{\cos \phi- \beta}{1-\beta \cos \phi}.
  \end{equation}
     \item[(e)] \underline{Transformation of the energy density, $\rho_E$, of a plane
            electromagnetic wave}
         \begin{equation}
       \rho'_{E'} = \rho_E\gamma(1- \beta \cos \phi)
   \end{equation}
        or, when $\phi = 0$,
          \begin{equation}
        \rho'_{E'} = \rho_E \sqrt{\frac{1-\beta}{1+\beta}}.
  \end{equation}
   These formulae are obtained by making the substitutions $E \rightarrow A^2/(8\pi) = \rho_E$, 
   $E' \rightarrow (A')^2/(8\pi) = \rho'_{E'}$ in the transformation formulae ({\rm E}8.4) and ({\rm E}8.5) of Einstein's
    `light complex'. They are identical to the transformation laws of the energy of a photon
      given by the replacements $ \rho_E \rightarrow E_{\gamma}$, 
      $ \rho'_{E'} \rightarrow E'_{\gamma}$
 
     \item[(f)] \underline{Radiation pressure of a plane electromagnetic
      wave incident on a moving mirror}
         \begin{equation}
          P = 2 \frac{A^2}{8 \pi} \frac{(\cos \phi-\beta)^2}{1-\beta^2}.
        \end{equation}

      \item[(g)] \underline{Transformation of the $y$-component of Amp\`{e}re's Law
          and of the Faraday-Lenz Law}
   \par  Here is assumed, as in Section 12, that the source charge moves with 
     velocity v along the $x$-axis in the frame S, and that the field point lies in
      the $x$-$y$ plane
   \par  In S:
   \begin{eqnarray}
    \frac{1}{c} \frac{\partial E_y}{\partial t} & = & -\frac{\partial B_z}{\partial x}, \\
  \frac{1}{c} \frac{\partial \vec{B}}{\partial t} & = & -\vec{\nabla} \times \vec{E}. 
   \end{eqnarray}
   In S':
   \begin{eqnarray}
   \frac{1}{c} \frac{\partial E'_{y'}}{\partial t'} & = & -\frac{\partial B'_{z'}}{\partial x'}, \\
  \frac{1}{c} \frac{\partial \vec{B}'}{\partial t'} & = & -\vec{\nabla}' \times \vec{E}'.
   \end{eqnarray}
      \item[(h)] \underline{Transformation of the charge and current 
                  densities}
    \par   In S:
       \begin{equation}
           (c \rho_Q; \vec{u} \rho_Q) = (c \rho_Q; u_x  \rho_Q, u_y  \rho_Q, u_z  \rho_Q). 
       \end{equation}
      In S':
       \begin{equation}
           (c \rho'_Q; \vec{u}' \rho'_Q) = (c \rho'_Q; u'_{x'}  \rho'_Q, u'_{y'}  \rho'_Q, u_{z'}  \rho'_Q). 
       \end{equation}
        where:
      \begin{eqnarray}
       u'_{x'} & = & \frac{u_x-v}{1-\frac{u_x v}{c^2}}, \\
       u'_{y'} & = & \frac{u_y}{\gamma(1-\frac{u_x v}{c^2})}, \\
      u'_{z'} & = & \frac{u_z}{\gamma(1-\frac{u_x v}{c^2})}, \\
       \rho'_Q & = & \rho_Q \gamma (1-\frac{u_x v}{c^2}).
  \end{eqnarray}

      \item[(i)] \underline{Charge motion in electric and magnetic fields}
       \par If the charge, $e$, of mass $m$ moves instantaneously parallel to the $x$-axis then:
        \begin{eqnarray}
        m \frac{d^2 x}{d t^2} & = & e E_x + O(\beta^2), \\
        m \frac{d^2 y}{d t^2} & = & e [E_y-\beta B_z] + O(\beta^2), \\
        m \frac{d^2 z}{d t^2} & = & e [E_z+\beta B_y] + O(\beta^2)
 \end{eqnarray}
   or, equivalently in 3-vector notation:
           \begin{equation}
        m \frac{d^2\vec{x}}{d t^2} = e[\vec{E}+ \vec{\beta} \times \vec{B}]  + O(\beta^2). 
     \end{equation}
       To all orders in $\beta$:
    \begin{equation}
     \frac{d p_x}{d t} = m \gamma^3 \frac{d^2 x}{d t^2}= e E_x.
     \end{equation}
        \item[(j)] \underline{The equivalence of mass and energy}
       \begin{equation}
       W \equiv {\rm~kinetic~energy~} = (\gamma-1)m c^2   
  \end{equation}
         or, equivalently,
      \begin{equation}
       E \equiv {\rm~relativistic~energy~} \equiv \gamma m c^2 = m c^2 + W.   
  \end{equation}
        When the object is at rest , $W = 0$, and $E  \rightarrow E_0$ where
   \begin{equation}
       E_0 \equiv {\rm~ relativistic~rest~energy~} =  m c^2.   
  \end{equation}
   \item[(k)] \underline{Motion of a charge in the plane perpendicular to a magnetic
     field $\vec{B}$}
   \begin{equation}
      R = {\rm~radius~of~orbit~} = \frac{m c^2}{e} \frac{\beta}{\sqrt{1-\beta^2}} \frac{1}{B}.   
   \end{equation}
   \end{itemize}

   \par In judging the value of any scientific work which satisfies the necessary pre-requisites
    of internal self-consistency and formal correctness, two questions are paramount: firstly,
     is it original? and secondly, is it important? These criteria will first be applied,
    point-by-point, to the list, (a)-(k), of correct results from Ref.~\cite{Ein1} just given, before,
    finally, discussing Einstein's methodology in more general terms, and giving a
     personal opinion about the merit of the work presented in Ref.~\cite{Ein1} from the perspective
     of the 21st Century. 
     \par Concerning (a), the LT in the form (15.1)-(15.4) was first given by Larmor in 1900
      ~\cite{Lar1900,CK}. The transformation given by Lorentz in 1904
    in his last pre-relativity paper ~\cite{Lor} was identical to that of Larmor~\cite{Lar1900,CK}.
  The transformation was given the name `Lorentz Transformation' by  Poincar\'{e} in 1905~\cite{HP1905}.
    The essential difference from earlier, or contemporary but independent,
    authors was that  Einstein claimed for the first time to {\it derive} the LT from some simple
    axioms. Although, as shown in Section 6 above, the `derivation' of Ref.~\cite{Ein1} was
    actually fallacious,
    the essential idea was taken up by many later authors, so that, at the present time, the 
   literature devoted to the derivation of the LT on the basis of various initial postulates
   is vast. See Refs.~\cite{BG,JHFLT1} for citations of only the derivations that
    do not employ Einstein's second postulate. 
    \par The time dilation prediction (15.5) in (a) is one of the two key
     predictions of Ref.~\cite{Ein1} that changed, for ever, a fundamental conception of
     physics. As a physical consequence of the relativistic
      transformation of time it was first pointed out by Larmor in 1897~\cite{Lar1897,CK}
      who noted that orbiting electrons in a uniformly moving inertial frame
      will appear, to a stationary observer, to be moving slow by the factor $1/\gamma$, or $1-\beta^2/2$ at 
      O($\beta^2$).
        In differential form, (15.5) is the basis for relativistic kinematics.
     The energy-momentum 4-vector, P, in the frame S, of a physical object of mass $m$, that is
      at rest in the frame S', is derived from  the space-time  4-vector
       $X  \equiv (c \tau; \vec{x})$ in S according to the relation:
        \begin{equation}
       P \equiv m \frac{d X}{d t'} = m \gamma \frac{d X}{d \tau} = m (\gamma c; \gamma \vec{v})
        \equiv m V
      \end{equation}
      where $V$ is the 4-vector velocity in the frame S. The differential form $d\tau = \gamma dt'$ of the
     time dilation formula (15.5) is essential for this definition of $P$ and the consequent formulae of
       relativistic kinematics.
      \par Although Lorentz had previously introduced the concept of `local time'
       as early as 1895~\cite{Lor1895}, and the differential form of (15.5) appears
      as an intermediate step (among many others) in the section of Poincar\'{e}'s 
      1906 relativity paper~\cite{HP1906} entitled `Contraction of Electrons', it seems
       reasonable to claim that Einstein
      was the first to publish the interpretation of the TD relation
      (15.5) as a universal property of the observed time 
       of {\it any} uniformly moving clock.
        \par The velocity addition formulae of (b) and (h) above were first given, 
         independently, by Einstein in Ref.~\cite{Ein1} and  Poincar\'{e}~\cite{HP1906}.
        \par The transformation laws (15.9)-(15.12) of transverse electric fields and
         magnetic fields in (c) were first published by Larmor in 1900~\cite{Lar1900,CK}.
         They were also given, independently of Einstein's work, by Lorentz in 1904~\cite{Lor}, and
         Poincar\'{e} in 1906~\cite{HP1906}.
          \par To the present writer's best knowledge, the formulae in (d) for the relativistic
            Doppler shift and aberration of light were given for the first time by
           Einstein in  Ref.~\cite{Ein1}. They contain implicitly, in view of
          the Planck-Einstein relation $E_{\gamma} = h \nu$, the relativistic kinematics
          of photons. Similarly, given the photon concept, the transformation
          formulae for the energy density of a plane electromagnetic wave, (15-16) and (15.17)
         in (e) also follow directly from (15.13) and the Planck-Einstein relation~\cite{JHFEJP}.
        \par The correct relativistic light pressure formula, (15.8), in (f) was obtained for 
         the first time by Einstein in Ref.~\cite{Ein1}, albeit as the result of an absurdly
        incorrect derivation (see Section 11 above). 
         \par The covariance of the transverse components of Amp\`{e}re's Law and of the
          Faraday-Lenz Law in (g) were independently demonstrated by Larmor in 1900~\cite{Lar1900,CK},
         Lorentz in 1904~\cite{Lor},
          and by Poincar\'{e} in 1906~\cite{HP1906}.
       \par The transformation of the charge and current densities in (15.23)-(15.28) of (h) 
          were also independently obtained by Poincar\'{e}~\cite{HP1906}. It is interesting
          to note that the 4-vector character of the charge and current densities in
          (15.23) and (15.24), also pointed out by  Poincar\'{e}~\cite{HP1906}, as well
          the 4-vector character of $U \equiv (\gamma_u c;\gamma_u \vec{u})$,
               are also implicit in Einstein's formulae (15.23)-(15.28).
            Choosing $\vec{u}$ parallel to the $x$-axis, (15.24), (15.25) and (15.28) give:
            \begin{eqnarray}
           \rho'_Q & = & \gamma (\rho_Q- \frac{(\rho_Q u) v}{c^2}), \\
           \rho'_Q u' & = & \gamma (\rho_Q u- v \rho_Q). 
             \end{eqnarray}
        Comparing (15.39) and (15.40) with (15.1) and (15.2), it can be seen that 
         $\rho_Q$ and  $\rho_Q u$ transform in the same manner as $\tau$ and $x$ respectively
         under the LT, and so constitute a 4-vector. Using the relation (15.28) to transform
         the charge density in the rest frame of the source charge, $\rho^{\ast}_Q$ into the frames
          S and S' gives the relations:
        \begin{eqnarray}
           \rho_Q & = & \gamma_u \rho^{\ast}_Q, \\
           \rho'_Q & = & \gamma_{u'}\rho^{\ast}_Q. 
             \end{eqnarray}
        Substitution of (15.41) and (15.42) into (15.39) and (15.40) gives:
        \begin{eqnarray}
           \gamma_{u'} & = & \gamma (\gamma_u - \frac{(\gamma_u  u) v}{c^2}), \\
          \gamma_{u'} u' & = & \gamma (\gamma_u\ u- v \gamma_u) 
             \end{eqnarray}

   showing that $ \gamma_u$ and  $ \gamma_u u$ also transform as $\tau$ and $x$, establishing
    the 4-vector character (already manifest for $V \equiv (\gamma c; \gamma \vec{v})$ in (15.38))
     for $U \equiv(\gamma_u c
; \gamma_u u,0,0)$.

    \par In (i) the right side of (15.32) gives the Lorentz force, the magnetic part of
      which also, as explained in \S 6:
    `...which {\it if we neglect the terms multiplied by second and higher powers of $v/c$}
        (my italics) is equal to the vector product of the velocity of the charge and the magnetic 
       force' (should be `magnetic field')`divided by the velocity of light'. In fact, the right
       side of (15.32) gives the Lorentz force correctly to all orders in $v/c$;  
       the left side, however, is the time derivative of the momentum, as in
       Eqs.~(14.17)-(14.19), not, as in Einstein's formulae `mass $\times$ acceleration'.
       This elegant derivation of the Lorentz force equation, directly from the transformation
        laws of electric and magnetic fields, although only valid
       to first order in $\beta$, gives a deep insight into what Einstein calls the `auxiliary'
        nature of the `magnetic field' concept relative to the physical effect ---the force on a test
        charge-- considered. It may be contrasted with Lorentz' derivation of his force formula
         based on a Lagrangian containing the electromagnetic potential
         and the Principle of Least Action. This, also discusssed in detail by Poincar\'{e}~\cite{HP1906},
         is  by comparison, lengthy and rich in mathematical detail, but weak in physical insight.
          A similar derivation may be found in the well-known text-book of Landau and
          Lifschitz~\cite{LLLF} as well as in Ref.\cite{JHFRCED} by the present author.
          \par Eq.~(15.33) gives the correct relativistic generalisation of Newton's Second Law, for
          the problem considered. The last member of this equation was independently given by 
           Poincar\'{e} in Section 7 of Ref.~\cite{HP1906}. However neither Einstein nor Poincar\'{e}
           introduced the concept of relativistic momentum implicit in the first member of
           (15.33). The second member of the equation is crucial for the derivation
           of the mass-energy equivalence relation (15.34) of (j).
           The most original and, in view of practical applications, most important, prediction
         of Ref.~\cite{Ein1} is that of (j) ---the equivalence of mass and energy. For this it
         is necessary to introduce the definition of relativistic energy of (15.35) by
          mathematical substitution in (15.34). The resulting formula, as well as the `$E = m c^2$'
         relation (15.36) are both of completely general validity. The paper, published
         by Einstein later in 1905~\cite{Ein2}, containing
          the (controversial\footnote{ In 1952 Ives claimed that the argument in 
              Ref.~\cite{Ein2} was circular~\cite{Ives}. Ives' arguments were contested
            by later authors. See Ref.~\cite{Fadner} and references therein.})
           thought experiment relating (to the lowest order in $\beta^2$
           only) the change of mass of an object with radiated eletromagnetic energy
          is then unnecessary to establish the mass-energy equivalence. Although  Poincar\'{e}
           had noted as early as 1900 that the momentum of an electromagnetic wave is is
            $1/c^2$ times the energy flux given by the Poynting vector~\cite{HP1900}, suggesting, in view
            of the classical definition of momentum, a mass density  $1/c^2$ times the energy
            density, Einstein's equation (15.34) shows clearly the general equivalence of
           mass and energy as well as demonstrating that $c$ is the limiting speed of any object
           that has a time-like energy momentum 4-vector.  
   \par Einstein's relation for the radius of curvature of the orbit of an electron in 
     a uniform magnetic field was was obtained previously by Lorentz~\cite{Lor} and even
      demonstrated by him in the same paper to to be in reasonable agreement with measurements
      of Kaufmann dating from 1902~\cite{Kauf1902}. This means that some experimental evidence for the
      non-Newtonian mechanics of SRT existed even before SRT was invented! However other predictions
       based, like that of Lorentz, on specfic electrodynamical models of the electron due to
        Abraham and Bucherer could not be excluded in 1904. It was not until 1908 that the SRT
        prediction of (15.36) was experimentally verified by Bucherer~\cite{Buch1908}. 
    \par Einstein is now universally considered to be the founding father of SRT. Is this a fair
      assesment in view of the previously published work of Larmor and Lorentz,
       and the contemporary, but independent, work of Poincar\'{e}?  It is sometimes claimed that
        it was Einstein who introduced the Special Relativity Principle into physics and that this 
      principle was not (or only partially) understood by the other authors just cited.
      The present writer finds it
      very difficult to understand this, since this Principle was clearly stated by Galileo in a graphic
      manner~\cite{Galileo} and was also well known to Newton. Two lines of algebra suffice to
      to demonstrate the invariance of Newton's Second Law under Galilean transformations.
      This Principle ---that the mechanical laws of physics are the same in all inertial frames--- is true
      in both classical Newtonian mechanics and in relativistic mechanics. It is sometimes
      claimed that Einstein was unique in extending the Special Relativity Principle, known to hold in
     classical mechanics, to the `rest of physics', in particular to classical electrodynamics.
      However, from a modern viewpoint, many aspects of the latter are most easily understood in terms
     of the relativistic kinematics of real or virtual photons~\cite{JHFRCED,JHFSTF}
     Also there is a deep correspondence between classical electrodynamics and quantum mechanics~\cite{JHFEJP} 
     as illustrated by Eq.~(11.2) above.
  Thus, at the fundamental level, given the existence of photons, 
      there is no real distinction between, on the one hand, electrodynamics, and on the other
 either classical (but relativistic) mechanics
      and quantum mechanics.
      \par One important difference
      between Einstein, on the one hand, and Larmor, Lorentz and  Poincar\'{e} on the other, was that
      the latter authors were all working on aether theories or electrodynamic models of electron
      structure whereas Einstein
       clearly stated that the  aether was `superflous' in his approach. However Einstein still
      clung to the concept of classical fields, and of wave motion in such fields, whereas in
      all other domains of physics such fields require some material support for their existence.
      The modern solution, after rejecting the aether, is to embrace the photon concept that
      renders also `superfluous' electric and magnetic force fields~\cite{JHFRCED} and identifies
      radiation fields as the classical limit of quantum mechanical probability
       amplitudes~\cite{JHFEJP,JHFSTF}. Einstein never did this, either in Ref.~\cite{Ein1} or
      in any later work.
      \par Pro-Einstein advocates often state, with some justification, that he was the first to
      understand that SRT deals with space-time geometry and kinematics, not with dynamics.
      This seems not to be borne out by titles of the papers in which the related work was published:
      Larmor ,`On a dynamical theory of the electric and luminiferous
      medium'~\cite{Lar1897} and `Aether and Matter'~\cite{Lar1900};
      Lorentz,~\cite{Lor} `Electromagnetic phenomena in a system moving with any velocity
       less than that of light', Einstein,~\cite{Ein1} `On the electrodynamics of moving
      bodies'; Poincar\'{e}~\cite{HP1906} `On the dynamics of the electron'.
      The only paper that does not specifically mention dynamics is that of
     Lorentz. Given that the `moving body' introduced in the mechanical section \S 10 
     of Ref.~\cite{Ein1} is an `electron' the titles of the Einstein and  Poincar\'{e}
     papers are very similar. However large parts of the latter are an attempt
     to describe electrons, considered from a modern viewpoint as just one type of
     elementary particle among many others, as a structure generated by the electromagnetic
     field and other forces of unspecified origin. This subject was not considered by Einstein.
     However, Einstein's caveat concerning the physical meaning  of the differential equations
    of motion and the longitudinal and transverse masses of his `electron' indicates that he did not 
     fully appreciate, at the time of writing Ref.~\cite{Ein1}, the purely kinematical nature of
     these formulae.
    \par Looking at the list, (a)-(h), of formally correct predictions, only the formulae
     for Doppler shift and aberration, the radiation pressure formula and the mass-energy
     equivalence relation are to be found in Einstein's work and not that of either Larmor and
       Lorentz, and/or Poincar\'{e}. In particular, the predictions in (a), (b), (c), (g), (h) and
       (j) were also all given (at least as formulae) by Poincar\'{e} in Ref.\cite{HP1906}.
     \par Some writers on the history of SRT have tended to be unfair to either Poincar\'{e}
      ~\cite{Pais} or to Einstein~\cite{Whittaker,Logunov}, while they, and many others, have largely
      ignored the work of Larmor. A notable exception is to be found in Pauli's monograph
      on SRT written in 1921 and re-issued in English translation in 1958~\cite{Pauli}.
      In particular, Larmor's priority for the LT is pointed out.
      Also the space-time geometric aspects of SRT, usually associated with
      the work of Minkowski~\cite{ Mink}, as well as the Group Theory of the LT,
      both of which were first developed by Poincar\'{e} in Ref.~\cite{HP1906}, are cited by Pauli.
     \par The major difference between the work of Einstein and that of his contemporaries
       lies, however, not in the mathematical results obtained but rather in style and
       methodology ---the introduction of thought experiments, not only, as previously
       in popular scientific, but also in research literature, and concentration on
       essential points using the minimal amount of mathematics to resolve the physical
       problems addressed. In common with Newton is the use of the minimum number of,
       and simplest possible, initial postulates, from which the conclusions are deduced by
       mathematical logic\footnote{This point was particularly stressed by Holton~\cite{Holton}}.
       Einstein saw very clearly that the space-time LT, if it was not to remain a mathematical
       abstraction, had to describe real, physical, clocks. This important concept
       is to be found nowhere in work of Larmor and Lorentz, or (in contrast to his popular 
       essays) in the research papers
      of Poincar\'{e}. Such clocks were an essential part of the thought experiments
      discussed in  \S 1, \S 2, \S 3 and \S 4 of Ref.~\cite{Ein1}. Unfortunately,
       as pointed out in Sections 5, 6 and 7 above, both the light signal clock synchronisation
       procedure of \S2 and the physical meaning of the second postulate defined in this
       section are misinterpreted. Crucially, the additive constants `to be placed on the right 
       side of each of these (LT) equations' were also neglected, leading to a fallacious
       `derivation' of the LT and spurious LC and RS effects, although only the LC effect
       was discussed in Ref.~\cite{Ein1};
 the RS effect discussed there, though also of a spurious
       nature (see Section 5 above) having a different origin.
        \par In fact, the work of the present author, pointing out the spurious nature of the
          LC and RS effects, is doing nothing more than interpreting the space-time LT in 
        the way that it is said in Ref.~\cite{Ein1} that it must be  ---introducing additive
       constants to correctly describe synchronised clocks at different spatial 
        locations. It is perhaps surprising that Einstein did not do himself
        what is clearly stated must be done, if clocks at different positions are
       considered, in the passage quoted in Section 2 above. Even more surprising is
       that it has also been overlooked, for more than a century, by historians of science
        and practitioners of SRT
   
    \par In conclusion, in the present writer's opinion,
       the most important message contained in  Ref.~\cite{Ein1} is the
        necessity to modify the mechanical theory established by Newton by 
         redefining the Newtonian momentum and energy of any physical object
         according to Eqs.~(13.22) and (13.26) above. These formulae were not given explicitly
        in  Ref.~\cite{Ein1}, but, as shown above, were implicit, in many places, in the results
     presented in the paper,
 such as in the TD relation (15.5) leading to the velocity 4-vector of Eq.~(15.38), in the 
   transformation laws of the frequency (15.13) and the energy density (15.16) of a plane
    electromagnetic wave (a parallel beam of monochromatic photons), in the transformation
     laws of charge densities and currents (15.23)-(15.28) and the mass-energy equivalence
      relations (15.34)-(15.36). For this alone, and in spite of its many flaws, both of physical
       principles and in the mathematical derivations presented,  Ref.~\cite{Ein1} still
      remains a serious candidate to be considered the most important single physics paper to be 
     published during the 20th Century.



\newpage

{\bf Appendix}
\renewcommand{\theequation}{E1.\arabic{equation}}
\setcounter{equation}{0}
    \par {\bf \S 1. Definition of Simultaneity}
            \begin{equation}
           t_B -t_A = t'_A -t_B
             \end{equation}
      \begin{equation}
           \frac{2 AB}{t'_A -t_B} = c
         \end{equation}

    \par {\bf \S 2. On the Relativity of Lengths and Times}
\renewcommand{\theequation}{E2.\arabic{equation}}
\setcounter{equation}{0}
 \begin{equation}
{\tt velocity} = \frac{{\tt light~path}}{{\tt time~interval}} = c
  \end{equation}
\begin{equation}
 t_B -t_A = \frac{r_{AB}}{c-v}
 \end{equation}
\begin{equation}
t'_A -t_B = \frac{r_{AB}}{c+v}
\end{equation} 
   \par {\bf \S 3. Theory of the Transformation of Co-ordinates and Times from a Stationary
 System to another System in Uniform Motion of Translation Relative to the Former}
\renewcommand{\theequation}{E3.\arabic{equation}}
\setcounter{equation}{0}
 \begin{equation}
x' \equiv x-vt = x(t=0) = {\rm~constant}
  \end{equation}
\begin{equation}
\frac{1}{2}(\tau_0 +\tau_2) = \tau_1
 \end{equation}
\begin{equation}
\frac{1}{2}[ \tau(0,0) + \tau(0,\frac{x'}{c-v}+ \frac{x'}{c+v})] = \tau(x',\frac{x'}{c-v})
 \end{equation}
\begin{equation}
  \frac{1}{2}\left(\frac{1}{c-v}+\frac{1}{c+v}\right) \frac{\partial \tau}{\partial t}
           =  \frac{\partial \tau}{\partial x'}+\frac{1}{c-v} \frac{\partial \tau}{\partial t'}
  \end{equation} 
\begin{equation}
 \frac{\partial \tau}{\partial x'} + \frac{v}{c^2-v^2}\frac{\partial \tau}{\partial x} = 0
 \end{equation}      
\begin{equation}
   \frac{\partial \tau}{\partial y} = 0,~~~\frac{\partial \tau}{\partial z} = 0
   \end{equation} 
\begin{equation}
\tau = a \left(t- \frac{v}{c^2-v^2}x'\right)
 \end{equation}
\begin{equation}
\xi = c \tau = a c\left(t_B- \frac{v x'_B}{c^2-v^2} \right)
 \end{equation}
\begin{equation}
 t_B = \frac{x'_B}{c-v}
 \end{equation} ]
\begin{equation}
  \xi_B = a\frac{c^2}{c^2-v^2} x'_B
 \end{equation} 
\begin{equation}
 \eta = c \tau = a c\left(t_B- \frac{v x'}{c^2-v^2} \right)
 \end{equation}
\begin{equation}
 \eta = \frac{a c}{\sqrt{c^2-v^2}} y
 \end{equation} 
\begin{equation}
\zeta = \frac{a c}{\sqrt{c^2-v^2}} z 
 \end{equation} 
  \begin{eqnarray}
    \tau & = & \phi(v) \gamma (t-\frac{v x}{c^2})  \\
         &   &  \nonumber  \\
    \xi & = & \phi(v) \gamma (x-vt)
    \end{eqnarray}  
\begin{equation}
\eta =  \phi(v) y
 \end{equation} 
\begin{equation}
\zeta = \phi(v) z
 \end{equation} 
\begin{equation}
\gamma = \frac{1}{\sqrt{1-(v/c)^2}}
\end{equation}
\begin{equation}
 x^2 + y^2 + z^2  = c^2 t^2
\end{equation}
 \begin{equation}
  \xi^2 + \eta^2 + \zeta^2  = c^2 \tau^2
 \end{equation}
  \begin{eqnarray}
    t'& = &  \phi(-v) \gamma(-v)(\tau+v \xi /c^2) = \phi(v) \phi(-v) t                       \\
 &   &  \nonumber  \\
    x'& = &  \phi(-v) \gamma(-v)(\xi+v \tau) = \phi(v) \phi(-v) x                       \\
 &   &  \nonumber  \\
   y'& = &  \phi(-v) \eta  = \phi(v) \phi(-v) y                     \\
 &   &  \nonumber  \\
  z'& = &  \phi(-v) \zeta  = \phi(v) \phi(-v)                        
   \end{eqnarray}  
 \begin{equation}
 \phi(v)\phi(-v) = 1
 \end{equation}
 \begin{equation}
\phi(v) =   \phi(-v)
 \end{equation}
     
  \begin{eqnarray}
    \tau & = & \gamma(t-\frac{v x}{c^2})   \\
 &   &  \nonumber  \\
    \xi & = & \gamma(x-vt)                  \\
 &   &  \nonumber  \\
    \eta & = & y                       \\
 &   &  \nonumber  \\
    \zeta & = & z                      
    \end{eqnarray}  

  \par {\bf \S 4. Physical Meaning of the Equations Obtained In Respect to Moving Rigid Bodies and Moving
   Clocks}
\renewcommand{\theequation}{E4.\arabic{equation}}
\setcounter{equation}{0}
 \begin{equation}
\xi^2+\eta^2+ \zeta^2 = R^2
 \end{equation}
   \begin{equation}
 \frac{x^2}{1-(v/c)^2} + y^2 +z^2 = R^2
 \end{equation}
 \begin{equation}
 \tau =\frac{1}{\sqrt{1-(v/c)^2}}(t-\frac{v x}{c^2})
 \end{equation}
   \begin{equation}
  \tau = t \sqrt{1-(v/c)^2} = t - (1- \sqrt{1-(v/c)^2})t
   \end{equation}

 \par {\bf \S 5. The Composition of Velocities}
\renewcommand{\theequation}{E5.\arabic{equation}}
\setcounter{equation}{0}
 \begin{equation}
 \xi = \omega_{\xi}\tau,~~~~ \eta =  \omega_{\eta}\tau,~~~~\zeta = 0
 \end{equation}
   \begin{eqnarray}
 x & = & \frac{(\omega_{\xi}+v)t}{1+\frac{v \omega_{\xi}}{c^2}} \\
 &   &  \nonumber  \\
 y & = & \omega_{\eta}t \frac{\sqrt{1-(v/c)^2}}{1+\frac{v \omega_{\xi}}{c^2}} \\
 &   &  \nonumber  \\
 z & = & 0
 \end{eqnarray}

   \begin{eqnarray}
 V^2 & = & \left(\frac{d x}{d t}\right)^2 + \left(\frac{d y}{d t}\right)^2 \\
 &   &  \nonumber  \\
 \omega^2 & = & \omega_{\xi}^2 + \omega_{\eta}^2 \\
 &   &  \nonumber  \\
 \alpha & = & \tan^{-1}(\omega_{\eta}/\omega_{\xi})
 \end{eqnarray}
\begin{equation}
V = \frac{\sqrt{v^2 + \omega^2 + 2 v \omega \cos \alpha-[(v \omega \sin \alpha)/c]^2}}
       {1 + \frac{v \omega \cos \alpha}{c^2}}
\end{equation}
\begin{equation}
V = \frac{v + \omega}{1+\frac{v  \omega}{c^2}}
\end{equation}
\begin{equation}
V = c \frac{2c-\kappa -\lambda}{2c-\kappa -\lambda - \kappa \lambda/c} < c
\end{equation}
\begin{equation}
V = \frac{c + \omega}{1+\frac{\omega}{c}} = c
\end{equation}

\par {\bf \S 6. Transformation of the Maxwell-Hertz Equations for Empty Space. On the Nature
    of the Electromotive Forces in a Magnetic Field During Motion}
\renewcommand{\theequation}{E6.\arabic{equation}}
\setcounter{equation}{0}
  \begin{eqnarray}
   \frac{1}{c} \frac{\partial E_x}{\partial t} & = &  \frac{\partial B_z}{\partial y}- \frac{\partial B_y}{\partial z},
   ~~~~ \frac{1}{c} \frac{\partial B_x}{\partial t} = 
   \frac{\partial E_y}{\partial z}- \frac{\partial E_z}{\partial y} \\
   \frac{1}{c} \frac{\partial E_y}{\partial t} & = &  \frac{\partial B_x}{\partial z}- \frac{\partial B_z}{\partial x},
   ~~~~ \frac{1}{c} \frac{\partial B_y}{\partial t} = 
   \frac{\partial E_z}{\partial x}- \frac{\partial E_x}{\partial z} \\
  \frac{1}{c} \frac{\partial E_z}{\partial t} & = &  \frac{\partial B_y}{\partial x}- \frac{\partial B_x}{\partial y},
   ~~~~ \frac{1}{c} \frac{\partial B_z}{\partial t} = 
   \frac{\partial E_x}{\partial y}- \frac{\partial E_y}{\partial x} 
    \end{eqnarray}

  \begin{eqnarray}
 & &\frac{1}{c} \frac{\partial E_x}{\partial \tau} = \frac{\partial ~}{\partial \eta}
 [\gamma(B_z-\beta E_y)]- \frac{\partial ~}{\partial \zeta}[\gamma(B_y+\beta E_z)] \\
 & & \frac{1}{c}\frac{\partial ~}{\partial \tau}[\gamma({\rm E}_y-\beta B_z)] =  \frac{\partial B_x}{\partial \zeta}
 - \frac{\partial ~}{\partial \xi}[\gamma(B_z-\beta E_y)] \\
 & & \frac{1}{c}\frac{\partial ~}{\partial \tau}[\gamma({\rm E}_z+\beta B_y)] = 
  \frac{\partial ~}{\partial \zeta}[\gamma(B_y+\beta E_z)]- \frac{\partial B_x}{\partial \eta} \\
 & &\frac{1}{c} \frac{\partial B_x}{\partial \tau} = \frac{\partial ~}{\partial \zeta}
 [\gamma({\rm E}_y-\beta B_z)]- \frac{\partial ~}{\partial \eta}[\gamma({\rm E}_z+\beta B_y)] \\
 & & \frac{1}{c}\frac{\partial ~}{\partial \tau}[\gamma(B_y+\beta E_z)] = 
  \frac{\partial ~}{\partial \xi}[\gamma({\rm E}_z+\beta B_y)]- \frac{\partial E_x}{\partial \zeta} \\
 & & \frac{1}{c}\frac{\partial ~}{\partial \tau}[\gamma(B_z-\beta E_y)] = \frac{\partial E_x}{\partial \eta}
  -\frac{\partial ~}{\partial \xi}[\gamma({\rm E}_y-\beta B_z)]           
   \end{eqnarray}

  \begin{eqnarray}
   \frac{1}{c} \frac{\partial E'_{\xi}}{\partial \tau} & = &  \frac{\partial B'_{\zeta}}{\partial \eta}
  - \frac{\partial B'_{\eta}}{\partial \zeta},
   ~~~~ \frac{1}{c} \frac{\partial B'_{\xi}}{\partial \tau} = 
   \frac{\partial E'_{\eta}}{\partial \zeta}- \frac{\partial E'_{\zeta}}{\partial \eta} \\
   \frac{1}{c} \frac{\partial E'_{\eta}}{\partial \tau} & = &  \frac{\partial B'_{\xi}}{\partial \zeta}
   - \frac{\partial B'_{\zeta}}{\partial \xi},
   ~~~~ \frac{1}{c} \frac{\partial B'_\eta}{\partial \tau} = 
   \frac{\partial E'_{\zeta}}{\partial \xi}- \frac{\partial E'_{\xi}}{\partial \zeta} \\
  \frac{1}{c} \frac{\partial E'_{\zeta}}{\partial \tau} & = &  \frac{\partial B'_{\eta}}{\partial \xi}
   - \frac{\partial B'_{\xi}}{\partial \eta},
   ~~~~ \frac{1}{c} \frac{\partial B'_{\zeta}}{\partial \tau} =    \frac{\partial E'_{\xi}}{\partial \eta}
   - \frac{\partial E'_{\eta}}{\partial \xi}
    \end{eqnarray}

    \begin{eqnarray}
  E'_{\xi} & = & \psi(v) E_x,~~~~ B'_{\xi} =\psi(v) B_x \\
  E'_{\eta} & = & \psi(v) \gamma [E_y-\beta B_z],~~~~B'_{\eta} =\psi(v) \gamma [B_y+\beta E_z] \\
   E'_{\zeta} & = & \psi(v) \gamma [E_z+\beta B_y],~~~~B'_{\zeta} =\psi(v) 
\gamma [B_z-\beta E_y]
  \end{eqnarray}
\begin{equation}
\psi(v)\psi(-v) = 1,~~~~\psi(v)= \psi(-v),~~~~\psi(v)= 1
\end{equation}
    \begin{eqnarray}
  E'_{\xi} & = & E_x,~~~~ B'_{\xi} = B_x \\
  E'_{\eta} & = & \gamma [E_y-\beta B_z],~~~~B'_{\eta} = \gamma [B_y+\beta E_z] \\
   E'_{\zeta} & = & \gamma [E_z+\beta B_y],~~~~B'_{\zeta} = \gamma [B_z-\beta E_y]
  \end{eqnarray}

\par {\bf \S 7. Theory of Doppler's Principle and Aberration}
\renewcommand{\theequation}{E7.\arabic{equation}}
\setcounter{equation}{0} 
   \begin{eqnarray}
     E_x & = & E_x^0 \sin \Phi,~~~~~ B_x = B_x^0 \sin \Phi \\  
    E_y & = & E_y^0 \sin \Phi,~~~~~ B_y = B_y^0 \sin \Phi \\  
    E_z & = & E_z^0 \sin \Phi,~~~~~ B_z = B_z^0 \sin \Phi 
   \end{eqnarray}
\begin{equation}
\Phi = \omega \{t-\frac{1}{c}(l x +m y + n z)\}
\end{equation}
  
       \begin{eqnarray}
  E'_{\xi} & = & E_x^0 \sin \Phi',~~~~ B'_{\xi} = B_x^0 \sin \Phi' \\
 &   &  \nonumber  \\
  E'_{\eta} & = & \gamma [E_y^0-\beta B_z^0] \sin \Phi',
 ~~~~B'_{\eta} = \gamma [B_y^0+\beta E_z^0]\sin \Phi' \\
 &   &  \nonumber  \\
   E'_{\zeta} & = & \gamma [E_z^0+\beta B_y^0]\sin \Phi',
 ~~~~B'_{\zeta} = \gamma [B_z^0-\beta E_y^0]\sin \Phi'
  \end{eqnarray}
\begin{equation}
\Phi' = \omega'\{\tau-\frac{1}{c}(l' \xi +m' \eta + n' \zeta)\}
\end{equation}

   \begin{eqnarray}
    \omega' & = & \omega \gamma(1-\beta l) \\
 &   &  \nonumber  \\
     l'& = & \frac{l-\beta}{1-\beta l} \\
 &   &  \nonumber  \\
     m'& = & \frac{m}{\gamma(1-\beta l)} \\
 &   &  \nonumber  \\
     n'& = & \frac{n}{\gamma(1-\beta l)} \\
 &   &  \nonumber  \\
      \nu' & = & \frac{\nu(1-\beta \cos \phi)}{\sqrt{1-\beta^2}} \\
 &   &  \nonumber  \\
      \nu' & = & \nu \sqrt{\frac{1-\beta}{1+\beta}} \\
 &   &  \nonumber  \\
      \cos \phi' & = & \frac{\cos \phi- \beta}{1- \beta \cos \phi} \\ 
 &   &  \nonumber  \\
       cos \phi' &  = & -\beta \\  
 &   &  \nonumber  \\
     (A')^2 & = & \frac{A^2(1-\beta \cos \phi)^2}{1-\beta^2} \\
 &   &  \nonumber  \\
    (A')^2 & = & \frac{A^2(1-\beta)}{1+\beta}
     \end{eqnarray}

  \par {\bf \S 8. Transformation of the Energy of Light Rays. Theory of the Pressure of Radiation Exerted on Perfect Reflectors}
\renewcommand{\theequation}{E8.\arabic{equation}}
\setcounter{equation}{0} 
\begin{equation}
(x-l c t)^2 + (y-m c t)^2+(z-n c t)^2 = R^2
 \end{equation}
\begin{equation}
 \xi^2\frac{(1-\beta l)^2}{1-\beta^2}+(\eta- m \gamma \beta \xi)^2 + (\zeta - n \gamma \beta \xi)^2
    = R^2
 \end{equation}
 \begin{equation}
\frac{V'}{V} = \frac{\sqrt{1-\beta^2}}{1- \beta \cos \phi}
 \end{equation}
 \begin{equation}
 \frac{E'}{E} = \frac{(A')^2 V'}{A^2 V} = \frac{1-\beta \cos \phi}{\sqrt{1-\beta^2}}
 \end{equation} 
  \begin{equation}
 \frac{E'}{E} = \sqrt{\frac{1-\beta}{1+\beta}}
 \end{equation} 
 \begin{equation}
A' =  A \frac{1-\beta \cos \phi}{\sqrt{1-\beta^2}}
 \end{equation} 
  \begin{equation}
  \cos \phi' = \frac{\cos \phi- \beta}{1- \beta \cos \phi}
 \end{equation} 
  \begin{equation}
\nu' = \nu \frac{1-\beta \cos \phi}{\sqrt{1-\beta^2}}
 \end{equation} 
      \begin{eqnarray}
       A'' & = & A' \\
 &   &  \nonumber  \\
   &  &  \nonumber \\
       \cos \phi'' & = &  -\cos \phi' \\ 
 &   &  \nonumber  \\
   &  &  \nonumber \\
       \nu'' & = & \nu'
      \end{eqnarray} 
       
   \begin{eqnarray}
       A''' & = &   A'' \frac{1+\beta \cos \phi''}{\sqrt{1-\beta^2}}
           =  A \frac{(1-2\beta \cos \phi+ \beta^2)}{1-\beta^2} \\
    &  &  \nonumber \\
         \cos \phi'''& = &  \frac{\cos \phi''+ \beta}{1+ \beta \cos \phi''}
          = \frac{2 \beta -(1+\beta^2) \cos \phi}{1-2\beta \cos \phi+ \beta^2} \\
    &  &  \nonumber \\
   \nu''' & = &   \nu'' \frac{1+\beta \cos \phi''}{\sqrt{1-\beta^2}}
           =   \nu \frac{(1-2\beta \cos \phi+ \beta^2)}{1-\beta^2}  
      \end{eqnarray}     
  \begin{equation}
  f_E = \frac{A^2(c \cos \phi - v)}{8 \pi}
  \end{equation}
   \begin{equation}
 f_{E'''} = \frac{(A''')^2(-c \cos \phi''' + v)}{8 \pi}
 \end{equation}
   \begin{equation}
   P = 2 \frac{A^2}{8 \pi} \frac{(\cos \phi- \beta)^2}{1-\beta^2}
 \end{equation}
 \begin{equation}
P = 2 \frac{A^2}{8 \pi} \cos ^2 \phi
 \end{equation}

  \par {\bf \S 9. Transformation of the Maxwell-Hertz Equations when Convection-Currents are
     Taken into Account}
\renewcommand{\theequation}{E9.\arabic{equation}}
\setcounter{equation}{0} 
  
  \begin{eqnarray}
   \frac{1}{c} \left\{ \frac{\partial E_x}{\partial t}+ u_x \rho \right\} & = &  \frac{\partial B_z}{\partial y}- \frac{\partial B_y}{\partial z},
   ~~~~ \frac{1}{c} \frac{\partial B_x}{\partial t}  = 
   \frac{\partial E_y}{\partial z}- \frac{\partial E_z}{\partial y} \\
   \frac{1}{c} \left\{ \frac{\partial E_y}{\partial t} + u_y \rho \right\} & = &  \frac{\partial B_x}{\partial z}- \frac{\partial B_z}{\partial x},
   ~~~~ \frac{1}{c} \frac{\partial B_y}{\partial t} = 
   \frac{\partial E_z}{\partial x}- \frac{\partial E_x}{\partial z} \\
  \frac{1}{c}\left\{ \frac{\partial E_z}{\partial t} + u_zx \rho \right\} & = &  \frac{\partial B_y}{\partial x}- \frac{\partial B_x}{\partial y},
   ~~~~ \frac{1}{c} \frac{\partial B_z}{\partial t} = 
   \frac{\partial E_x}{\partial y}- \frac{\partial E_y}{\partial x} 
    \end{eqnarray}
 \begin{equation}
  \rho  = \frac{\partial E_x}{\partial x} +  \frac{\partial E_y}{\partial y}
      +  \frac{\partial E_z}{\partial z}
 \end{equation}

  \begin{eqnarray}
   \frac{1}{c}\left\{ \frac{\partial E'_{\xi}}{\partial \tau} + u_{\xi} \rho' \right\} & = &  \frac{\partial B'_{\zeta}}{\partial \eta}
  - \frac{\partial B'_{\eta}}{\partial \zeta},
   ~~~~ \frac{1}{c} \frac{\partial B'_{\xi}}{\partial \tau} = 
   \frac{\partial E'_{\eta}}{\partial \zeta}- \frac{\partial E'_{\zeta}}{\partial \eta} \\
   \frac{1}{c}\left\{ \frac{\partial E'_{\eta}}{\partial \tau} + u_{\eta} \rho' \right\} & = &  \frac{\partial B'_{\xi}}{\partial \zeta}
   - \frac{\partial B'_{\zeta}}{\partial \xi},
   ~~~~ \frac{1}{c} \frac{\partial B'_\eta}{\partial \tau} = 
   \frac{\partial E'_{\zeta}}{\partial \xi}- \frac{\partial E'_{\xi}}{\partial \zeta} \\
  \frac{1}{c}\left\{ \frac{\partial E'_{\zeta}}{\partial \tau} + u_{\zeta} \rho' \right\} & = &  \frac{\partial B'_{\eta}}{\partial \xi}
   - \frac{\partial B'_{\xi}}{\partial \eta},
   ~~~~ \frac{1}{c} \frac{\partial B'_{\zeta}}{\partial \tau}  =    \frac{\partial E'_{\xi}}{\partial \eta}
   - \frac{\partial E'_{\eta}}{\partial \xi} 
    \end{eqnarray}
   
 \begin{eqnarray}
  u'_{\xi} &  =  & \frac{u_x-v}{1-u_x v /c^2} \\   
 u'_{\eta} &  =  & \frac{u_y}{\gamma(1-u_x v /c^2)} \\ 
u'_{\zeta} &  =  & \frac{u_z}{\gamma(1-u_x v /c^2)} 
\end{eqnarray} 
\begin{equation}
\rho'  = \frac{\partial E'_{\xi}}{\partial \xi} +  \frac{\partial E'_{\eta}}{\partial \eta}
      +  \frac{\partial E'_{\zeta}}{\partial \zeta} = \gamma(1-u_x v /c^2) \rho
\end{equation}

\par {\bf \S 10. Dynamics of the Slowly Accelerated Electron} 
\renewcommand{\theequation}{E10.\arabic{equation}}
\setcounter{equation}{0} 
  \begin{eqnarray}
   m\frac{d^2 x}{d t^2} & = & e E_{x} \\
   m\frac{d^2 y}{d t^2} & = & e E_{y} \\
 m\frac{d^2 z}{d t^2} & = & e E_{z}
 \end{eqnarray}
   \begin{eqnarray}
   m\frac{d^2\xi}{d \tau^2} & = & e E'_{\xi} \\
   m\frac{d^2\eta}{d \tau^2} & = & e E'_{\eta} \\
 m\frac{d^2\zeta}{d \tau^2} & = & e E'_{\zeta}
 \end{eqnarray}
\begin{equation}
 \xi = \gamma(v-vt),~~~\eta = y,~~~\zeta = z,~~~\tau = \gamma(t-vx/c^2)
\end{equation}
\begin{equation}
E'_{\xi} = E_x,~~~ E'_{\eta} = \gamma(E_y-\beta B_z),~~~E'_{\zeta} = \gamma(E_z+\beta B_y)
\end{equation}
   \begin{eqnarray}
  \frac{d^2 x}{d t^2} & = & \frac{e}{m\gamma^3} E_x \\  
  \frac{d^2 y}{d t^2} & = & \frac{e}{m\gamma}({\rm E}_y-\beta B_z) \\ 
 \frac{d^2 z}{d t^2} & = & \frac{e}{m\gamma}({\rm E}_z+\beta B_y) \\
  m\gamma^3\frac{d^2 x}{d t^2} & = & e E_x = e E'_{\xi} \\  
 m\gamma^2\frac{d^2 y}{d t^2} & = & e \gamma(E_y-\beta B_z) = e E'_{\eta} \\ 
 m\gamma^2\frac{d^2 z}{d t^2} & = & e \gamma(E_z+\beta B_y) = e E'_{\zeta}
 \end{eqnarray}
\begin{equation}
 {\rm Longitudinal~mass~} = \frac {m}{(\sqrt{1-\beta^2})^3}
\end{equation}
\begin{equation}
{\rm~~~~~ Transverse~mass~} = \frac {m}{1-\beta^2}
\end{equation}
\begin{equation}
W = \int e E_x d x = m \int_0^v \gamma^3 v dv =
    mc^2\left\{\frac{1}{\sqrt{1-\beta^2}}-1\right\}
\end{equation}
\begin{equation}
\frac{A_m}{A_e} = \frac{v}{c}
\end{equation}
\begin{equation}
{\cal P} = \int E_x d x = \frac{m}{e}c^2\left\{\frac{1}{\sqrt{1-\beta^2}}-1\right\}
\end{equation}
\begin{equation}
-\frac{d^2 y}{d x^2} = \frac{v^2}{R} = \frac{e}{m}\beta B_z \sqrt{1-\beta^2}
\end{equation}
\begin{equation}
R = \frac{m c^2}{e}\frac{\beta}{\sqrt{1-\beta^2}} \frac{1}{B_z}
\end{equation}
\pagebreak

\end{document}